\documentclass[12pt]{iopart}
\usepackage{iopams}  
\usepackage{graphicx,xcolor}
\usepackage{bbm}
\usepackage[bottom=2.cm]{geometry}

\DeclareMathAlphabet\mathbfcal{OMS}{cmsy}{b}{n}

\expandafter\let\csname equation*\endcsname\relax 
\expandafter\let\csname endequation*\endcsname\relax
\usepackage{amsmath}
\usepackage[colorlinks=true,linkcolor=blue]{hyperref}
\usepackage[titletoc]{appendix}
\begin{document}

\title[Divergence-free cQED]{Quantum Networks in Divergence-free Circuit QED}

\author{A Parra-Rodriguez$^{1}$, E Rico$^{1,2}$, E Solano$^{1,2,3}$ and I L Egusquiza$^{4}$}

\address{$^1$Department of Physical Chemistry, University of the Basque Country UPV/EHU, Apartado 644, 48080 Bilbao, Spain\\
		$^2$IKERBASQUE, Basque Foundation for Science, Maria Diaz de Haro 3, 48013 Bilbao, Spain,\\
		$^3$Department of Physics, Shanghai University, 200444 Shanghai, China\\
		$^4$Department of Theoretical Physics and History of Science, University of the Basque Country UPV/EHU, Apartado 644, 48080 Bilbao, Spain}

\begin{abstract}
Superconducting circuits are one of the leading quantum platforms for quantum technologies. With growing system complexity, it is of crucial importance to develop scalable circuit models that contain the minimum information required to predict the behaviour of the physical system. Based on microwave engineering methods, divergent and non-divergent Hamiltonian models in circuit quantum electrodynamics have been proposed to explain the dynamics of superconducting quantum networks coupled to infinite-dimensional systems, such as transmission lines and general impedance environments. Here, we study systematically common linear coupling configurations between networks and infinite-dimensional systems. The main result is that the simple Lagrangian models for these configurations present an intrinsic natural length that provides a natural ultraviolet cutoff. This length is due to the unavoidable dressing of the environment modes by the network. In this manner, the coupling parameters between their components correctly manifest their natural decoupling at high frequencies.  Furthermore, we show the requirements to correctly separate infinite-dimensional coupled systems in local bases. We also compare our analytical results with other analytical and approximate methods available in the literature. Finally, we propose several applications of these general methods to analog quantum simulation of multi-spin-boson models in non-perturbative coupling regimes.

\end{abstract}

\maketitle
\tableofcontents
\newpage
\pagestyle{plain}
\section[Introduction]{Introduction}
Circuit Quantum Electrodynamics (cQED) has been proposed as an implementation for developing quantum technologies. In particular, quantum computation algorithms have already been developed in this platform, where qubits of information are encoded in artificial atoms relying on  Josephson junctions, and bosonic electromagnetic modes in resonators play the role of quantum buses. In addition, cQED has already met most of the DiVincenzo's criteria for an universal quantum computer, with the coherence time as the critical factor to improve in order to mantain computation fidelities with increasing number of quantum gates. A special characteristic of these systems is that light (bosonic modes) and matter (qubits) can be designed to interact ultra-strongly, thus behaving as natural analog quantum simulators. However, this characteristic has a flip side, since  noise channels are also proportional to strength of this coupling. Different approaches have been suggested to overcome this issue such as  quantum control or quantum filtering. However, most of these approaches rely on accurate circuital models that allow the prediction and engineering of the coupling of networks of partite quantum subsystems. 

When dealing with networks of superconducting qubits, several methods have been used to derive first-principle quantum Hamiltonian descriptions to describe the effective dynamics and statistical properties observed in the experiments. The two seminal works in this field correspond to the ``Quantum Network Theory'' derived by Yurke and Denker \cite{YurkeDenker_1984_QNT} and the systematic Hamiltonian description of Devoret ``Quantum Fluctuations in Superconducting Circuits'' \cite{Devoret_1995_QFluct}. In the former, the basic rules for first principles circuit quantization of linear and non-linear elements was presented in order to derive input-ouput relations of charge operators. However, no general Hamiltonian description was derived and thus the dynamics of the conjugate flux operators were not shown. On the other hand, dissipative elements were introduced in an analog manner to the Caldeira-Leggett model \cite{CaldeiraLeggett_1983} with semi-infinte transmission lines, and equivalent results to those of Caves \cite{Caves_1980} on the noise-amplification relation were obtained. The second crucial reference \cite{Devoret_1995_QFluct} provided us with general rules to derive Hamiltonians of lumped electrical elements, and made use of the Caldeira-Leggett model to describe a closed Hamiltonian for an LC-oscillator inductively coupled to the impedance environment, which was replaced by infinite harmonic oscillators. These general rules for quantizing lumped-element circuits were later extended with systematic approaches for commonly-used classes of circuits \cite{BKD_2004,Burkard_2005}.

In \cite{ChakravartySchmid_1986}, Chakravarty and Schmid presented for the first time the action of a transmission line in the form that is used nowadays, when describing the input-output impedance of an open transmission line with a Josephson impurity. They also used the path integral formulation to derive the spectral density, akin to the results  of Leggett \cite{Leggett_1984}. They obtained for that case a quadratic behaviour for small frequencies and linear growth for large frequency.

Quite some time later, in 2004, Blais et al. [9] quantized the transmission line following the canonical quantization procedure of Devoret [2] for the continuum of harmonic oscillators. In an appendix they considered multiple modes, and pointed out that there exists a physical ultraviolet cutoff in the modes because the system cannot be exactly one dimensional. It is implicit in their work  that the coupling of the modes to the qubit scale in the form $g_k\sim\sqrt{\omega_k}$. This approximate approach to quantize circuits with tranmission lines that requires frequency cutoffs for the multi-mode Rabi Hamiltonian in cQED has routinely been used in theory \cite{Bourassa_2009} and experiments \cite{Houck_2008,Filipp_2011,Sundaresan_2015}. Without the phenomenological introduction of several types of mode truncations, such models, in which the scaling  $g_k\sim\sqrt{\omega_k}$ is present, would have predicted divergent Lamb-shifts \cite{Lamb_1947} or effective qubit-qubit couplings in the dispersive approximation. Notice that mode truncation can be circumvented for   qubit decay constants, in that they are computed to have a finite value  by carrying out a Markov approximation, as in the result of Wigner and Weisskopf \cite{Weisskopf_1930}. In this regard, see \cite{Moein_2017_CutFree} for an explicit presentation. Yet again, one needs adjustments and further approximations in order to recover finite predictions for these physical quantities

The study of different classes of circuits resulted in other  models that would not present divergent predictions for observables. That is the case in the work of Bourassa et al. \cite{Bourassa_2012}, where a transmission line with an inline transmon was studied. Although not explicitly shown, the modified normal modes of that system would be coupled with monotonically decaying constants above certain saturation frequency. There  was however no strict separation of anharmonic and harmonic degrees of freedom and thus a simple multi-mode quantum Rabi model could not be recovered  \cite{Bourassa_2012}. Similar approaches have been used to describe linear electromagnetic environments with impedance black-boxes connected to non-linear elements at their ports \cite{Nigg_2013,Solgun_2014,Solgun_2015} in what has been termed ``black-box quantization''. Again, in \cite{Nigg_2013}, Nigg et al. generalized the concept introduced in \cite{Bourassa_2012} where the degree of freedom corresponding to the flux/phase differences across Josephson junctions could effectively interact with general electromagnetic normal modes. This method has proved to be the best way to describe the physics of Josephson junctions inside  3D cavities, see e.g. \cite{Kirchmair_2013,Shankar_2013,Vlastakis_2013}, due to the fact that the interaction between the harmonic modes and the non-linear element cannot be assumed to be local. The black-box techniques do also take into account that the electromagnetic modes have finite bandwidth, because the cavity has open ports from which  energy flows away. Such procedures require finding a discrete equivalent lumped element circuit to simulate the linear response of the cavity from the point of the junctions. An approximate Foster decomposition of the impedance was used in \cite{Nigg_2013}, whereas a more involved and accurate Brune decomposition was used in \cite{Solgun_2014,Solgun_2015} to describe one and multiport general non-reciprocal passive electromagnetic environments. However, the quantization methods were constrained to the reciprocal set of networks, leaving the quantum description of the general non-reciprocal ones as an open problem. In \cite{Nigg_2013}, the Josephson fluxes were included in the linear description of the whole system, while non linear couplings appeared on expanding the cosine potentials  in the normal mode basis. However, in \cite{Solgun_2014,Solgun_2015}, such Josephson variables where kept independent, thus reaching Hamiltonian models with both linear capacitive and inductive couplings between the normal modes of the environment and the anharmonic variables.

Although the black-box methods above have been very successful in describing experiments, it has not been hitherto clear how such linear systems can be later coupled to other systems. Thus, in this paper we focus on
techniques that involve coupling linear systems with infinite modes to lumped element quantum circuits. As is only to be expected, this entails some dressing of the infinite modes, that are ineluctably modified by the coupling. 

The first quantization of a general impedance, modelled as an infinite series of harmonic oscillators, capacitively coupled to Josephson junctions was derived by Paladino et al. \cite{Paladino_2003}. A complete Hamiltonian without the diamagnetic $A^2$-term was derived by using the correct basis of harmonic variables. It was also noted that there is no need to add counterterms to the Hamiltonian because in the experiment there is only access to renormalized parameters. In fact, the model developed there allows the  engineering of the system, as the Hamiltonian is written in terms of the bare parameters. It was however not shown that the coupling of the anharmonic to the harmonic degrees of freedom would decay above certain frequency. Using similar techniques and based on Paladino et al., it was later noted by Bergenfeldt and Samuelsson \cite{Bergenfeldt_2012_DivFreeQDot}, that using the continuous Lagrangian and canonical quantization procedure for a system with a 1D transmission line resonator capacitively coupled to a quantum dot could be bipartite-diagonalized and that the capacitive linear coupling constants would monotonically decay for high-frequency modes. Other methods to derive non-divergent quantum-Rabi Hamiltonians for tranmission line resonators coupled to Josephson junctions where also developed by Malekakhlagh and T\"ureci in \cite{Moein_2016} and by Mortensen at al. \cite{Mortensen_2016_NM_TL}, although the non-divergent characteristic of the coupling constants was not then explored and explained. Recently, two works \cite{Gely_2017_DivFree,Moein_2017_CutFree} have independently been able to explain the mechanisms by which the infinite degree of freedom in a transmission line resonator decouple above certain mode when they are capacitively connected to a Josephson junction, without making any assumptions on the circuit parameters. The first method uses the Foster mode decomposition to describe the impedance of a transmission line resonator, in an similar manner to \cite{Paladino_2003}, whereas the second used the previous results achieved in \cite{Moein_2016}. Finally, we remark that non-divergent but approximate methods to describe Josephson junctions capacitively coupled to transmission lines have been studied by Koch et al. \cite{Koch_2010} and Peropadre et al. \cite{Peropadre_2013} among others.

In this article, we generalize the ideas introduced by Paladino et al. \cite{Paladino_2003}, Bourassa et al. \cite{Bourassa_2012} and Malekakhlagh et al. in \cite{Moein_2016,Moein_2017_CutFree}, following and extending well based mathematical machinery \cite{Walter_1973_EVP} previously used by some of us \cite{Adrian_2016_MT}. This is done to describe general networks of superconducting circuits that include circuital elements supporting infinite modes, such as  transmission lines of finite, semi-infinite and infinite length, and general impedances, coupled to lumped-element networks capacitively, inductively and galvanically. Using the theory of eigenvalue problems with the eigenvalue in the boundary condition, for whose expansion theorems we develop a new proof,  we recover the results achieved in \cite{Gely_2017_DivFree} with the Foster-decomposition method and in \cite{Moein_2016,Moein_2017_CutFree} with a regularization technique on the space-local interactions. We identify and solve the sources of the technical problems by those presentations.

As was surmised in most related works, the main source of complications in the quantization of such systems lies  in the need to invert an infinite dimensional kinetic matrix, in different guises and origins.

For systems modeled directly from Lagrangian densities for the subsystem that presents  infinite modes,  the usual functional techniques available for continuous linear fields are suspect in the present context because their coupling to discrete variables complicate the issue.  It is therefore imperative to use an alternative approach. One such is to perform the Legendre transformation for the discrete system and then take the continuous limit in the Hamiltonian, such as in the approach of Malekaklagh et al. \cite{Moein_2016}. An alternative (which we follow in this work) is to expand in modes and then obtain a canonical Hamiltonian for the whole system, as has been done, for instance, by \cite{Bourassa_2009} and many others. In this second approach, we signal and clarify the issues involved in the choice of modes, and explicitly compute the intrinsic cutoff for the coupling constants. The crucial point is that the separate identification of lumped element network, on the one hand, and transmission line, on the other, that is used in the Lagrangian presentation, cannot persist when passing on to the required Hamiltonian formalism, and  proper dressing of the infinite continuous modes with the discrete modes is necessary. This also requires the correct identification of the degrees of freedom. For instance, when several transmission lines are present and coupled the same network, it is not always possible to separate modes as pertaining only to one transmission line: the presence of the network forces the modes to be distributed on several transmission lines.

Modelling a system with infinite degrees of freedom coupled to a network with a finite number of modes can be done in a number of ways. A Lagrangian density is not the only possibility, quite evidently. In the context of linear passive non dissipative electrical circuits an alternative is given by an analysis of immittances, be they impedances or admittances, with infinite poles, which are then translated into lumped element circuits, with infinite capacitances and inductances. In order to write down a Hamiltonian it is again necessary to invert an infinite dimensional kinetic matrix.  A possible approach is to consider a truncation in the number of modes associated with the impedance to a finite number $N$, to proceed with an inversion and then to  the limit $N\to \infty$. In many cases of interest, this procedure leads to the uncoupling of the impedance modes and the finite network. This comes about because some coupling vector in the Lagrangian  has infinite norm in that limit. More precisely, because that coupling vector has infinite norm with respect to a specific inner product, determined by the inverse capacitance matrix of the infinite dimensional system. We give two solutions to this problem, and then show their equivalence. The first one, in parallel to the presentation for transmission lines, consists in a canonical transformation in the line of that presented by Paladino et al. \cite{Paladino_2003}. That is, a rearrangement of the degrees of freedom, dressing the impedance modes with network modes. We present this formalism for the first Foster form of an impedance coupled capacitively to a network. The second solution comes from the identification of the proper normal modes of the impedance in the Hamiltonian, by the standard canonical transformation only in impedance modes. This is first done for finite $N$ and then taking the limit $N\to\infty$.   We extend this analysis to multiport impedances.

After this introductory section, we present a catalogue of configurations with transmission lines in the formalism of Lagrangian density, in which we study systematically mode expansions, counting of degrees of freedom, and separability of modes. We defer to an appendix the relevant mathematical apparatus used in this section. In the following section we turn to the coupling of networks to canonical impedances. We first study the reassignment of modes, dressing the impedance  with network modes, and then the diagonalization of impedance modes in the Hamiltonian to avoid the uncoupling in the infinite mode limit. In the fourth section we retake transmission lines. We use the previous analysis to provide explicit analytically computable examples, after a general discussion on spin-boson models as derived from these capacitive couplings. We finish with a recapitulation and a summary of conclusions and proposals for future work.

Regarding notation, we use boldface italic or boldface for column vectors, both finite and infinite dimensional, as in $\bi{a}$ and $\boldsymbol{\phi}$. We are dealing with real vector spaces (except in example \ref{sec:an-example-galvanic}), and duality is given by transpose, $\bi{a}^T$. This is justified also for infinite dimensional vectors since we only consider Hilbert spaces. Sans serif is used for matrix and operators. Througout we make use of block-matrix presentations. In the infinite dimensional case this is due to the underlying direct sum structures. 

\section[Networks with Transmission Lines]{Networks with Transmission Lines}
\label{sec:netw-with-transm}
Quantum networks of superconducting qubits make use of transmission lines to either carry information away from the computational system with open boundaries or to store and manipulate it in the form of resonators. As any conducting box does, superconducting transmission lines theoretically support an infinite number of electromagnetic modes as bosonic degrees of freedom. Typically, the dynamics of the whole system is well described in terms of controlled anharmonic subsystems, e.g. qubits, interacting with a countable, possibly infinite, number of bosonic modes, i.e. harmonic oscillators. A typical requirement for such effective models to be valid is that the coupling strength between subsystems is small compared to the energy defining the subsystems themselves. However, even in this small-coupling limit, multi-mode effects can have crucial effects on the predicted effective coupling between two separated information units \cite{Filipp_2011}. 

In this section, we develop the tools required to write exact quantum Hamiltonians of systems of general anharmonic subsystems linearly coupled to transmission lines with closed or open boundary conditions, keeping the multi-mode feature of the lines, and verifying that ultraviolet divergenceless predictions are a natural consequence of the canonical quantization procedure.

Our viewpoint   issues from a Lagrangian description. We then  write the transmission lines in terms of an infinite set of modes, and  carefully proceed to a Hamiltonian formulation in which to perform canonical quantization. As usual, the description in terms of modes is not unique. We make explicit use of this freedom to identify the most adequate choice, where the criterion is that the network modes are uncoupled from each other in the Hamiltonian formalism. In particular, given the capacitive coupling scheme we study, $\lambda/4$ or $\lambda/2$ resonator modes expansions are seen to be inadequate: they would lead to a description in which network and transmission line are uncoupled. The correct set of modes necessarily is dressed by the parameters of the coupling. This dressing means that the mode form functions are \emph{not} eigenfunctions of a Sturm--Liouville operator. In fact, carrying out a na\"{\i}ve separation of variables, we see that the mode form functions are determined by a boundary condition differential equation singular value problem, in which the singular value also enters the boundary condition. This kind of singular value problem is hugely different from the Sturm--Liouville case, and standard textbook material does not cover it. We provide mathematical details and a new proof of the expansion theorems one requires in order for these functions to be indeed mode functions in Appendix \ref{Walter_appendix}. Applying these techniques we obtain definite predictions for the couplings, with a natural intrinsic cutoff frequency. This cutoff frequency comes about because the dressing of the transmission lines requires a length parameter, $\alpha$, that provides us with a natural ultraviolet cutoff. It is important to stress that this cutoff is intrinsic to the model, with no need to argue about the validity of the model itself for it to appear.
\subsection{Linear coupling to lumped-element Networks}
\label{subsec:chap2_linear_cupling}

In this subsection, we study common linear coupling configurations between a finite set of degrees of freedom and transmission lines, namely mixed inductive and capacitive ``point-like'' coupling to (i) a network or to (ii) multiple networks, and (iii) mixed inductive and capacitive galvanic coupling.

\subsubsection{Mixed linear coupling}
\label{subsubsec:chap2_mixed_linear_coupling}

\begin{figure*}[h]
	\centering{\includegraphics[width=0.65\textwidth]{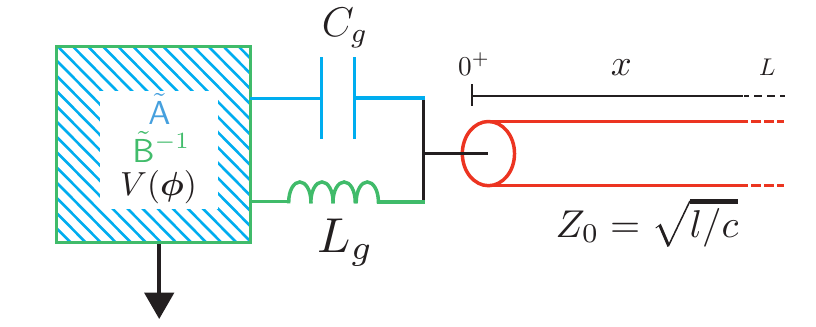}}
	\caption{\label{fig:TL_LCcoup_Network} \textbf{Transmission line inductively and capacitively coupled to a finite network}. The network has internal flux degrees of freedom $\boldsymbol{\phi}_{i}$, with capacitance $\mathsf{A}$ and inductive $\mathsf{B}^{-1}$ matrices and general non-linear potential $V(\boldsymbol{\phi})$.}
\end{figure*}

We consider the circuit in Fig. \ref{fig:TL_LCcoup_Network} with a network of degrees of freedom $\boldsymbol{\phi}$ linearly coupled through capacitor $C_g$ and inductor $L_g$ to a transmission line at one end. Given that our non-linear network has a nonlinear potential in flux variables, as it is the case when there are Josephson junctions, it facilitates the analysis to choose flux variables as our set of position-like coordinates. 

Following standard microwave theory \cite{Blais_2004,Bourassa_2009,Pozar_2009}, the Lagrangian of this circuit can be written in terms of a discrete set of flux variables describing the network, collected in the column vector $\boldsymbol{\phi}$, and a flux field $\Phi(x,t)$,
\begin{eqnarray}
\fl \qquad\qquad L= \frac{1}{2}\dot{\boldsymbol{\phi}}^T \mathsf{A}\dot{\boldsymbol{\phi}}-\frac{1}{2}\boldsymbol{\phi}^T \mathsf{B}^{-1}\boldsymbol{\phi}-V(\boldsymbol{\phi})+\int_{0}^{L}dx\,\left[\frac{c}{2}\dot{\Phi}(x,t)^2-\frac{1}{2l}(\Phi'(x,t))^2\right]\nonumber\\
 +C_g\left(\frac{\dot{\Phi}(0,t)^2}{2}-\dot{\boldsymbol{\phi}}^T\bi{a} \dot{\Phi}(0,t)\right)-\frac{1}{L_g}\left(\frac{\Phi(0,t)^2}{2}-\boldsymbol{\phi}^T\bi{b}\Phi(0,t)\right),\label{eq:Lag_TL_LCcoup_Network}
\end{eqnarray}
where $\mathsf{A}=\tilde{\mathsf{A}}+C_g \bi{a}\bi{a}^T$ and $\mathsf{B}^{-1}=\tilde{\mathsf{B}}^{-1}+\bi{b}\bi{b}^T/L_g$ are the capacitance and inductance submatrices of the network respectively, and $\bi{a}$ and $\bi{b}$ are coupling vectors to the network from the transmission line with finite norm. Notice here that we do not assume any specific description of the network in terms of branch or node flux variables. We do nonetheless emphasize that the network has to be connected non-trivially to the common ground in order for current to circulate through $C_g$ and $L_g$. We remark that in the whole analysis we can take the limits of $C_g\rightarrow0$ and $L_g\rightarrow\infty$ to disconnect the transmission line from the network through its corresponding element. The classical equations of motion of this system read
\begin{eqnarray}
\fl \quad\qquad\qquad\qquad\quad c\ddot{\Phi}(x,t)&=&\frac{\Phi''(x,t)}{l}\label{eq:EulLag_TL_LCcoup_Network11},\\
\fl \quad\qquad\qquad\qquad \quad \frac{\Phi'(0,t)}{l}&=&C_g\left(\ddot{\Phi}(0,t)-\bi{a}^T\ddot{\boldsymbol{\phi}}\right)+\frac{1}{L_g}\left({\Phi}(0,t)-\bi{b}^T{\boldsymbol{\phi}}\right),\label{eq:EulLag_TL_LCcoup_Network12}\\
\fl \,\qquad\qquad\quad\quad \mathsf{A}\ddot{\boldsymbol{\phi}}+\mathsf{B}^{-1}{\boldsymbol{\phi}}&=&C_g\bi{a}\ddot{\Phi}(0,t)+\frac{1}{L_g}\bi{b}\Phi(0,t)-\frac{\partial V(\boldsymbol{\phi})}{\partial \boldsymbol{\phi}}\label{eq:EulLag_TL_LCcoup_Network13}.
\end{eqnarray}
Let us first assume, for simplicity, that the transmission line has finite length $L$ (see Appendix \ref{subsec:infin-length-transm} for infinite length transmission lines, and explicit computation in \ref{sec:an-example:-half} and \ref{sec:an-example-galvanic}). A textbook analysis would carry out separation of variables, that is, it would introduce a decomposition of the flux field in a countable basis of functions $\Phi(x,t)=\sum_n \Phi_n(t)u_n(x)$, justified physically as normal modes. There is an issue in this case, however, in that there is a coupling at the endpoint $x=0$ with the network that involves the second derivative with respect to time of the flux field. Even if all the network variables were set to zero, we would still have, from Eq. (\ref{eq:EulLag_TL_LCcoup_Network12}), a boundary condition that would involve the separation constant ($-\omega_n^2$ in (\ref{eq:EVP_TL_LCcoup_Network_eq0}) below). Furthermore, setting the network variables to zero would not be consistent, since the transmission line sources the network in equation (\ref{eq:EulLag_TL_LCcoup_Network13}).

Here we will take the following approach: we shall retain the dependence of the boundary condition on the separation constant, by introducing a length parameter $\alpha$ that will later be set to an optimal value, according to a precise optimality criterion. Namely, that in the Hamiltonian presentation there be no coupling amongst the transmission line modes.

In this manner, the  field equations for the line yield the following homogeneous eigenvalue problem
\begin{eqnarray}
\qquad\qquad \ddot{\Phi}_n(t)=-\omega_n^2\Phi_n(t),\label{eq:EVP_TL_LCcoup_Network_eq0}\\
\qquad\qquad u_n''(x)=-k_n^2u_n(x),\label{eq:EVP_TL_LCcoup_Network_eq1}\\
\qquad\qquad u_{n}'(0)=-k_n^2\alpha u_n(0)+\frac{1}{\beta}u_n(0),\label{eq:EVP_TL_LCcoup_Network_eq2}\\
\qquad\qquad u_n(L)=0,\label{eq:EVP_TL_LCcoup_Network_eq3}
\end{eqnarray}
where the frequencies are related to the wavenumbers through $\omega_n^2=k_n^2/lc$, and we have assumed for concreteness a short to ground boundary condition at $x=L$. Notice that this choice is not a restriction of our method, and other boundary conditions can be considered at $x=L$, i.e. the general case as at the other end $x=0$.

As we have already pointed, this form of Eq. (\ref{eq:EVP_TL_LCcoup_Network_eq2}) can be derived by setting to zero the network fluxes in (\ref{eq:EulLag_TL_LCcoup_Network12}), in which case the parameter $\alpha$ would be given by $C_g/c$. It can also be obtained by solving $\ddot{\boldsymbol{\phi}}$ in (\ref{eq:EulLag_TL_LCcoup_Network13}), substituting it in (\ref{eq:EulLag_TL_LCcoup_Network12}) and consistently imposing  $\boldsymbol{\phi}=-\mathsf{B}\partial_{\boldsymbol{\phi}}V(\bi{\phi})$. In this case the parameter $\alpha$ would be given as $(C_g/c)\left(1-C_g\bi{a}^T\mathsf{A}^{-1}\bi{a}\right)$, which, as we will see, is  optimal from our point of view.  Indeed, and 
as previously envisaged in \cite{Moein_2016}, the second approach uses the information about the network capacitance matrix $\mathsf{A}$ and its coupling vector $\bi{a}$ to derive a Hamiltonian without mode-mode coupling in the purely harmonic sector. The physical reason for this choice is that in this manner the inhomogeneous source term  corresponds to the current through the anharmonic potential.

So far we have concentrated on the more crucial parameter $\alpha$. The second length parameter, $\beta$, is more easily determined. Nonetheless we also allow it to be free, and its value will also be fixed a posteriori. We remark that Dirichlet,  Neumann, and Robin homogeneous boundary conditions are included in the analysis, with corresponding sets of parameters $\beta_D=0$ for Dirichlet (line ended in open circuit), $(\alpha_N,\beta_N)=(0,\infty)$ for Neumann (line ended in short circuit), and $\alpha_R=0$ for Robin (pure inductive coupling). Analogously, the pure capacitive coupling boundary condition corresponds to $\beta=\infty$.

For fixed parameters $\alpha$ and $\beta$, the system of equations
from 
Eq.
(\ref{eq:EVP_TL_LCcoup_Network_eq1}) through to (\ref{eq:EVP_TL_LCcoup_Network_eq3}) define a generalized eigenvalue problem, with an easily determined secular equation and generalized eigenfunctions.
Furthermore, those eigenfunctions satisfy the following orthogonality conditions
\begin{eqnarray}
\fl \quad\quad\quad \langle u_n,u_m\rangle_{\alpha}=c \left(\int_{0}^{L}dx\, u_n(x) u_m(x)+  \alpha u_n(0) u_m(0)\right)=N_{\alpha}\delta_{nm},\label{eq:TL_LCcoup_Network_ortho_1}\\
\fl \quad\quad\quad\langle u_n,u_m\rangle_{1/\beta}=\frac{1}{l}\left(\int_{0}^{L}dx\, u_n'(x) u_m'(x)+ \frac{1}{\beta} u_n(0) u_m(0)\right)=\omega_n^2 N_{\alpha}\delta_{nm},\label{eq:TL_LCcoup_Network_ortho_2}
\end{eqnarray}
where $N_{\alpha}$ is a free normalization constant in capacitance units.

From these considerations, a number of authors have used these generalised eigenfunctions and orthogonality for an expansion in modes. We should note however that the possiblility of expanding a function in these eigenfunctions, i.e. that they form a basis in a suitable space of functions, is by no means deducible from standard Sturm--Liouville theory. Fortunately, the topic has been examined in the mathematical literature (see, inter alia, \cite{Walter_1973_EVP}), and it is indeed the case that an expansion theorem does hold. We provide more mathematical details, and a new proof of the expansion theorem, in Appendix \ref{Walter_appendix}.

Now, knowing that we can expand in this generalized eigenbasis, we write the Lagrangian (\ref{eq:Lag_TL_LCcoup_Network}) as 
\begin{equation}	L=\frac{1}{2}\dot{\bi{X}}^T\mathsf{C}\dot{\bi{X}}-\frac{1}{2}\bi{X}^T \mathsf{L}^{-1}\bi{X} -V(\boldsymbol{\phi}),
\end{equation}
where we have defined the vector of fluxes
\begin{equation}
  \label{eq:xvector}
  \bi{X}=
  \begin{pmatrix}
    \boldsymbol{\phi}\\
    \boldsymbol{\Phi}
  \end{pmatrix},
\end{equation}
and the capacitance and inductance matrices
\begin{eqnarray}
	\,\,\qquad \mathsf{C}&=&\begin{pmatrix}
	\mathsf{A}& -C_g \bi{a}\bi{u}^T\\
	-C_g \bi{u}\bi{a}^{T} & N_{\alpha} \mathbbm{1} + d\bi{u} \bi{u}^T
	\end{pmatrix},\label{eq:C_LCcoup_Network}\\
	\qquad \mathsf{L}^{-1}&=&\begin{pmatrix}
	\mathsf{B}^{-1}& - \bi{b}\bi{u}^T/L_g\\
	- \bi{u}\bi{b}^{T}/L_g & N_{\alpha} (\omega_n^2) + e\bi{u} \bi{u}^T
	\end{pmatrix}.\label{eq:Linv_LCcoup_Network}
\end{eqnarray}
with $\bi{u}\equiv (u_0(0), u_1(0),...u_n(0),\ldots)^T$ being the coupling vector (of infinite dimensionality), the parameters $d \equiv C_g-c \alpha$ and $e\equiv 1/L_g - 1/\beta l$, $\mathbbm{1}$ the infinite-dimensional identity matrix and $(\omega_n^2)=\mathrm{diag}(\omega_0^2,\omega_1^2,...)$ the diagonal matrix of squared frequencies of the eigenvalue problem. Notice that $\bi{u}$ is generically normalizable. Even more importantly, the quantity $\bi{u}^T\left[N_\alpha \mathbbm{1}+d\bi{u}\bi{u}^T\right]^{-1}\bi{u}=1/C_g$ is finite unless $C_g$ is zero.  The vector $\bi{u}$ is in fact an element of the $l^2$ sequence Hilbert space, by the construction of Appendix \ref{sec:finite-length-transm}, and its norm depends directly on the parameter $\alpha$, namely $\left|\bi{u}\right|^2=\bi{u}^T\bi{u}=N_\alpha/\alpha c$. The dimensionful parameter $N_\alpha$ was introduced so that this norm be adimensional.

We can now invert the capacitance matrix and derive the Hamiltonian 
\begin{equation}
H=\frac{1}{2}{\bi{P}}^T\mathsf{C}^{-1}{\bi{P}}+\frac{1}{2}\bi{X}^T \mathsf{L}^{-1}\bi{X}+V(\boldsymbol{\phi}),\label{eq:Ham_LCcoup_Network}
\end{equation}
where the conjugate charge variables to the fluxes are $\bi{P}=\partial L/\partial \bi{X}=(\bi{q}^T,\bi{Q}^T)^T$, and the inverse capacitance matrix is 
\begin{equation}
	\mathsf{C}^{-1}= \begin{pmatrix}
	\mathsf{A}^{-1}+ \frac{C_g^2 |\bi{u}|^2}{D} \mathsf{A}^{-1}\bi{a}\bi{a}^T \mathsf{A}^{-1}&   \frac{C_g}{D} \mathsf{A}^{-1} \bi{a}\bi{u}^T\\
	 \frac{C_g}{D} \bi{u}\bi{a}^T\mathsf{A}^{-1}& \frac{1}{N_{\alpha}}\mathbbm{1}+ \frac{1}{|\bi{u}|^2}\left( \frac{1}{D}- \frac{1}{N_{\alpha}}\right) \bi{u} \bi{u}^T
	\end{pmatrix},\nonumber 
\end{equation}
with $D= N_{\alpha}+|\bi{u}|^2(d- C_g^2 \bi{a}^T \mathsf{A}^{-1}\bi{a})$. It now behoves us to insert the requirement that there be no mode-mode coupling in the description of the transmission line.  Recalling that $d$ and $e$ depend on the parameters $\alpha$ and $\beta$, which we have so far left undetermined, we can choose these parameters $\alpha$ and $\beta$ to satisfy the equations $D=N_{\alpha}$ and $e=0$, thus removing the harmonic mode-mode couplings, with the result 
\begin{eqnarray}
	\qquad\qquad\qquad \alpha&=&\frac{C_g(1-C_g\bi{a}^T \mathsf{A}^{-1}\bi{a})}{c},\label{eq:TL_LCcoup_Network_alpha_fix}\\
	\qquad\qquad\qquad \beta&=&L_g/l.\label{eq:TL_LCcoup_Network_beta_fix}
\end{eqnarray}
Next, in order to find the frequencies $\omega_n$, we have to solve the eigenvalue problem (\ref{eq:EVP_TL_LCcoup_Network_eq1}-\ref{eq:EVP_TL_LCcoup_Network_eq3}) with the values of $\alpha$ and $\beta$ presented in (\ref{eq:TL_LCcoup_Network_alpha_fix}) and (\ref{eq:TL_LCcoup_Network_beta_fix}), and the final Hamiltonian will be
\begin{eqnarray}
\fl \quad\quad\quad H= \frac{1}{2}\bi{q}^T(\mathsf{A}^{-1}+ \frac{C_g^2}{\alpha c} \mathsf{A}^{-1}\bi{a}\bi{a}^T \mathsf{A}^{-1})\bi{q}+\frac{1}{2}\boldsymbol{\phi}^T\mathsf{B}^{-1}\boldsymbol{\phi}+V(\boldsymbol{\phi})+\sum_n \frac{Q_n^2}{2N_{\alpha}}+\frac{N_{\alpha}\omega_n^2\Phi_n^2}{2}\nonumber\\
+\frac{C_g}{N_{\alpha}} (\bi{q}^T\mathsf{A}^{-1} \bi{a})\sum_n Q_n u_n(0)-\frac{1}{L_g} (\boldsymbol{\phi}^T\bi{b})\sum_n \Phi_n u_n(0),\label{eq:Ham_LCcoup_Network2}
\end{eqnarray}
where we have used the normalization equality $|\bi{u}|^2=N_{\alpha}/\alpha c$. 

To complete the process of quantization, we promote the conjugate variables to operators with the commutator $[\hat{X}_i,\hat{P}_j]=i\hbar \delta_{ij}$.  Finally the quantum Hamiltonian in terms of annihilation and creation operators, related to flux and charge variables by $\hat{\Phi}_n=i\sqrt{\hbar/2\omega_n N_\alpha}(a_n-a_n^{\dagger})$ and $\hat{Q}_n=\sqrt{\hbar\omega_n N_\alpha/2}(a_n+a_n^{\dagger})$,
\begin{eqnarray}
\fl \quad\quad\quad\quad \tilde{H}= \frac{1}{2}\hat{\bi{q}}^T(\mathsf{A}^{-1}+ \frac{C_g^2}{\alpha c} \mathsf{A}^{-1}\bi{a}\bi{a}^T \mathsf{A}^{-1})\hat{\bi{q}}+\frac{1}{2}\hat{\boldsymbol{\phi}}^T\mathsf{B}^{-1}\hat{\boldsymbol{\phi}}+V(\hat{\boldsymbol{\phi}})+\sum_n \omega_n a_n^\dagger a_n\nonumber\\
+C_g\sqrt{\frac{\hbar}{2 N_\alpha}} (\hat{\bi{q}}^T\mathsf{A}^{-1} \bi{a})\sum_n (a_n+a_n^\dagger)\sqrt{\omega_n} u_n(0) \nonumber\\
-\frac{i}{L_g}\sqrt{\frac{\hbar}{2 N_\alpha}} (\hat{\boldsymbol{\phi}}^T\bi{b})\sum_n (a_n-a_n^{\dagger}) \frac{u_n(0)}{\sqrt{\omega_n}}.\nonumber 
\end{eqnarray}
This Hamiltonian is as exact as the starting point, the Lagrangian (\ref{eq:Lag_TL_LCcoup_Network}), and here we can see a first result: the (capacitive) coupling constants $g_n\propto \sqrt{\omega_n}u_n(0)$ do not grow without bound. As we discuss in detail in  \ref{sec:generic-behaviour}, the large $n$ behaviour of $u_n(0)$ is $1/n$, while that for $\omega_n$ is $n$. It follows that $g_n\sim n^{-1/2}$. There is no need for an ultraviolet cutoff extrinsic to the model (\ref{eq:Lag_TL_LCcoup_Network}); rather,  the correct choice of modes to expand in has provided us with a natural length scale, intrinsic to the model, that translates into  an intrinsic ultraviolet cutoff.
\subsubsection{Linearized galvanic coupling}
\label{sec:line-galv-coupl}
Another very common circuit configuration that has been used in cQED is the so called {\it galvanic} coupling between harmonic modes and non-harmonic variables, see \cite{Bourassa_2009}. Indeed, such a configuration has proved to be the most efficient way thus far to reach the ultrastrong coupling regime in light-matter interactions \cite{Niemczyk_2010,Pol_2017,Yoshihara_2017}.
\begin{figure*}[ht]
	\centering{\includegraphics[width=0.85\textwidth]{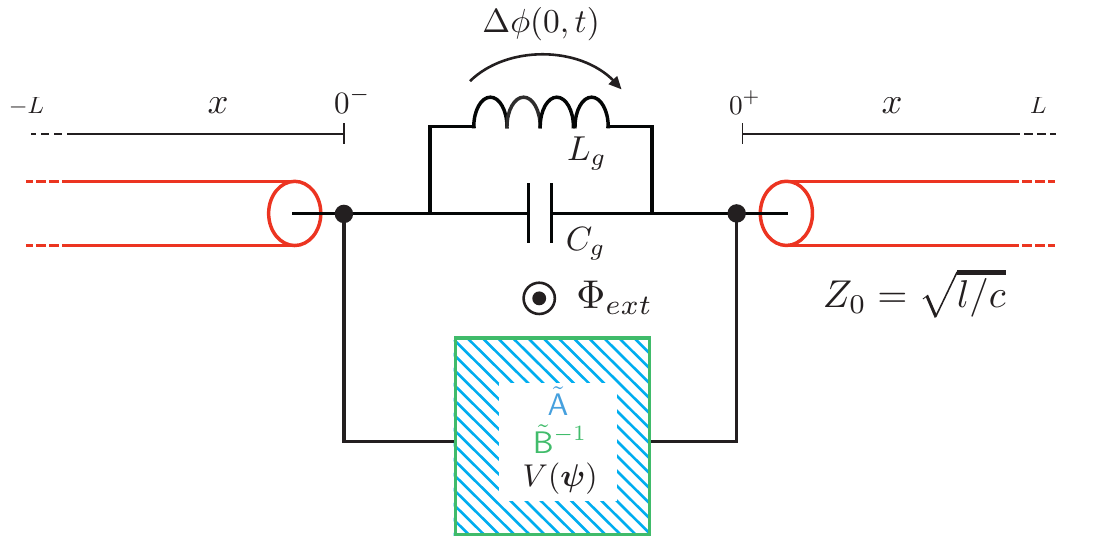}}
	\caption{\label{fig:TL_LCgalvcoup_Network} \textbf{Transmission line galvanically coupled to a finite network}. The network has internal flux degrees of freedom $\boldsymbol{\phi}_{i}$, with capacitance $\mathsf{A}$ and inductive $\mathsf{B}^{-1}$ matrices and general non-linear potential $V(\boldsymbol{\phi}_{i})$.}
\end{figure*}

The Lagrangian describing a generalized galvanic configuration, see Fig. \ref{fig:TL_LCgalvcoup_Network}, can be written as 
\begin{eqnarray}
\fl \quad\quad\quad\quad L= \frac{1}{2}\dot{\boldsymbol{\psi}}^T \tilde{\mathsf{A}}\dot{\boldsymbol{\psi}}-\frac{1}{2}\boldsymbol{\psi}^T \tilde{\mathsf{B}}^{-1}\boldsymbol{\psi}-V(\boldsymbol{\psi})+\int_{-L}^{L}dx\,\left[\frac{c}{2}\dot{\Phi}(x,t)^2-\frac{1}{2l}(\Phi'(x,t))^2\right]\nonumber\\
+\frac{C_g}{2}\Delta\dot{\Phi}(0,t)^2-\frac{1}{2L_g}\Delta\Phi(0,t)^2,\label{eq:Lag_TL_LCgalvcoup_Networks}
\end{eqnarray}
where the set of internal variables is collected in a column vector $\boldsymbol{\psi}=(\psi_1,...,\psi_N)^T$, and $\Delta\Phi(0,t)$ is the flux difference in the line. The first order of business is  to identify a good set of independent variables. In order to achieve that,  we  impose  Kirchoff's laws in the connection, a constraint that fixes at least one of the degrees of freedom in the network,
\begin{equation}
	\psi_N(t)=\Delta \Phi(0,t)+\Phi_{\mathrm{ext}}+ \bi{g}^T\boldsymbol{\phi}(t),\label{eq:TL_LCgalvcoup_Networks_Vtransf}
\end{equation}
where the new truncated set of variables is $\boldsymbol{\phi}\equiv(\psi_1, \psi_2,...\psi_{N-1})^T$, and $\bi{g}$ is a constant vector on that reduced subspace. We reduce the number of variables and find that the Lagrangian has both capacitive and inductive coupling to the flux difference in the line $\Delta \Phi(0,t)$
\begin{eqnarray}
\fl \quad\quad\quad L= \frac{1}{2}\dot{\boldsymbol{\phi}}^T \mathsf{A}\dot{\boldsymbol{\phi}}-\frac{1}{2}\boldsymbol{\phi}^T \mathsf{B}^{-1}\boldsymbol{\phi}-V(\boldsymbol{\phi},\Phi_{\mathrm{ext}})+\int_{-L}^{L}dx\,\left[\frac{c}{2}\dot{\Phi}(x,t)^2-\frac{1}{2l}(\Phi'(x,t))^2\right]\nonumber\\
+\frac{C_{gA}}{2}\Delta\dot{\Phi}(0,t)^2-\frac{1}{2L_{gB}}\Delta\Phi(0,t)^2+\frac{1}{2L_B}(2 \Delta \Phi(0,t)\Phi_{\mathrm{ext}}+\Phi_{\mathrm{ext}}^2)\nonumber\\
-C_A(\dot{\boldsymbol{\phi}}^T\bi{a})\Delta\dot{\Phi}(0,t)+\frac{1}{L_B}(\boldsymbol{\phi}^T\bi{b})(\Delta\Phi(0,t)+\Phi_{\mathrm{ext}}),\label{eq:Lag_TL_LCgalvcoup_Networks_2}
\end{eqnarray}
where $C_{gA}=(C_g+C_A)$ and $L_{gB}=L_g L_B/(L_g + L_B)$, with $C_A$ and $L_B$ the coupling capacitance and inductance parameters coming out of the transformation (\ref{eq:TL_LCgalvcoup_Networks_Vtransf}) in (\ref{eq:Lag_TL_LCgalvcoup_Networks}). In this reduction, we decompose the matrix $\tilde{\mathsf{A}}$ as
\begin{equation}
\tilde{\mathsf{A}}=\begin{pmatrix}
\mathsf{A}_1 & \bi{a}_1\\
\bi{a}_1^T & C_A
\end{pmatrix}.
\end{equation}
It follows that $\mathsf{A}=\mathsf{A}_1+\bi{a}_1\bi{g}^T+\bi{g}\bi{a}_1^T+C_A\bi{g}\bi{g}^T$ and $\bi{a}=-\bi{g}-\bi{a}_1/C_A$ in (\ref{eq:Lag_TL_LCgalvcoup_Networks_2}). An analogous procedure provides us with matrix $\mathsf{B}$ and coupling vector $\bi{b}$. The equations of motion for this Lagrangian are 
\begin{eqnarray}
\fl \,\,\qquad\qquad\qquad\qquad\quad\,\, c\ddot{\Phi}(x,t)&=&\frac{\Phi''(x,t)}{l}\label{eq:EulLag_TL_LCgalvcoup_Networks11},\\
\fl \qquad\qquad\qquad\qquad\quad\, \frac{\Phi'(0^-,t)}{l}&=&\frac{\Phi'(0^+,t)}{l},\\
\fl \,\quad\,\,\,\,  C_A\bi{a}^T\ddot{\boldsymbol{\phi}} +\frac{1}{L_B}\left(\bi{b}^T{\boldsymbol{\phi}} + \Phi_{\mathrm{ext}}\right)&=&\frac{\Phi'(0^-,t)}{l}+C_{gA}\Delta\ddot{\Phi}(0,t) + \frac{1}{L_{gB}}\Delta{\Phi}(0,t),\label{eq:EulLag_TL_LCgalvcoup_Networks12}\\
\fl \qquad \mathsf{A}\ddot{\boldsymbol{\phi}}+\mathsf{B}^{-1}{\boldsymbol{\phi}}+\frac{\partial V(\boldsymbol{\phi},\Phi_{\mathrm{ext}})}{\partial \boldsymbol{\phi}}&=&C_A\bi{a}\Delta\ddot{\Phi}(0,t)+\frac{1}{L_B}\bi{b}\left(\Delta\Phi(0,t)+\Phi_{\mathrm{ext}}\right)\label{eq:EulLag_TL_LCgalvcoup_Networks13}.
\end{eqnarray}
We decompose again the flux field in a countable basis of functions $\Phi(x,t)=\sum_n \Phi_n(t)u_n(x)$ (given that we assumed the line of finite length) and the field equations for the line yield the following homogeneous eigenvalue problem
\begin{eqnarray}
\ddot{\Phi}_n(t)=-\omega_n^2\Phi_n(t),\\
u_n''(x)=-k_n^2u_n(x),\label{eq:EVP_TL_LCgalvcoup_Networks_eq1}\\
u_{n}'(0^-)=u_{n}'(0^+)=k_n^2\alpha \Delta u_n(0)-\frac{1}{\beta}\Delta u_n(0),\label{eq:EVP_TL_LCgalvcoup_Networks_eq2}\\
u_n(-L)=u_n(L)=0,\label{eq:EVP_TL_LCgalvcoup_Networks_eq3}
\end{eqnarray}
where $\Delta u_n(0)=u_n(0^-)-u_n(0^+)$. Again, Eqs. (\ref{eq:EVP_TL_LCgalvcoup_Networks_eq1}-\ref{eq:EVP_TL_LCgalvcoup_Networks_eq3}) define a generalized eigenvalue problem with eigenvalue-dependent boundary conditions, see Sec. \ref{sec:galvanic-coupling}, whose eigenfunctions satisfy the following orthogonality conditions
\begin{eqnarray}
\fl \quad\quad \langle u_n,u_m\rangle_{\alpha}=c \left(\int_{0}^{L}dx\, u_n(x) u_m(x)+  \alpha \Delta u_n(0) \Delta u_m(0)\right)=N_{\alpha}\delta_{nm},\label{eq:TL_LCgalvcoup_Networks_ortho_1}\\
\fl \quad\quad \langle u_n,u_m\rangle_{1/\beta}=\frac{1}{l}\left(\int_{0}^{L}dx\, u_n'(x) u_m'(x)+ \frac{1}{\beta} \Delta u_n(0) \Delta u_m(0)\right)=\omega_n^2 N_{\alpha}\delta_{nm},\label{eq:TL_TL_LCgalvcoup_Networks_ortho_2}
\end{eqnarray}
where $N_{\alpha}$ is a free normalization constant in capacitance units. Notice that in this case we can choose real eigenfunctions, and we have done so. Making use of the above equations we can rewrite the Lagrangian (\ref{eq:Lag_TL_LCgalvcoup_Networks_2}) as 
\begin{equation}
L=\frac{1}{2}\dot{\bi{X}}^T\mathsf{C}\dot{\bi{X}}^T-\frac{1}{2}\bi{X}^T \mathsf{L}^{-1}\bi{X} -V(\boldsymbol{\phi},\Phi_{\mathrm{ext}}),
\end{equation}
with $\bi{X}=({\boldsymbol{\phi}}^T,
{\boldsymbol{\Phi}}^T)^T$ and 
\begin{eqnarray}
\qquad \mathsf{C}&=&\begin{pmatrix}
\mathsf{A}& -C_A \bi{a}\Delta\bi{u}^T\\
-C_A \Delta\bi{u}\bi{a}^{T} & N_{\alpha} \mathbbm{1} + d\Delta\bi{u} \Delta\bi{u}^T
\end{pmatrix},\label{eq:C_TL_LCgalvcoup_Networks}\\
\qquad \mathsf{L}^{-1}&=&\begin{pmatrix}
\mathsf{B}^{-1}& - \bi{b}\Delta\bi{u}^T/L_B\\
- \Delta\bi{u}\bi{b}^{T}/L_B & N_{\alpha} (\omega_n^2) + e\Delta\bi{u} \Delta\bi{u}^T
\end{pmatrix},\label{eq:Linv_TL_LCgalvcoup_Networks}
\end{eqnarray}
where we have defined the coupling vector $\Delta\bi{u}\equiv (\Delta u_0(0), \Delta u_1(0),...)^T$ and the parameters $d \equiv C_{gA}-c \alpha$ and $e\equiv 1/L_{gB} - 1/\beta l$. As usual, $\mathbbm{1}$ stands for the infinite-dimensional identity matrix, and $(\omega_n^2)$ is the diagonal matrix of squared frequencies. Following the same steps as in the previous section we derive the Hamiltonian
\begin{eqnarray}
\fl \quad H= \frac{1}{2}\bi{q}^T(\mathsf{A}^{-1}+ \frac{C_A^2}{\alpha c} \mathsf{A}^{-1}\bi{a}\bi{a}^T \mathsf{A}^{-1})\bi{q}+\frac{1}{2}\boldsymbol{\phi}^T\mathsf{B}^{-1}\boldsymbol{\phi}+V(\boldsymbol{\phi})+\sum_n \frac{Q_n^2}{2N_{\alpha}}+\frac{N_{\alpha}\omega_n^2\Phi_n^2}{2}\nonumber\\
\fl \qquad\qquad\qquad +\frac{C_A}{N_{\alpha}} (\bi{q}^T\mathsf{A}^{-1} \bi{a})\sum_n Q_n \Delta u_n(0)-\frac{1}{L_B} (\boldsymbol{\phi}^T\bi{b})\sum_n \Phi_n \Delta u_n(0),\,\label{eq:Ham_TL_LCgalvcoup_Networks}
\end{eqnarray}
from which canonical quantization can be done. Again the criterion has been the elimination of the harmonic mode-mode couplings, and the solution for the parameters reads
\begin{eqnarray}
\qquad\qquad \alpha&=&\frac{C_{gA}-C_A^2\bi{a}^T \mathsf{A}^{-1}\bi{a}}{c},\nonumber\\
\qquad\qquad \beta&=&L_{gB}/l.\nonumber
\end{eqnarray}
\subsubsection{Multiple networks coupled to line}
\label{sec:mult-netw-coupl}
\begin{figure*}[ht]	\centering{\includegraphics[width=0.85\textwidth]{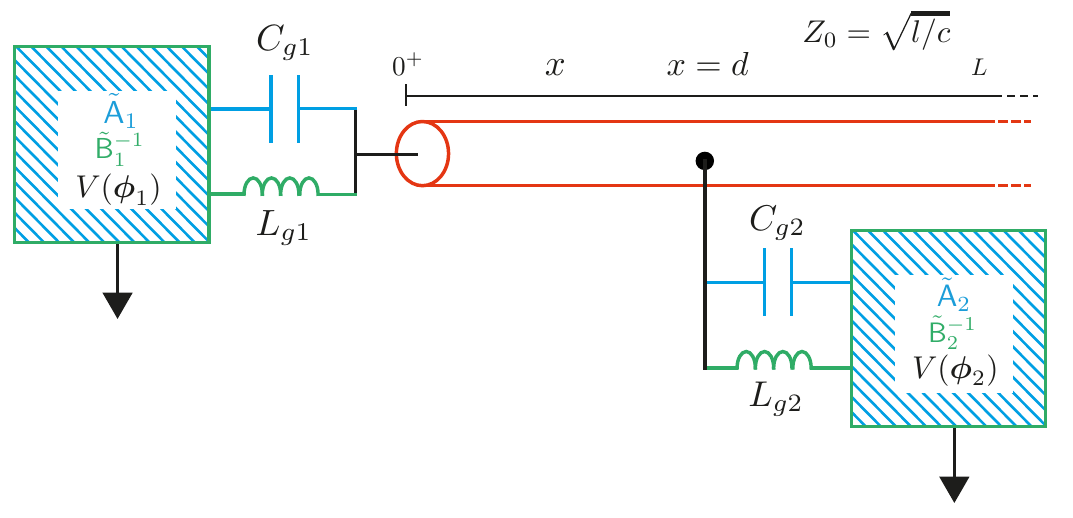}}
	\caption{\label{fig:TL_LCcoup_2Networks} \textbf{Transmission line linearly coupled to two finite networks}. The networks have internal flux degrees of freedom $\boldsymbol{\phi}_{i}$, with capacitance $\tilde{\mathsf{A}}_{i}$ and inductive $\tilde{\mathsf{B}}_i^{-1}$ matrices and general non-linear potential $V(\boldsymbol{\phi}_{i})$, with $i\in \{1,2\}$.}
\end{figure*}

We consider now the generalization of \ref{sec:line-galv-coupl} with a number $M$ of networks linearly coupled to a common transmission line, e.g. the circuit in Fig. \ref{fig:TL_LCcoup_2Networks} has two networks of degrees of freedom $\boldsymbol{\phi}_i$ coupled through capacitors $C_{gi}$ and inductors $L_{gi}$ to a transmission line at positions $\vec{x}=({0,d})$.

A Lagrangian of the generalized circuit, with $M$ networks, can be written as 
\begin{eqnarray}
\fl L= \sum_i^M\frac{1}{2}\dot{\boldsymbol{\phi}}_i^T \mathsf{A}_i\dot{\boldsymbol{\phi}}_i-\frac{1}{2}\boldsymbol{\phi}_i^T \mathsf{B}_i^{-1}\boldsymbol{\phi}_i-V(\boldsymbol{\phi}_i)
+ C_{gi}\left(-(\dot{\boldsymbol{\phi}}_i^T \bi{a}_i)\dot{\Phi}(x_i,t)+\frac{\dot{\Phi}(x_i,t)^2}{2}\right)\nonumber\\
\fl \,\, -\sum_i^M \frac{1}{L_{gi}}\left(-(\boldsymbol{\phi}_i^T \bi{b}_i)\Phi(x_i,t)+
\frac{\Phi(x_i,t)^2}{2}\right)
+\int_{0}^{L}dx\,\left[\frac{c}{2}\dot{\Phi}(x,t)^2-\frac{1}{2l}(\Phi'(x,t))^2\right],\nonumber 
\end{eqnarray}
where $\mathsf{A}_i=\tilde{\mathsf{A}}_i+C_{gi} \bi{a}_i\bi{a}_i^T$ and $\mathsf{B}_i^{-1}=\tilde{\mathsf{B}}_i^{-1}+\bi{b}_i\bi{b}_i^T/L_{gi}$ are the capacitance and inductance submatrices of the network respectively and $\bi{a}_i$ and $\bi{b}_i$ are coupling vectors to the finite networks from the transmission line. Following the same procedure as in last section,  we expand the flux field in an eigenbasis $\Phi(x,t)=\sum_n \Phi_n(t)u_n(x)$ and derive the wave equations 
\begin{eqnarray}
\qquad\qquad\qquad\qquad \ddot{\Phi}_n(t)=-\omega_n^2\Phi_n(t),\nonumber\\
\qquad\qquad\qquad\qquad u_n''(x)=-k_n^2u_n(x),\nonumber
\end{eqnarray}
and a number of boundary conditions of two possible forms, namely 
\begin{eqnarray}
u_{n}'(x_i)=-k_n^2\alpha_i u_n(x_i)+\frac{1}{\beta_i}u_n(x_i),\,\,\,\,\,\,\forall x_i \in \{0,L\}\quad\mathrm{and}\label{eq:EVP_TL_LCcoup_2Networks_eq22}\\
\Delta u_{n}'(x_i)=-k_n^2\alpha_i u_n(x_i)+\frac{1}{\beta_i}u_n(x_i),\,\,\,\,\,\,\forall x_i \notin \{0,L\},\label{eq:EVP_TL_LCcoup_2Networks_eq23}
\end{eqnarray}
depending on whether the $i^{th}$-network is connected at one end of the line or inbetween, respectively. Here $\Delta u_{n}'(x_i)\equiv u_{n}'(x_i^+)-u_{n}'(x_i^-)$, and for networks connected with boundary conditions of Eq. (\ref{eq:EVP_TL_LCcoup_2Networks_eq23}), we further require continuity of $u_{n}(x)$ at $x_i$ . Regardless of the position of connection, the new inner products for the eigenfunctions are
\begin{eqnarray}
\fl \qquad\langle u_n,u_m\rangle_{\{\alpha_i\}}=c \left(\int_{0}^{L}dx\, u_n(x) u_m(x)+  \sum_{x_i}  \alpha_i u_n(x_i) u_m(x_i)\right)=N_{\alpha}\delta_{nm},\nonumber\\ 
\fl \qquad\langle u_n,u_m\rangle_{\{1/\beta_i\}}=\frac{1}{l}\left(\int_{0}^{L}dx\, u_n'(x) u_m'(x)+ \sum_{x_i}\frac{1}{\beta_i} u_n(x_i) u_m(x_i)\right)=\omega_n^2 N_{\alpha}\delta_{nm},\nonumber 
\end{eqnarray}
The Lagrangian can thus be rewritten as 
\begin{equation}
L=\frac{1}{2}\dot{\bi{X}}^T\mathsf{C}\dot{\bi{X}}^T-\frac{1}{2}\bi{X}^T \mathsf{L}^{-1}\bi{X} -V(\boldsymbol{X}),\nonumber
\end{equation}
where $\bi{X}=(\boldsymbol{\phi}_1^T,\boldsymbol{\phi}_2^T,...,\boldsymbol{\phi}_M^T,
{\boldsymbol{\Phi}}^T)^T$ with the new capacitance and inductance matrices  
\begin{eqnarray}
\mathsf{C}=\begin{pmatrix}
\mathsf{A}_1&0&\dots&-C_{g1} \bi{a}_1\bi{u}_1^T\\
0&\mathsf{A}_2&&-C_{g2} \bi{a}_2\bi{u}_2^T\\
\vdots&&\ddots&\vdots\\
-C_{g1} \bi{u}_1\bi{a}_1^{T} & -C_{g2} \bi{u}_2\bi{a}_2^{T} &\dots& N_{\alpha} \mathbbm{1} + \sum_i^M d_i\bi{u}_i \bi{u}_i^T
\end{pmatrix},\label{eq:C_LCcoup_2Networks}\\
\mathsf{L}^{-1}=\begin{pmatrix}
\mathsf{B}_1^{-1}&0&\dots&-\bi{b}_1\bi{u}_1^T/L_{g1}\\
0&\mathsf{B}_2^{-1}&&-\bi{b}_2\bi{u}_2^T/L_{g2}\\
\vdots&&\ddots&\vdots\\
-\bi{b}_1\bi{u}_1^T/L_{g1} & -\bi{b}_2\bi{u}_2^T/L_{g2} &\dots&  N_{\alpha} (\omega_n^2) + \sum_i^M e_i\bi{u}_i \bi{u}_i^T
\end{pmatrix},\label{eq:Linv_LCcoup_2Networks}
\end{eqnarray}
where we have defined the coupling vectors to the $i^{th}$ network as  $\bi{u}_i \equiv\bi{u}(x_i)=(u_0(x_i),u_1(x_i),...)^T$.

Let us for now assume the invertibility of the capacitance matrix $\mathsf{C}$ (we examine this assumption critically in the next subsection, \ref{subsec:invertibility_variable_counting}).
Using the property that the coupling vectors are orthogonal $\langle \bi{u}_i,\bi{u}_j\rangle=\delta_{ij}N_{\alpha}/\alpha_i c$, see Eq. (\ref{eq:genericsumrule}) of Appendix \ref{Walter_appendix}, we determine
\begin{eqnarray}
\fl \mathsf{C}^{-1}=
\begin{pmatrix}
\mathsf{A}_1^{-1}+ s_1 \mathsf{A}_1^{-1}\bi{a}_1\bi{a}_1^T \mathsf{A}_1^{-1}&0&\dots&t_1\mathsf{A}_1^{-1}\bi{a}_1\bi{u}_1^T\\
0&\mathsf{A}_2^{-1}+ s_2 \mathsf{A}_2^{-1}\bi{a}_2\bi{a}_2^T \mathsf{A}_2^{-1}&&t_2 \mathsf{A}_2^{-1}\bi{a}_2\bi{u}_2^T\\
\vdots&&\ddots&\vdots\\
t_1 \bi{u}_1\bi{a}_1^{T}\mathsf{A}_1^{-1} & t_2 \bi{u}_2\bi{a}_2^{T}\mathsf{A}_2^{-1} &\dots& \frac{1}{N_{\alpha}} \mathbbm{1} + \sum_i^M r_i\bi{u}_i \bi{u}_i^T
\end{pmatrix},\nonumber
\end{eqnarray}
where we have defined  paramaters $s_i \equiv -C_{gi}^2 |\bi{u}_i|^2/D_i$, $t_i= C_{gi}/D_i$,  $r_i=1/|\bi{u}_i|^2(1/D_i-1/N_{\alpha})$ and $D_i\equiv N_{\alpha}+ |\bi{u}_i|^2(d_i-C_{gi}^2\bi{a}_i^T \mathsf{A}_i^{-1} \bi{a}_i)$. Finally, we can choose the relevant coefficients of the eigenvalue problem $(\alpha_i, \beta_i)$ such that $r_i=e_i=0$, $\forall i$. That is, we solve the equations $D_i = N_\alpha$ for $\alpha_i$ and $\beta_i = L_{gi}/l$, in order to arrive to a Hamiltonian with a well defined infinite harmonic set
\begin{eqnarray}
\fl \quad\quad H= \frac{1}{2}\sum_i\hat{\bi{q}}_i^T(\mathsf{A}_i^{-1}+ \frac{C_{gi}^2}{\alpha_i c} \mathsf{A}_i^{-1}\bi{a}_i\bi{a}_i^T \mathsf{A}_i^{-1})\hat{\bi{q}}_i+\frac{1}{2}\hat{\boldsymbol{\phi}}_i\mathsf{B}_i^{-1}\hat{\boldsymbol{\phi}}_i+V(\hat{\boldsymbol{\phi}}_i)+\sum_n \omega_n a_n^\dagger a_n\nonumber\\
+\sum_i C_{gi}\sqrt{\frac{\hbar}{2 N_\alpha}} (\hat{\bi{q}}_i^T\mathsf{A}_i^{-1} \bi{a}_i)\sum_n (a_n+a_n^\dagger)\sqrt{\omega_n} u_n(x_i) \nonumber\\-\sum_i\frac{i}{L_{gi}}\sqrt{\frac{\hbar}{2 N_\alpha}} (\hat{\boldsymbol{\phi}}_i^T \bi{b}_i)\sum_n (a_n-a_n^{\dagger})\frac{u_n(x_i)}{\sqrt{\omega_n}},\label{eq:Ham_TL_LCcoup_2Networks3}
\end{eqnarray}
where we have promoted conjugate variables to operators as in previous sections. Again, the coupling coefficients of the capacity part are governed by $\sqrt{\omega_n}u_n(x_i)$, and thus have a large $n$ behaviour of the form $n^{-1/2}$.
\subsection{Invertibility and variable counting}
\label{subsec:invertibility_variable_counting}
In the previous section we have assumed that the capacitance matrix $\mathsf{C}$ has inverse, and thus there is no overcounting of velocity degrees of freedom. However, this assumption does not always hold. Fortunately, it can be easily checked, by determining the conditions for the existence of a zero eigenvalue. Let us first examine the simple case of the network connected to the transmission line, under the assumption that the capacitance submatrix $\mathsf{A}$ is invertible. The condition for the invertibility of $\mathsf{C}$ in (\ref{eq:C_LCcoup_Network}) is determined by analyzing the possible existence of a zero eigenvalue, for which
\begin{equation}
\mathsf{C}\begin{pmatrix}
\bi{y}\\
\bi{z}
\end{pmatrix}=0.\nonumber
\end{equation}
The above matrix equation reduces to 
\begin{eqnarray}
	\quad\quad\quad\quad\quad\quad\quad\quad\quad\mathsf{A}\bi{y}&=&C_g \bi{a} (\bi{u}^T \bi{z}),\label{eq:C_LCcoup_Network_invert_1}\\
	\quad\quad\quad\quad\quad\quad C_g \bi{u} (\bi{a}^T \bi{y})&=&N_\alpha \bi{z}+d\bi{u}(\bi{u}^T \bi{z}).\label{eq:C_LCcoup_Network_invert_2}
\end{eqnarray}
Solving $\bi{y}$ in (\ref{eq:C_LCcoup_Network_invert_1}) and substituting in (\ref{eq:C_LCcoup_Network_invert_2}) we can derive the following equation
\begin{equation}
	(\bi{u}^T \bi{z})\frac{C_g}{\alpha c}\left(1- C_g \bi{a}^T \mathsf{A}^{-1}\bi{a}\right)=0,\nonumber
\end{equation}
where we have used the sum rule $|\bi{u}|^2=N_{\alpha}/\alpha c$ and $d=C_g-\alpha c$.  Notice furthermore that if we were to assume $\bi{u}^T\bi{z}$ were zero, equation (\ref{eq:C_LCcoup_Network_invert_2}) would tell us that $\bi{u}$ and $\bi{z}$ are parallel, and we would be forced to have $\bi{z}$ of zero norm, so we can conclude that  $\bi{u}^T\bi{z}\neq0$ if the eigenvector is not trivial. It follows  that, unless $\alpha$ is zero or infinity (in which case there is no  capacitive connection), a non-trivial solution can appear only when $\left(1- C_g \bi{a}^T \mathsf{A}^{-1}\bi{a}\right)=0$. Thus, unless  $\left(1- C_g \bi{a}^T \mathsf{A}^{-1}\bi{a}\right)$ is zero, $\mathsf{C}$ is invertible.

Having a non invertible capacitance matrix means that at least one combination of the initial variables will not be dynamical, and will be frozen in a value determined by the potential part. For our purposes, namely the provision of quantum mechanical models, this is a complication that can readily be eliminated by a good choice of variables, in which this frozen variable is discarded.

A different analysis corresponds to the inductance matrix. In this case the question at hand is the presence of zero modes. For a linear network where the potential $V(\boldsymbol{\phi})=0$, the condition for the invertibility of the inductance matrix $\mathsf{L}^{-1}$ can also be examined. In particular, consider $\mathsf{L}^{-1}$  given by Eq. (\ref{eq:Linv_LCcoup_Network}). Solving the equation $\mathsf{L}^{-1} (\bi{y}, \bi{z})^T=0$, and using the second sum rule  $\bi{u}^T (N_\alpha (\omega_n^2))^{-1} \bi{u}=\sum_n u_n(0)^2/N_\alpha \omega_n^2=\beta l$, see equation (\ref{eq:secondsumrule}) in Appendix \ref{Walter_appendix}, we can derive 
\begin{equation}
	(\bi{u}^T \bi{z})\frac{\beta l}{L_g}\left(1- \frac{1}{L_g} \bi{b}^T \mathsf{B}\bi{b}\right)=0.\nonumber
\end{equation}
Similarly to the capacitance coupling case, for $\beta \neq \{0,\infty\}$ the inductance matrix $\mathsf{L}^{-1}$ is  not invertible when $\left(1- \frac{1}{L_g} \bi{b}^T \mathsf{B}\bi{b}\right)=0$. In contrast to the capacitive case, given that there is no general potential,  $V(\boldsymbol{\phi})=0$, the description with such set of degrees of freedom can be used but a zero-mode will appear. 

The generalization to the $M$-networks connected to the transmission line is straightforward. In order for the capacitance matrix $\mathsf{C}$ and the inductance matrix $\mathsf{L}^{-1}$ in Eqs. (\ref{eq:C_LCcoup_2Networks}, \ref{eq:Linv_LCcoup_2Networks}) respectively to be non-invertible we require that 
\begin{eqnarray}
\quad\quad\quad\quad\quad\quad \frac{C_{gi}}{\alpha_i c}\left(1- C_{gi} \bi{a}_i^T \mathsf{A}_i^{-1}\bi{a}_i\right)&=&0,\nonumber\\
\quad\quad\quad\quad\quad\quad \frac{\beta_i l}{L_{gi}}\left(1- \frac{1}{L_{gi}} \bi{b}_i^T \mathsf{B}_i\bi{b}_i\right)&=&0,\nonumber
\end{eqnarray}
for all the networks, i.e. $ \forall i\in \{1,...,M\}$.

As we will see presently,  a frequent approach in the field of superconducting circuits is to truncate the number of modes to a finite quantity $N$. In so doing, the possibility exists that in the large $N$ limit the model presents non-dynamical modes, even if it is not the case for finite $N$, and some computations can present inadequate behaviours in that limit.

There is an additional reason, which will be exemplified in the next subsection, and is directly in line with the results presented above for transmission lines. In transmission lines capacitively coupled to networks we have seen that the choice of expansion modes is not free if we demand that the Hamiltonian description of the complete system  be understood as  being given by an infinite number of independent modes with no coupling among themselves, together with a network Hamiltonian, and coupling of this network to the independent modes. This basic idea of coupling of otherwise independent subsystems is essential in phenomenological model building. It is the case, however, that in some circumstances a na\"{\i}ve separation of subsystems will lead to non invertibility of the capacitance operator. That is, the separation in subsystems has led to overcounting of variables. One has to identify the structure of modes that precisely accounts for the proper amount of independent variables, by assessing how the couplings restrict our freedom in the choice of expansion.

In the next section, we are going to discuss this problem for the  particular case of the connection of two transmission lines via a very simple network. We will also provide an alternative solution and explanation.

\subsection{Linear coupling between Transmission Lines}
\label{subsec:TL_LCcoup_TL}

\begin{figure*}[h]
	\centering{\includegraphics[width=0.85\textwidth]{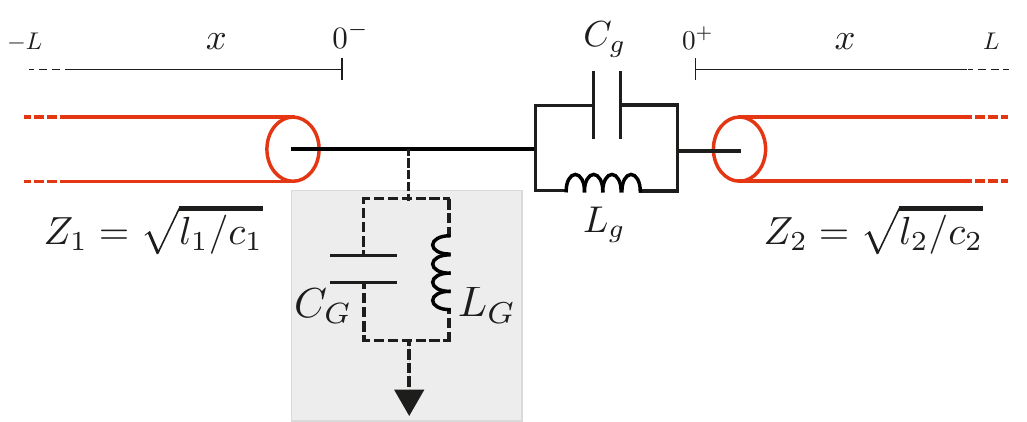}}
	\caption{\label{fig:TL_LCcoup_TL} \textbf{Transmission line inductively and capacitively coupled to a finite network}. Two transmission lines characterized by their capacitance $c_i$ and inductance $l_i$ per unit length are coupled and grounded through $LC$-resonators. Given that the flux field is described in partite bases, ground capacitor $C_G$ (inductor $L_G$) becomes necessary when the lines are capacitively (inductively) coupled for the $\mathsf{C}$ ($\mathsf{L}$) matrix to be invertible.}
\end{figure*}

Let us now consider the circuit in Fig. \ref{fig:TL_LCcoup_TL}, in which two transmission lines are coupled via a simple network. It is apparent that there are two subsystems, namely the left and right transmission lines. We shall see, however, that the description in those terms would be wrong if  either of the capacities $C_G$ and $C_g$ were absent. In such a situation, $C_GC_g=0$, there is a need for considering the whole system to identify the proper expansion in modes. Notice the difference with respect to the example of galvanic coupling, subsection \ref{sec:line-galv-coupl}, in that here the two endpoints are only connected by one oscillator. 

In this example we see that this essentially means that for there to be a description in terms of separate subsystems  there needs to be an endpoint variable for each transmission line that suitably dresses the transmission line modes, and that those endpoint variables be independent among themselves.

The Lagrangian for the circuit  reads
\begin{eqnarray} 
\fl \quad L = \int_{-L}^{0^-}dx\,\left[\frac{c_1}{2}\dot{\Phi}_1(x,t)^2 -\frac{1}{2l_1}(\Phi_1'(x,t))^2\right]+ \int_{0^+}^{L}dx\,\left[\frac{c_2}{2}\dot{\Phi}_2(x,t)^2-\frac{1}{2l_2}(\Phi_2'(x,t))^2\right]\nonumber\\
	+\frac{C_{g}}{2}(\dot{\Phi}_1(0^-,t)-\dot{\Phi}_2(0^+,t))^2-\frac{1}{2L_{g}}({\Phi}_1(0^-,t)-{\Phi}_2(0^+,t))^2\nonumber\\
	+\frac{C_{G}}{2}\dot{\Phi}_1(0^-,t)^2-\frac{1}{2L_{G}}{\Phi}_1(0^-,t)^2, \label{eq:Lag_TL_LCcoup_TL}
\end{eqnarray}
giving rise to wave equations for each of the transmission lines,
and  Kirchhoff's equations
\begin{eqnarray}
\fl \quad\quad\quad -\frac{1}{l}\Phi_1(0^-,t)=C_{G}\ddot{\Phi}(0^-,t)+\frac{1}{L_{G}}{\Phi}(0^-,t)-\frac{1}{l}\Phi_2(0^+,t),\nonumber\\
\fl \quad\quad\quad -\frac{1}{l}\Phi_2(0^+,t)=	C_{g}(\ddot{\Phi}_1(0^-,t)-\ddot{\Phi}_2(0^+,t)) + \frac{1}{L_{g}}({\Phi}_1(0^-,t)-\Phi_2(0^+,t)).\nonumber 
\end{eqnarray}

As has been our approach all along, we now look for expansions  $\Phi_1(x,t)=\sum_n \Phi_n(t)u_n(x)$ and $\Phi_2(x,t)=\sum_n \Psi_n(t)v_n(x)$ that provide us with a good description of the system.  Following the arguments presented in previous examples and in Appendix \ref{Walter_appendix}, we introduce one pair of free parameters,  $\alpha_i$ and $\beta_i$, for each boundary condition equation (\ref{eq:EVP_TL_LCcoup_TL_eq2}, \ref{eq:EVP_TL_LCcoup_TL_eq3}), and achieve separation of variables with eigenvalue-dependent boundary conditions,
\begin{eqnarray}
\ddot{\Phi}_n(t)=-\omega_n^2\Phi_n(t),\,\,\,
u_n''(x)=-k_n^2u_n(x),\label{eq:EVP_TL_LCcoup_TL_eq11}\\
\ddot{\Psi}_n(t)=-\Omega_n^2\Psi_n(t),\,\,\,
v_n''(x)=-\chi_n^2v_n(x),\label{eq:EVP_TL_LCcoup_TL_eq12}\\
u_{n}'(0^-)=k_n^2\alpha_1 u_n(0^-)-\frac{1}{\beta_1} u_n(0^-),\label{eq:EVP_TL_LCcoup_TL_eq2}\\
v_{n}'(0^+)=-\chi_n^2\alpha_2 v_n(0^+)+\frac{1}{\beta_2} v_n(0^+).\label{eq:EVP_TL_LCcoup_TL_eq3}
\end{eqnarray}
These generalized singular value problems fall in the class studied in Appendix \ref{Walter_appendix}, and the expansion theorems are guarantee of our approach. 
 Again, the eigenfunctions fulfill the orthogonality relations 
\begin{eqnarray}
\fl \quad\quad\quad \langle u_n,u_m\rangle_{\alpha_1}=c_1 \left(\int_{-L}^{0^-}dx\, u_n(x) u_m(x)+  \alpha_1 u_n(0^-)  u_m(0^-)\right)=N_{\alpha_1}\delta_{nm},\nonumber\\
\fl \quad\quad\quad \langle u_n,u_m\rangle_{1/\beta_1}=\frac{1}{l_1}\left(\int_{-L}^{0^-}dx\, u_n'(x) u_m'(x)+ \frac{1}{\beta_1}  u_n(0^-) u_m(0^-)\right)=\omega_n^2 N_{\alpha_1}\delta_{nm},\nonumber\\
\fl \quad\quad\quad \langle v_n,v_m\rangle_{\alpha_2}=c_2 \left(\int_{0^+}^{L}dx\, v_n(x) v_m(x)+  \alpha_2 v_n(0^+)  v_m(0^+)\right)=N_{\alpha_2}\delta_{nm},\nonumber\\
\fl \quad\quad\quad \langle v_n,v_m\rangle_{1/\beta_2}=\frac{1}{l_2}\left(\int_{0^+}^{L}dx\, v_n'(x) v_m'(x)+ \frac{1}{\beta_2}  v_n(0^+) v_m(0^+)\right)=\Omega_n^2 N_{\alpha_2}\delta_{nm},\nonumber
\end{eqnarray}
where $N_{\alpha_i}$ are free normalization constants with dimensions of capacitance. The Lagrangian (\ref{eq:Lag_TL_LCcoup_TL}) is now rewritten in terms of modes as 
\begin{equation}
L=\frac{1}{2}\dot{\bi{X}}^T\mathsf{C}\dot{\bi{X}}^T-\frac{1}{2}\bi{X}^T \mathsf{L}^{-1}\bi{X}\nonumber
\end{equation}
where the full flux vector is $\bi{X}=({\boldsymbol{\Phi}},
{\boldsymbol{\Psi}})^T$ and the capacitance and inverse inductance matrices are
\begin{eqnarray}
\mathsf{C}=\begin{pmatrix}
N_{\alpha_1} \mathbbm{1} + d_1\bi{u} \bi{u}^T& -C_{g} \bi{u}\bi{v}^T\\
-C_{g} \bi{v}\bi{u}^{T} & N_{\alpha_2} \mathbbm{1} + d_2\bi{v} \bi{v}^T
\end{pmatrix},\label{eq:C_TL_LCcoup_TL}\\
\mathsf{L}^{-1}=\begin{pmatrix}
N_{\alpha_1} (\omega_n^2) + e_1\bi{u} \bi{u}^T& - \bi{u}\bi{v}^T/L_{g}\\
- \bi{v}\bi{u}^{T}/L_{g} & N_{\alpha_2} (\Omega_n^2) + e_2\bi{v} \bi{v}^T
\end{pmatrix}.\label{eq:Linv_TL_LCcoup_TL}
\end{eqnarray}
Following the notational conventions we have used previously, 
the coupling vectors are named  $\bi{u}= ( u_0(0^-),  u_1(0^-),...)^T$ and  $\bi{v}= ( v_0(0^+),  v_1(0^+),...)^T$. We introduce  parameters $d_1 = C_{\Sigma}-c_1 \alpha_1$ and $d_2 = C_{g}-c_2 \alpha_2$ where $C_{\Sigma}=C_{g}+C_{G}$ is the total capacitance,  and $e_1= 1/L_{\Sigma} - 1/\beta_1 l_1$ and $e_2= 1/L_{g} - 1/\beta_2 l_2$ with $L_{\Sigma }=L_{g}L_{G}/(L_{g}+L_{G})$ being  the equivalent parallel inductance.

\subsubsection{Derivation of the Hamiltonian}
\label{sec:deriv-hamilt}
It is easy to calculate the inverse of the capacitance matrix (Legendre transformation) in this basis of modes
\begin{equation}
	\mathsf{C}^{-1}=\begin{pmatrix}
	\frac{1}{N_{\alpha_1}} \mathbbm{1} + \delta_1\bi{u} \bi{u}^T& \beta \bi{u}\bi{v}^T\\
	\beta \bi{v}\bi{u}^{T} & \frac{1}{N_{\alpha_2}} \mathbbm{1} + \delta_2\bi{v} \bi{v}^T
	\end{pmatrix},\nonumber
\end{equation}
Inserting the definitions of $d_1$ and $d_2$ and the normalization of vectors $|\bi{u}|^2=N_{\alpha_1}/c_1 \alpha_1$ and $|\bi{v}|^2=N_{\alpha_2}/c_2 \alpha_2$, we can check that the parameters are
\begin{eqnarray}
\qquad\qquad	\delta_1&=&\frac{c_1\alpha_1(c_1 \alpha_1 - C_G)}{C_G N_{\alpha_1}^2},\nonumber\\
\qquad\qquad	\delta_2&=&\frac{c_2 \alpha_2(c_2\alpha_2 C_{\Sigma}-C_g C_G)}{C_gC_G N_{\alpha_2}^2},\nonumber\\
\qquad\qquad	\beta&=&\frac{c_1 c_2 \alpha_1 \alpha_2}{C_G N_{\alpha_1}N_{\alpha_2}}.\nonumber
\end{eqnarray}
Let us now insert the condition that the modes in a transmission line have no direct coupling among themselves, i.e. $\delta_1=\delta_2=0$. This criterion determines the coefficients $\alpha_1$ and $\alpha_2$ to be $C_G/c_1$ and $C_GC_g/c_2(C_G+C_g)$ respectively. These are clearly the natural capacity length scales for each of the transmission lines, since they are given in both cases by the ratio of the total capacity from the endpoint of the line to the reference zero potential divided by the capacity density of the line. The coupling strength $\beta$ simplifies then to $C_GC_g/N_{\alpha_1}N_{\alpha_2}(C_G+C_g)$.

Let us now examine possible pathological situations. First of all, bear in mind that fixing the length scales $\alpha_1$ and $\alpha_2$ as above is necessary according the criterion we presented. Nonetheless, any value other than 0 or infinity would provide us with a description of the system, for general values of the parameters of the lumped elements.
 The value zero is excluded because it would not provide us with the description of the coupling. Such a condition entails there being no current at the endpoint. As to infinity, this would fix the value of the potential at the endpoint, again inhibiting coupling.

 There are two other pathological cases, given by $C_gC_G=0$. First, $C_g=0$. In this situation, $\delta_2$ blows up unless $\alpha_2$ is set to 0. But we are then in a case in which there is proper coupling of the transmission lines with our Hamiltonian  description. Let us therefore examine this case in more detail directly in the capacitance matrix itself, assuming that $\alpha_2\neq0$. In this  case with $C_g=0$, $d_2=-c_2 \alpha_2$ and, using the general result that $c_2\alpha_2=N_{\alpha_2}/|\bi{v}|^2$, the bottom-right submatrix in (\ref{eq:C_TL_LCcoup_TL}) becomes proportional to the projector $\mathbbm{1}-\bi{v}\bi{v}^T/|\bi{v}|^2$. This projector spans the vector space orthogonal to $\bi{v}$, and gives 0 when acting on $\bi{v}$. It follows that the column vector $(0^T,\bi{v}^T)^T$ is an eigenvector of $\mathsf{C}$ with eigenvalue 0. We have a nondynamical variable in our description.

 Passing now to the case $C_G=0$, one can write the capacitance matrix in this situation in the form
 \begin{equation}
   \nonumber 
   \mathsf{C}\to
   \begin{pmatrix}
     N_{\alpha_1}\left(\mathbbm{1}-\frac{\bi{u}\bi{u}^T}{|\bi{u}|^2}\right)&0\\ 0&  N_{\alpha_2}\left(\mathbbm{1}-\frac{\bi{v}\bi{v}^T}{|\bi{v}|^2}\right)
   \end{pmatrix}+
   C_g\begin{pmatrix}
     \bi{u}\bi{u}^T&  -\bi{u}\bi{v}^T\\ - \bi{v}\bi{u}^T & \bi{v}\bi{v}^T
   \end{pmatrix},
 \end{equation}
whence the zero eigenvalue vector $\left(\bi{u}^T/|\bi{u}|^2,\bi{v}^T/|\bi{v}|^2\right)^T$ is readily computed.
  Again, (\ref{eq:C_TL_LCcoup_TL}) becomes singular and we have an overcounting of the degrees of freedom. 

This analysis has provided us with an understanding of the issue beyond the purely algebraic treatment, in that we can only partition usefully the subsystem into its two composants if indeed there are enough degrees of freedom. In particular, we must have enough kinetic terms. This also suggests that in systems for which the partitioning fails in terms of non-invertibility of the capacitance matrix, it is sensible to study the possibility that our modelling is lacking some additional capacitances, which would solve the problem.
  
Regarding the rank of the inductance matrix (\ref{eq:Linv_TL_LCcoup_TL}),  and following the  section above, \ref{subsec:invertibility_variable_counting}, we study the equation $\mathsf{L}^{-1} (\bi{y}^T, \bi{z}^T)^T=0$. Introducing the parameters $e_1$, $e_2$ and the sum rules
$\bi{u}^T (N_{\alpha_1} (\omega_n^2))^{-1} \bi{u}=\sum_n u_n(0)^2/N_{\alpha_1} \omega_n^2=\beta_1 l_1$ and $\bi{v}^T (N_{\alpha_2} (\Omega_n^2))^{-1} \bi{v}=\sum_n v_n(0)^2/N_{\alpha_2} \Omega_n^2=\beta_2 l_2$, we derive the equation
\begin{equation}
	\frac{\beta_2 l_2}{L_g}\left(1- \frac{L_\Sigma}{L_g} \right)=0.\nonumber
\end{equation} 
Again, we can distinguish a few interesting cases. If there is inductive coupling between the lines, i.e. $L_g$ has a finite value, the limit of open ground inductor $L_G\rightarrow \infty$ makes the inductive matrix singular. On the other hand, if we disconnect the lines $L_g\rightarrow \infty$, we can find orthogonal complete bases as long as we develope the right field mode in a basis with $\beta_2 l_2\rightarrow \infty$.

\subsection{Exact and Approximate Quantization Methods}
\label{subsec:ch2_approximate_quantization}

Up to here we have studied exact methods to derive quantum Hamiltonians of circuits with transmission lines linearly coupled to networks of finite variables or other transmission lines. Thus, insofar as the starting point, namely the Lagrangian, is a good description of the system under study, so is the Hamiltonian, with no further approximation. Additionally, there are no divergences intrinsic to these models, since there is a natural cutoff.

\begin{figure*}[h]
	\centering{\includegraphics[width=0.65\textwidth]{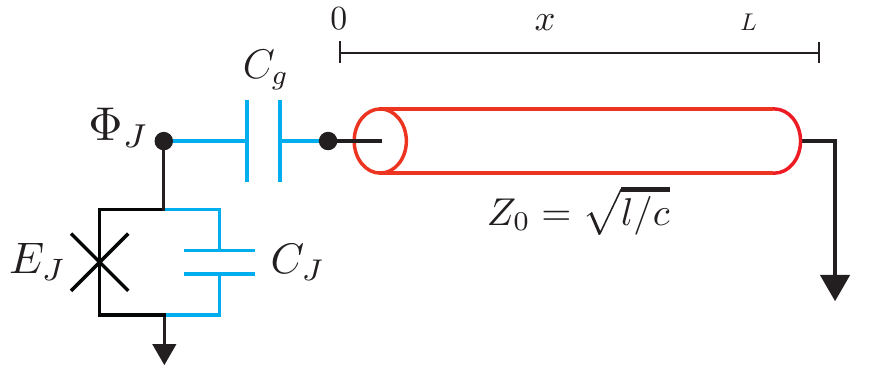}}
	\caption{\label{fig:TL_Ccoup_CQubit} \textbf{Transmission line capacitively coupled to a network composed of a unique Josephson junction}. The exact coupling parameters $g_n$ have a finite cutoff frequency due to the Josephson junction capacitance $C_J$, see Eqs. (\ref{eq:TL_Ccoup_CQubit_gn}) and (\ref{eq:TL_Ccoup_CQubit_gn_approx}), and Fig. \ref{fig:TL_Ccoup_CQubit_gn}.}
\end{figure*}

In this section we shall first use the previous techniques in a particularly simple example, for which we shall later portray some of the approximations present in the literature, with a view to clarifying how those approximations are the actual source of the divergences there encountered. The  example is that of a charge qubit capacitively coupled to a finite length transmission line resonator ended in a short to ground, see Fig. \ref{fig:TL_Ccoup_CQubit}.
Following \ref{subsubsec:chap2_mixed_linear_coupling}, the Lagrangian of the circuit in Fig. \ref{fig:TL_Ccoup_CQubit} can be written, once the field has been expanded in modes, as 
\begin{equation}
L=\frac{1}{2}\dot{\bi{X}}^T\mathsf{C}\dot{\bi{X}}-\frac{1}{2}\bi{X}^T \mathsf{L}^{-1}\bi{X} +E_J\cos(2\pi\phi_J/\Phi_0),\nonumber
\end{equation}
where we have defined the vector of fluxes $\bi{X}=(\phi_J,\boldsymbol{\Phi}^T)^T$, $\Phi_0$ is the magnetic flux quantum, and the capacitance and inductance matrices are
\begin{eqnarray}
\qquad\qquad \mathsf{C}=\begin{pmatrix}
C_J+C_g& -C_g \bi{u}^T\\
-C_g \bi{u} & N_{\alpha} \mathbbm{1} + d\bi{u} \bi{u}^T
\end{pmatrix},\nonumber\\ 
\qquad\qquad \mathsf{L}^{-1}=\begin{pmatrix}
0& 0^T\\
0 & N_{\alpha} (\omega_n^2)
\end{pmatrix},\nonumber 
\end{eqnarray}
where $d=C_g-\alpha c$. For definiteness, we rewrite here the homogeneous eigenvalue problem, Eqs. (\ref{eq:EVP_TL_LCcoup_Network_eq1}-\ref{eq:EVP_TL_LCcoup_Network_eq3}), that must be solved in order to find the eigenfrequencies $\omega_n$ and generalized eigenfunctions $u_n(x)$,
\begin{eqnarray}
\qquad\qquad\qquad	u_n''(x)&=&-k_n^2u_n(x),\label{eq:EVP_App_Quant_eq1}\\
\qquad\qquad\qquad	u_{n}'(0)&=&-k_n^2\alpha u_n(0),\label{eq:EVP_App_Quant_eq2}\\
\qquad\qquad\qquad	u_n(L)&=&0.\label{eq:EVP_App_Quant_eq3}
\end{eqnarray}
\subsubsection{Exact Legendre transformation }
We can solve the eigenvalue problem with eigenfunctions normalized by $\langle u_n,u_m\rangle_{\alpha}$ as in  (\ref{eq:TL_LCcoup_Network_ortho_1}). The eigenfunctions are readily seen to be $u_n(x)=A_n \sin(k_n(L-x))$, where the amplitude for  the $n^{\mathrm{th}}$ mode is
\begin{equation}
	A_n=\sqrt{\frac{2 N_\alpha}{c L}} \sqrt{\frac{1+(\alpha k_n)^2}{1+(\alpha/ L) + (\alpha k_n)^2}},\nonumber
\end{equation}
given the choice of normalization in (\ref{eq:TL_LCcoup_Network_ortho_1}). As to the wavenumbers of the modes, they are the positive non trivial solutions of  the transcendental equation
\begin{equation}
	\alpha k = \cot (k L).\nonumber
\end{equation}
We can calculate the components of the coupling vector $\bi{u}=(u_1(0),u_2(0),...)^T$ ,
\begin{equation}
	u_n(0)=\sqrt{\frac{2 N_\alpha}{c L}} \sqrt{\frac{1}{1+(\alpha/ L) + (\alpha k_n)^2}},\nonumber
\end{equation}
from which one can directly infer the finiteness of the norm, since $k_n\sim n$, whence $u_n(0)\sim n^{-1}$. Thanks to the results of Appendix \ref{Walter_appendix}, we know that this finite norm is $|\bi{u}|^2=N_\alpha/\alpha c$. Using Eq. (\ref{eq:TL_LCcoup_Network_alpha_fix}), we find that the choice $\alpha=C_gC_J/c(C_g+C_J)$  results in an exact Hamiltonian
\begin{equation}
H=\frac{1}{2}{\bi{P}}^T\mathsf{C}^{-1}{\bi{P}}-\frac{1}{2}\bi{X}^T \mathsf{L}^{-1}\bi{X} +E_J\cos(2\pi\phi_J/\Phi_0),\nonumber
\end{equation}
without mode-mode coupling, where the inverse of the capacitance matrix is
\begin{equation}
\mathsf{C}^{-1}=\begin{pmatrix}
\frac{1}{C_J}& \frac{C_g}{N_\alpha C_\Sigma} \bi{u}^T\\
\frac{C_g}{N_\alpha C_\Sigma} \bi{u} & \frac{1}{N_{\alpha}} \mathbbm{1}
\end{pmatrix},\nonumber\\
\end{equation}
and the charge conjugate variables are $\bi{P}=\partial L/\partial \dot{\bi{X}}=(q_J,\bi{Q}^T)^T$. The parameter $C_\Sigma$ is again $C_J+C_g$. Promoting the conjugate variables to operators and introducing annihilation and creation operators for the harmonic sector, we derive the Hamiltonian
\begin{equation}
H = 4 E_C \hat{n}_J^2-E_J \cos(2\pi\hat{\phi}_J/\Phi_0)+ \sum_n \hbar g_n \hat{n}_J(a_n+a_n^\dagger)+ \sum_n \hbar \omega_n a_n^\dagger a_n,\nonumber
\end{equation}
with the charge energy of the qubit $E_C =e^2/2 C_J$ and coupling defined as 
\begin{equation}
g_n= v_{p}\frac{C_g}{C_\Sigma} \sqrt{\frac{Z_0}{R_Q}}\sqrt{\frac{2\pi k_n}{L(1+(\alpha/L) + (\alpha k_n)^2)}}.\label{eq:TL_Ccoup_CQubit_gn}
\end{equation}
Here $R_Q= h/(2e)^2$ is the quantum of resistance and $v_{p}=1/\sqrt{lc}$ the velocity of propagation in the line. As  shown in \cite{Gely_2017_DivFree,Moein_2017_CutFree} the coupling grows as $g_n\propto \sqrt{k_n}$ for low frequency modes, where the wavenumbers resemble those of an open line, i.e. $k_n\approx (2n+1)\pi/2L$, and decays as $g_n\propto 1/\sqrt{k_n}$ for large $n$, when the wavenumbers tend to those of a line ended in a short, $k_n\approx n \pi/L$.  Observe that $g_n\sim n^{-1/2}$ for large $n$.

\begin{figure*}[h]
\centering{\includegraphics[width=0.6\textwidth]{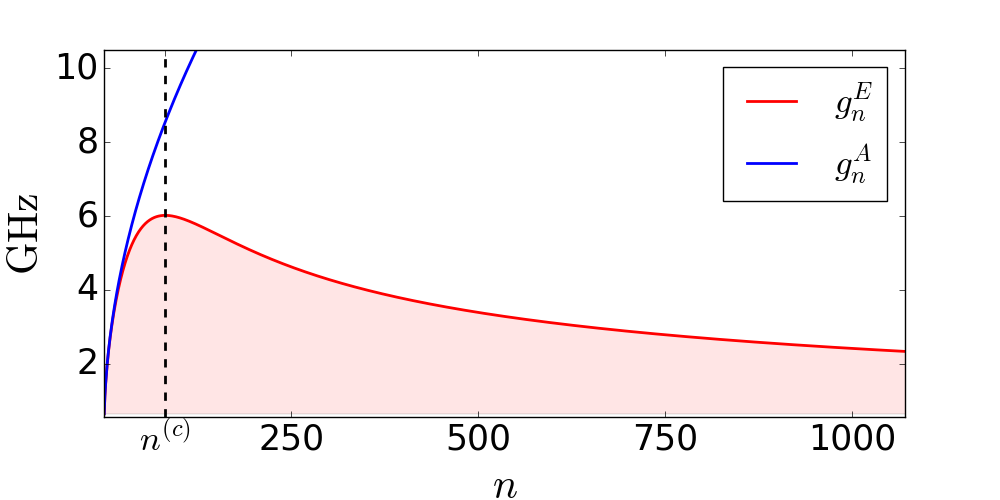}}
\caption{\label{fig:TL_Ccoup_CQubit_gn} \textbf{Capacitive coupling $g_n$ per mode for the circuit in Fig. \ref{fig:TL_Ccoup_CQubit}}. We apply the exact and approximate formulae to the experiment of Device A in \cite{Bosman_2017_MMUSC}, with parameters $C_g=40.3\,\mathrm{fF}$, $C_J=5.13 \,\mathrm{fF}$, $c=249\,\mathrm{pF/m}$, $l=623\,\mathrm{nH/m}$ and $L=4.7\,\mathrm{mm}$. In red, $g_n$ for the exact derivation, Eq. (\ref{eq:TL_Ccoup_CQubit_gn}): a natural cutoff of the coupling constant appears at mode number $n^{(c)}=81$ with frequency $\omega_{81}=702.5\,\mathrm{GHz}$ (dashed line). In blue, the approximate divergent coupling constant from Eq. (\ref{eq:TL_Ccoup_CQubit_gn_approx}).}
\end{figure*}

Given that, for large $n$, $(k_{n+1}-k_n)/k_n\sim n^{-1}$, if the maximum of $g_n$ is given at high $n$ we can treat the wavenumber as a continuous variable, and predict at which mode number the coupling saturates by solving the equation $\partial g_n /\partial k_n=0$, which yields as a result $k_n^{(c)}\approx\sqrt{\frac{1+\alpha/L}{\alpha^2}}$. For typical experimental values where the capacitance of the network is smaller than the total capacitance of the transmission line $\alpha/L\ll1$, such a device has many modes with frequencies close to a $\lambda/4$-resonator and the saturation point appears in a high frequency mode, see Fig. \ref{fig:TL_Ccoup_CQubit_gn}.

Clearly, $1/\alpha$ is a natural ultraviolet smooth cutoff. The frequencies in the model go all the way to infinity, but the coupling is indeed moderated by this cutoff. In our search of the literature we have not identified situations in which the maximum of the coupling has been detected, since the modes under study have lain rather below it. We submit this as a prediction for future experiments. 

\subsubsection{Approximations and introduction of a cutoff}
\label{sec:appr-intr-cutoff}
Let  us now compare the exact results above with some presentations in the literature which rely on some widely used  approximations.

As stated above, for low lying modes, such that $\alpha k_n\ll 1$, the couplings $g_n$ will scale with $\sqrt{k_n}$. The issue is that for many of the experimental setups the wavenumber for maximal coupling is very large in comparison with accessible wavenumbers.

Let us examine the situation in which  the total capacitance external to the transmission line, $C_gC_J/(C_g+C_J)$, is much smaller than the total capacitance of the transmission line, $cL$. In such a case, $\alpha \ll L$. Then two consistent approximations can be made. In the first place, the secular equation is best rewritten as $(\alpha/L) (kL)=\cot(kL)$. For $\alpha\ll L$ and low-lying modes, the equation is approximately $\cot(kL)=0$. In fact, a perturbative analysis shows that this is consistent for $k_n$ as long as $\alpha\ll L/n$. Secondly, the boundary condition (\ref{eq:EVP_App_Quant_eq2}) in such a case is well approximated by $u_n'(0)$, again with the same restriction for $n$, namely $n\ll L/\alpha$.
This approximation $u_n'(0)\approx 0$ gives us eigenfunctions $u_n(x)=N_c \cos(k_n x)$, where $N_c$ is a normalization constant, and wavenumbers $k_n=(2n+1)\pi/2L$. In contrast with the exact Hamiltonian in the  section above, this approximation, equivalent to setting $\alpha$ to zero,  results in a coupling vector $\bi{u}=(u_1(0),u_2(0),...)$ with infinite norm $|\bi{u}|^2=\sum_{n=1}^{\infty} N_c^2\rightarrow\infty$. This basis of modes is orthogonal with respect to the inner product $\langle u_n,u_m\rangle_{\alpha=0}$ (\ref{eq:TL_LCcoup_Network_ortho_1}), such that the two normalization constants are related through $N_c=\sqrt{2N_\alpha/cL}$.

The issue now is that the capacitance matrix $\mathsf{C}$ for the case of $\alpha=0$, while formally simple, presents vectors $\bi{u}$ all of whose components equal $N_c$, and the formulae for inversion cannot be applied. Let us therefore introduce a truncation of modes of the transmission line to $N$. The vector $\bi{u}$ has in this case norm squared $|u|^2= 2 N_\alpha N/cL$. Define the vector of unit length $\bi{e}=\bi{u}/|\bi{u}|$. The capacitance matrix reads
\begin{eqnarray}
  \mathsf{C}&=&
  \begin{pmatrix}
    C_\Sigma&- C_g \sqrt{\frac{2N N_\alpha}{cL}}\bi{e}^T\\ - C_g \sqrt{\frac{2N N_\alpha}{cL}}\bi{e}& N_\alpha\left(\mathbbm{1}+\frac{2 C_g N}{cL}\bi{e}\bi{e}^T\right)
  \end{pmatrix}\nonumber\\
  &=&\begin{pmatrix}
    C_J&0\\0&N_\alpha
  \end{pmatrix}+C_g
  \begin{pmatrix}
    1& - \sqrt{\frac{2N N_\alpha}{cL}}\bi{e}^T\\- \sqrt{\frac{2N N_\alpha}{cL}}\bi{e}& \frac{2N_\alpha N}{cL}\bi{e}\bi{e}^T
  \end{pmatrix}
\,\nonumber 
\end{eqnarray}
Assume that $C_g\ll cL$ (which entails the smallness of the total external capacitance in comparison to the capacitance of the transmission line). As long as the number of modes under consideration, $N$, is not too large, we can consider that the terms with $C_g$ are perturbative with respect to the others. Given two matrices $\mathsf{A}$ and $\mathsf{B}$ such that $\mathsf{B}$ can be understood as very small with respect to the invertible $\mathsf{A}$, we have the approximate expression $\left(\mathsf{A}+\mathsf{B}\right)^{-1}\approx\mathsf{A}^{-1}-\mathsf{A}^{-1}\mathsf{B}\mathsf{A}^{-1}$. Thus, to order $\left(C_gN/cL\right)^{1/2}$ we have an approximate inverse capacitance matrix
\begin{equation}
  \nonumber
  \mathsf{C}^{-1}\approx
  \begin{pmatrix}
    C_\Sigma^{-1}& \frac{C_g}{C_\Sigma}\sqrt{\frac{2N}{N_\alpha cL}}\bi{e}^T\\  \frac{C_g}{C_\Sigma}\sqrt{\frac{2N}{N_\alpha cL}}\bi{e} & N_\alpha^{-1}\mathbbm{1}
  \end{pmatrix}\approx \begin{pmatrix}
    C_\Sigma^{-1}& \frac{C_g}{C_\Sigma N_\alpha}\bi{u}^T\\  \frac{C_g}{C_\Sigma N_\alpha}\bi{u} & N_\alpha^{-1}\mathbbm{1}
  \end{pmatrix}.
\end{equation}
It is important to insist that this approximation is only valid if the number of modes taken into account is such that indeed $C_g N/cL\ll1$.  Nonetheless, the rightmost expression does not portray explicitly the truncation in modes. 

We now use this approximate inverse capacitance matrix. Promoting the conjugate variables $\bi{P}=\partial L/\partial \dot{\bi{X}}=(q_J, \bi{Q}^T)^T$ to operators and changing to annihilation and creation operators for the harmonic sector, we derive the Hamiltonian
\begin{equation}
	H \approx 4 E_C \hat{n}_J^2-E_J \cos(2\pi \hat{\phi}_J/\Phi_0)+ \sum_n \hbar g_n \hat{n}_J(a_n+a_n^\dagger)+ \sum_n \hbar \omega_n a_n^\dagger a_n\nonumber
\end{equation}
with the charge energy of the qubit $E_C =e^2/2 C_\Sigma$ and the coupling defined as 
\begin{equation}
	g_n= v_{p}\frac{C_g}{C_\Sigma} \sqrt{\frac{Z_0}{R_Q}}\sqrt{\frac{2\pi k_n}{L}},\label{eq:TL_Ccoup_CQubit_gn_approx}
\end{equation}
where $R_Q= h/(2e)^2$ is the quantum of resistance and $v_{p}=1/\sqrt{lc}$ the velocity of propagation in the line.
As repeatedly stated this Hamiltonian is only an approximation, valid for the low lying modes. The number of modes for which it applies can be large, if the experimental parameters are adequate. However, were we to take this Hamiltonian as the starting point, and $N\to\infty$, we would have divergences in the spectral function and other relevant quantities. Their origin is that the approximations that prominently feature in its derivation are incompatible with the ultraviolet limit, here represented by  $N\to\infty$.

Thus, in the literature, where frequently this Hamiltonian has indeed been taken as the phenomenological model to describe the system at hand, the introduction from outside of the model of an ultraviolet cutoff has been proposed almost systematically. From our point of view this is unnecessary, since this phenomenological Hamiltonian is only a good approximation to the lower modes, and there is a natural length that provides us with the cutoff, namely $\alpha$.

\section{Networks with Canonical Impedances}
The quantization techniques described above are useful to obtain Hamiltonian descriptions of circuit networks with transmission lines, starting from  first principles. More general passive environments, e.g. 3D superconducting cavities, have been used to design high-coherent qubits \cite{Paik_2011}. Lumped-element descriptions of the response function of such environments have been used within the {\it black-box} paradigm to derive Hamiltonians \cite{Nigg_2013,Solgun_2014,Solgun_2015}. This technique relies on a lumped-element description with numerable modes, such that its impedance response $Z(s)$ agrees with that of an electromagnetic environment either simulated with a computer solving Maxwell's equations or directly measured in an experiment.

The separation of system and environment degrees of freedom was not possible in \cite{Nigg_2013}, because the lumped-element circuit   expansion of the impedance was approximated with the first Foster form, and the linear part of the system was incorporated into the impedance. Thus, the Josephson-junction phase-drop degree of freedom had to be written in terms of all the harmonic variables, resulting in mode-mode couplings to all orders, see Eq. (6) in \cite{Nigg_2013}. On the other hand, other lumped-element descriptions, such as the second Foster expansion \cite{Devoret_1995_QFluct} and the Brune expansion \cite{Solgun_2014,Solgun_2015}, presented in an in-built way separation of the environment degrees of freedom and the ones of the network it is attached to through its ports. As mentioned in \cite{Nigg_2013} and \cite{Solgun_2014,Solgun_2015}, such descriptions have intrinsic convergent properties.

For historical reasons, we first review the derivation of Paladino et al. \cite{Paladino_2003} where a flux variable capacitively is coupled to a one port general lossless passive and reciprocal impedance $Z(s)$ expanded in an infinite series of harmonic oscillators, i.e. the first Foster form. Recall that such description with a stage of a lone capacitor without inductor would correspond to a total impedance $Z_T(s)=\frac{1}{s C_B} + Z(s)$, as  seen by the anharmonic variable, with a pole at $s=0$. The rest of the expansion must be an electromagnetic environment whose impedance response at frequency $s=0$ has a zero, i.e. $Z(0)=0$. The generalization to the coupling of the general impedance to a more complex network can be easily done using the results of the sections above.

In the second section, we extend the multi-network case that we studied in previous section \ref{subsec:TL_LCcoup_TL}, where the infinite dimensional subsystem was a transmission line, to a general multi-mode infinite-dimensional lossless passive and reciprocal environment that couples linearly to finite-dimensional networks. We also show how this analysis can be applied for example to simplify the quantization of the $1^{\mathrm{st}}$-Foster circuit done by Paladino et al. Mathematical details and other particular circuit cases as the $2^{\mathrm{nd}}$-Foster expansion are left for an Appendix section. In this section, we have restricted ourselves to the analysis of infinite-dimensional environments capacitively coupled to networks to lighten up the proofs, as the combined case with inductive coupling is an easy extension of this problem.

\subsection{$1^{\mathrm{st}}$ Foster-Form Impedance Quantization}
\label{sec:1st-foster-expansion}

Equivalently to the analysis in  section \ref{sec:netw-with-transm}, the main goal is to find a Hamiltonian where the infinite set of canonical variables are coupled to a finite set of them without mode-mode couplings  and without divergence issues. In the literature these two points are frequently related by referring to the $A^2$-term. In the present context, as will become clearer in the next section, divergences are physical inasmuch as they impact either on an infinite mass renormalization for the finite set of variables or in quantities determined by the spectral density, that codifies the effect of the environment on the reduced dynamics of the network. In both cases, the divergences can be traced back to the normalizability of the coupling vectors/matrices in the block diagonal decomposition, with respect to the proper inner product. Similarly to the case of network lines (see in particular \ref{sec:appr-intr-cutoff}), it might well be the case that the root of the divergence is that an approximation that is valid for a truncation to a finite number $N$ of impedance modes is not valid in the limit $N\to\infty$. An alternative problem arises when some transformations are carried out in the finite $N$ case, and intermediate computations become invalid in the infinite limit.  This has caused difficulties in the literature that have led several authors to convoluted arguments to be able to discard such divergences.

Here we consider an example that can, with a special choice of parameters $C_\alpha$ and $L_\alpha$, also be used to describe a transmission line. This will prove convenient to relate  both approaches.

We first follow, with a number of simplifications and generalizations, an analysis proposed by Paladino et al. \cite{Paladino_2003}, and we prove that it corresponds to a particular canonical transformation to diagonalize impedance modes. The focus is to allow us some freedom in rescaling and reorganizing the impedance modes, with the criterion that the final Hamiltonian presents no mode-mode coupling. This freedom is the analogue to the freedom in $\alpha$ and $\beta$ parameters in the transmission line case.

\begin{figure*}[ht]
	\centering{\includegraphics[width=0.85\textwidth]{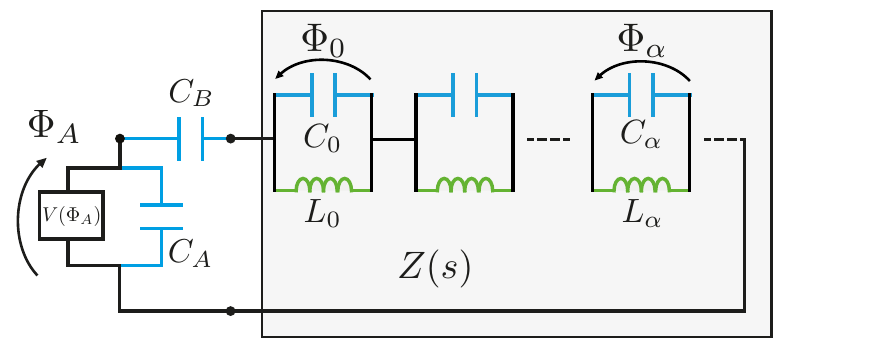}}
	\caption{\label{fig:1st_Foster_form} \textbf{$1^{\mathrm{st}}$-Foster form capacitively coupled to an anharmonic flux variable}. One port impedance $Z(s)$ modelled by a series of LC oscillators ($C_\alpha,L_\alpha$) is capacitively coupled to an anharmonic variable $\phi_A$ with capacitance $C_A$ and potential $V(\Phi_A)$ through capacitor $C_B$.}
      \end{figure*}

      Let us consider a family of circuits described by  Fig. \ref{fig:1st_Foster_form}. The corresponding Lagrangian     can be written choosing as variables the branch flux differences at the capacitors $C_\alpha$ and $C_A$,
\begin{eqnarray}
L&=& \frac{C_A}{2} \dot{\Phi}_A^2+\frac{C_B}{2} \left(\dot{\Phi}_A-\sum_{\alpha}\dot{\Phi}_{\alpha}\right)^2+\sum_{\alpha}\left[\frac{C_{\alpha}}{2}\dot{\Phi}_{\alpha}^2-\frac{1}{2L_{\alpha}}\Phi_{\alpha}^2\right]-V(\Phi_A)\nonumber\\
&=&\frac{1}{2}\dot{\bi{\Phi}}^T \mathsf{C}\dot{\bi{\Phi}}-\frac{1}{2}\bi{\Phi}^T \mathsf{L}^{-1}\bi{\Phi} - V(\Phi_A),\label{eq:Lag_Paladino_circuit}
\end{eqnarray}
where $\bi{\Phi}=(\Phi_A,\bi{\Phi}_{\alpha}^T)^T=(\Phi_A, \Phi_0, \Phi_1, ...)^T$. The capacitance matrix reads
\begin{equation}
\mathsf{C}=\begin{pmatrix}
C_{\Sigma} & -C_B\bi{e}_\alpha^T\\
-C_B\bi{e}_\alpha & \mathsf{C}_{\alpha}+C_B \bi{e}_\alpha\bi{e}_\alpha^T
\end{pmatrix},\label{eq:Paladino_Cmat}
\end{equation}
where we define $C_{\Sigma}=C_A+C_B$, the variable $\Phi_A$ couples equally  to all $\Phi_\alpha$, and thus to $\bi{\Phi}_{\alpha}$ through the vector of ones $\bi{e}_\alpha=(1, 1, 1, ...)^T$, and the inductance matrix is 
\begin{equation}
\mathsf{L}^{-1}=\begin{pmatrix}
0 & 0\\
0 & \mathsf{L}_\alpha^{-1}
\end{pmatrix}.\nonumber
\end{equation}

This coupling vector has an $l^2$ norm $|\bi{e}_\alpha|^2=N$, with $N$ the number of stages, that diverges as $N$ tends to infinity. In this limit, then, one must be careful in assigning meaning while inverting the capacitance matrix when $N\rightarrow\infty$. In particular, and as signalled in Appendix \ref{App:1st_Foster}, we will have a pathological case if, in the limit $N\to\infty$, we have that $\bi{e}_\alpha^T\mathsf{C}_\alpha^{-1}\bi{e}_\alpha\to\infty$.  In what follows we are assuming $C_A\neq0$, such that the total capacitance matrix is full rank even in that limit, something that can be checked using the same arguments as in Sec.  \ref{subsec:invertibility_variable_counting} and taking the limit, see again  Appendix \ref{App:1st_Foster}.

Our objective is to carry out changes of variables that allow us to write a Hamiltonian with no mode-mode coupling that still mantains a coupling between the external variable and the impedance variables. In order to do so, let us start by dressing the external coordinate with mode coordinates, allowing ourselves some freedom in the amount of dressing. That is, we introduce a new coordinate $\eta$ as a linear combination of the old coordinates  $\Phi_A= \eta + t \bi{e}_\alpha^T\bi{\Phi}_{\alpha}$ with a free parameter $t$, and write the modified Lagrangian
\begin{equation}
L_I=\frac{1}{2}\dot{\bi{x}}^T \mathsf{C}_{x}\dot{\bi{x}}-\frac{1}{2}\bi{x}^T \mathsf{L}^{-1}\bi{x} - V(\eta + t  \bi{e}_\alpha^T\bi{\Phi}_{\alpha})\label{eq:Lag_Paladino_circuit_I}
\end{equation}
where $\bi{x}=(\eta,\bi{x}_{\alpha}^T)^T$, with $\bi{x}_{\alpha}=\bi{\Phi}_\alpha$ and
\begin{equation}
\mathsf{C}_{x}=\begin{pmatrix}
a & b\bi{e}_\alpha^T\\
b\bi{e}_\alpha & \mathsf{M}_{\alpha}
\end{pmatrix}.\label{eq:Paladino_diag_C_I}
\end{equation}
The block capacitance matrix $\mathsf{M}_{\alpha}$ equals $\mathsf{C}_\alpha+d \bi{e}_\alpha\bi{e}_\alpha^T$, and we introduce parameters  $a=C_{\Sigma}$, $b=(tC_{\Sigma}-C_B)$ and $d = C_B - 2 C_B t + C_{\Sigma} t^2$.

In the second step, we rescale the coordinates to diagonalize the capacitance matrix $\mathsf{M}_{\alpha}$ as $\bi{x}_{\alpha}=M_0^{1/2} \mathsf{M}_{\alpha}^{-1/2}\bi{y}_{\alpha}$, where $\bi{y}=(\eta, \bi{y}_{\alpha}^T)^T$ and $M_0$ is a constant with dimensions of capacitance. That is, 
\begin{eqnarray}
\mathsf{C}_{y}=\begin{pmatrix}
a & b\bi{f}_\alpha^T\\
b\bi{f}_\alpha & M_0\mathbbm{1}
\end{pmatrix},\quad \mathrm{and}\label{eq:Paladino_diag_C_II} \quad 
\mathsf{L}_{y}^{-1}=\begin{pmatrix}
0 & 0\\
0 & (\mathsf{L}_{\alpha}^{y})^{-1}
\end{pmatrix}.
\end{eqnarray}
The inductance block submatrix reads $(\mathsf{L}_{\alpha}^{y})^{-1}=M_0 \mathsf{M}_{\alpha}^{-1/2}\mathsf{L}_{\alpha}^{-1}\mathsf{M}_{\alpha}^{-1/2}$, while (and this is the crucial point) the new coupling matrix is given by $\bi{f}_\alpha=M_0^{1/2}\mathsf{M}_{\alpha}^{-1/2}\bi{e}_\alpha$. The Lagrangian in this second step is therefore
\begin{equation}
L_{II}=\frac{1}{2}\dot{\bi{y}}^T \mathsf{C}_{y}\dot{\bi{y}}-\frac{1}{2}\bi{y}^T \mathsf{L}_y^{-1}\bi{y} - V(\eta + t  \bi{e}_\alpha^T\bi{\Phi}_{\alpha}),\nonumber 
\end{equation}
where we have kept the old variables in the anharmonic potential for simplicity. It can be easily checked that the new coupling vectors $\bi{f}_\alpha$ have finite norm in the limit of infinite oscillators even if  $\lim\limits_{N\rightarrow\infty}\bi{e}_\alpha^T \mathsf{C}_\alpha^{-1} \bi{e}_\alpha=\infty$, see Appendix \ref{App:1st_Foster} for the complete proof. In this special but very common case (see for example  \cite{Gely_2017_DivFree}), 
\begin{equation}
\lim\limits_{N \rightarrow\infty} M_0^{-1}|\bi{f}_\alpha|^2=\lim\limits_{N \rightarrow\infty} \bi{e}_\alpha^T\mathsf{M}_{\alpha}^{-1}\bi{e}_\alpha=1/d.\nonumber
\end{equation}
In the third step we undo the initial point transformation through $\eta = \Phi_A - t \bi{e}_\alpha^T\bi{\Phi}_{\alpha}$, in order to remove the interaction from the general potential $V(\Phi_A)$,
\begin{equation}
L_{III}=\frac{1}{2}\dot{\bi{z}}^T \mathsf{C}_{z}\dot{\bi{z}}-\frac{1}{2}\bi{z}^T \mathsf{L}_{z}^{-1}\bi{z} -V(\Phi_A)\nonumber
\end{equation}
with $\bi{z}=(\Phi_A,\bi{z}_\alpha)$, $\mathsf{L}_{z}^{-1}=\mathsf{L}_{y}^{-1}$, and where the capacitance matrix has trasformed to
\begin{eqnarray}
\mathsf{C}_{z}&=&\begin{pmatrix}
C_{\Sigma} & (b+tC_{\Sigma})\bi{f}_\alpha^T\\
(b+tC_{\Sigma})\bi{f}_\alpha & M_0\mathbbm{1}+C_{\Sigma}t^2\bi{f}_\alpha\bi{f}_\alpha^T
\end{pmatrix}.\nonumber
\end{eqnarray}
Now that we have finite-norm coupling vectors, we can invert the capacitance matrix 
\begin{eqnarray}
\mathsf{C}_{z}^{-1}&=&\begin{pmatrix}
\alpha & \beta\bi{f}_\alpha^T\\
\beta\bi{f}_\alpha & M_0^{-1}\mathbbm{1}+\delta\bi{f}_\alpha\bi{f}_\alpha^T
\end{pmatrix},\nonumber 
\end{eqnarray}
where we have defined the parameters
\begin{eqnarray}
\qquad\qquad \alpha&=&\frac{M_0 + C_{\Sigma}t^2|\bi{f}_\alpha|^2}{D_z},\nonumber\\
\qquad\qquad \beta &=& \frac{-(b+C_{\Sigma}t)}{D_z},\nonumber\\
\qquad\qquad \delta &=&\frac{(b+C_{\Sigma}t)^2-C_{\Sigma}^{2}t^{2}} {M_0D_z},\nonumber\\
\qquad\qquad D_z&=&M_0 C_{\Sigma}+((b+C_{\Sigma}t)^2-C_{\Sigma}^{2}t^{2})|\bi{f}_\alpha|^2,\nonumber
\end{eqnarray}
and derive the Hamiltonian
\begin{equation}
H=\frac{1}{2}\bi{p}^T \mathsf{C}_{z}^{-1}\bi{p}+\frac{1}{2}\bi{z}^T \mathsf{L}_{z}^{-1}\bi{z}+V(\Phi_A).\nonumber 
\end{equation}
We denote with $\bi{p}=\partial L_{III}/\partial \dot{\bi{z}}=(q_A,\bi{p}_\alpha^T)^T$ the charge variables conjugate to the $\bi{z}$ fluxes. In order to simplify the Hamiltonian and remove the mode-mode coupling ($A^{2}$-like term) in the capacitance sector, we use one of the solutions to the equation $\delta=0$, i.e. $t=C_B/C_{\Sigma}$ with the condition that $b=0$, and we obtain $d=C_s=C_A C_B/C_{\Sigma}$, which is the series capacitance seen by the impedance. Finally, we remove the mode-mode coupling in the inductance matrix with a canonical unitary transformation $\mathsf{U}$, such that $(\bar{\mathsf{L}}_\alpha^{z})^{-1}=\mathsf{U}(\mathsf{L}_\alpha^{z})^{-1}\mathsf{U}^T$ be diagonal. The variables are then rotated through $\bar{\bi{p}}_\alpha=\mathsf{U} \bi{p}_\alpha$ and $\bar{\bi{z}}_\alpha=\mathsf{U}^T \bi{z}_\alpha$. All the expressions can be simplified in the limit of infinite oscillators, and for this specific case where $\lim\limits_{N\rightarrow \infty}|\bi{f}_\alpha|^2\rightarrow M_0/d$, the Hamiltonian reduces to
\begin{equation}
H=\frac{q_A^{2}}{2C_A}+V(\Phi_A)-\frac{C_B}{M_0 C_{\Sigma}}q_A\sum_{\alpha}\bar{f}_{\alpha}\bar{p}_{\alpha}+\sum_{\alpha}\frac{1}{2}\left[\frac{\bar{p}^2_{\alpha}}{M_0}+M_0 \Omega_{\alpha}^{2}\bar{z}_{\alpha}^{2}\right].\nonumber 
\end{equation}
with the frequencies $\Omega_{\alpha}\equiv (M_0 \bar{L}_{\alpha}^{z})^{-1/2}$ and the rotated coupling vectors $\bar{\bi{f}}_\alpha=\mathsf{U}\bi{f}_\alpha$, which preserve the same norm as the old ones $|\bi{\bar{f}}_\alpha|^2=|\bi{f}_\alpha|^2$. As previously commented, the canonical quantization procedure 
directly goes  through, by promoting the canonical variables to operators with canonical commutation relations  $[\Phi_A,q_A]=i\hbar$ and $[\bar{z}_\alpha,\bar{p}_\beta]=i\hbar \delta_{\alpha\beta}$.

This procedure has followed the structure of that presented in \cite{Paladino_2003}. We have been explicit about each step, in order to dispel some misconceptions that have arisen in the literature. We shall see in the next section, by extending it to the multiport case, that it can be replaced by the introduction of a single canonical transformation that only pertains to the impedance modes. The transformation, parametrized by an operator $M_\alpha$, incorporates the freedom we gave us via the $t$ parameter, which disappears, and will be determined by the requirement of no mode-mode coupling.

\subsection{Multiport Impedance Quantization}
\label{sec:multi-port-impedance}

\begin{figure*}[ht]
\centering{\includegraphics[width=0.5\textwidth]{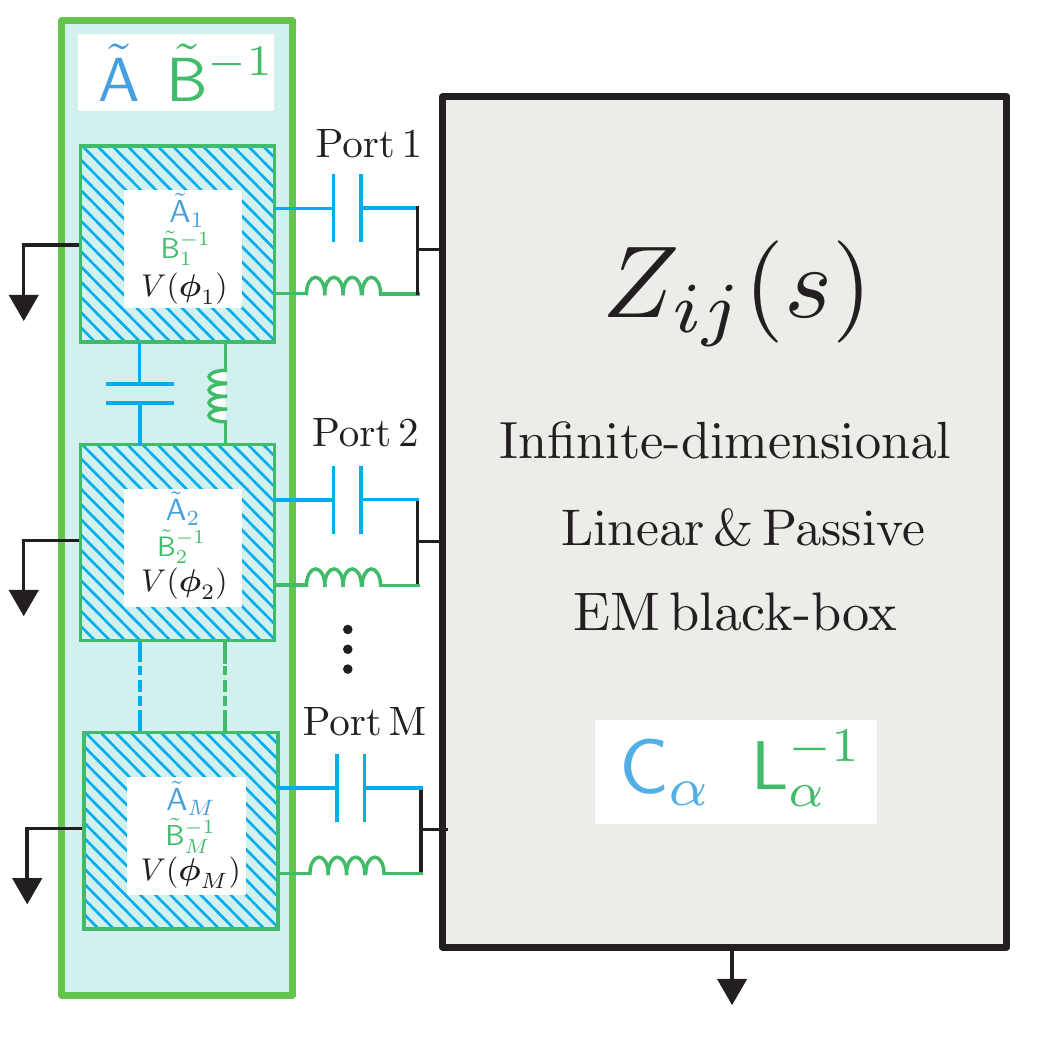}}
\caption{\label{fig:MPort_Z_LCcoup_Networks} \textbf{Infinite-dimensional multiport impedance connected to finite sized anharmonic network.} A linear and passive, electromagnetic (EM) environment is modeled as a multiport impedance $Z_{ij}(s)$, fitted to a general lumped-element circuit with capacitance and inductance matrices ($\mathsf{C}_\alpha, \mathsf{L}^{-1}_\alpha$). The general infinite dimensional system is linearly coupled to anharmonic networks ($\mathsf{A}_i,\mathsf{B}_i^{-1},V(\boldsymbol{\phi}_i)$) with finite degrees of freedom that are also directly connected.}
\end{figure*}

We have already encountered a case of a multiport impedance in section \ref{sec:mult-netw-coupl}. Since that was a case involving a transmission line, the methods presented above were better suited for the analysis. Nonetheless, it can also be analyzed from the perspective of this section; see figure \ref{fig:MPort_Z_LCcoup_Networks} for a general multiport circuit linearly coupled to M
non-linear networks. We concentrate, as always, on capacitive coupling.

Let us consider thus a capacitance matrix of the form
\begin{equation}
  \label{eq:multiportcapacitance}
  \mathsf{C}=
  \begin{pmatrix}
    \mathsf{A}& -\sum_i\bi{a}_i\bi{u}^T_i \\ -\sum_i \bi{u}_i\bi{a}_i^T &  \mathsf{C}_\alpha+\sum_i\bi{u}_i\bi{u}_i^T
  \end{pmatrix},
\end{equation}
where we assume that $\mathsf{A}$ and $\mathsf{C}_\alpha$ are symmetric and positive, and that the vectors $\left\{\bi{a}_i\right\}_{i=1}^M$ on one side and $\left\{\bi{u}_i\right\}_{i=1}^M$ on another side are separately linearly independent. 
The notation used here is reminiscent but not completely equivalent to that in \ref{sec:mult-netw-coupl}. Namely, what were presented there as $\bi{a}_i$ give rise to vectors here with the same notation, after padding with zeroes. Furthermore, we have chosen a different normalisation in order to unclutter formulae.

The matrix $\mathsf{C}$ is a  block diagonal matrix perturbed by off-diagonal blocks each of rank $M$. This is the correct description for an $M$
port circuit, and is amenable to the inversion given in Appendix \ref{sec:multiport-impedanceapp}. The general formula presented there is however not very illuminating, and in the particular case of $\mathsf{C}$ we can present the inverse in a much cleaner way, as follows:
\begin{equation}
  \label{eq:inverse-multiport-c}
  \mathsf{C}^{-1}=
  \begin{pmatrix}
    \mathsf{A}^{-1}+\beta_{ij}\mathsf{A}^{-1}\bi{a}_i\bi{a}_j^T\mathsf{A}^{-1} & \rho_{ij} \mathsf{A}^{-1}\bi{a}_i\bi{u}_j^T\mathsf{C}_\alpha^{-1}\\
    \gamma_{ij} \mathsf{C}_\alpha^{-1}\bi{u}_i\bi{a}_j^T \mathsf{A}^{-1}& \mathsf{C}_\alpha^{-1}+\lambda_{ij}\mathsf{C}_\alpha^{-1}\bi{u}_i\bi{u}_j^T\mathsf{C}_\alpha^{-1}
  \end{pmatrix},
\end{equation}
where Einstein summation convention has been used, as it will henceforward. On demanding that this matrix indeed be the inverse we obtain four linear equations for the matrices $\mathsf{\beta}$, $\mathsf{\rho}$, $\mathsf{\gamma}$ and $\mathsf{\lambda}$. The solution to this system of matrix equations is
\begin{eqnarray}
  \label{eq:inverse-cap-coeffmatrices}
  \beta&=&\mu\cdot\left(\mathbbm{1}+\mu-\nu\cdot\mu\right)^{-1},\nonumber\\
  \gamma&=&  \left(\mathbbm{1}+\mu-\nu\cdot\mu\right)^{-1},\nonumber\\
  \lambda&=& \left(\mathbbm{1}+\mu-\nu\cdot\mu\right)^{-1}\cdot \left(\nu-\mathbbm{1}\right),\nonumber\\
  \rho&=& \mathbbm{1}+\mu\cdot\left(\mathbbm{1}+\mu-\nu\cdot\mu\right)^{-1}\cdot\left(\nu-\mathbbm{1}\right).
\end{eqnarray}
Here, the matrices $\mu$ and $\nu$ are given by
\begin{eqnarray}
  \qquad \quad \mu_{ij}=\bi{u}_i^T\mathsf{C}_\alpha^{-1}\bi{u}_j\quad \mathrm{and} \quad 
  \nu_{ij}=\bi{a}_i^T\mathsf{A}^{-1}\bi{a}_j.\nonumber
\end{eqnarray}
Assume that the multiport impedance presents infinite modes. A possible approach is to cutoff the number of modes in the impedance to a finite number $N$. 
Now, the issue, as pointed above for the single port case, is that the matrix $\mu$ can blow up in an $N\to\infty$ limit. That, by itself, might not be so pernicious. However, in such a situation $\lambda$ would tend to zero, and the coupling matrix norm, defined as $\rho_{ij}\gamma_{ki}\bi{u}_j^T\mathsf{C}_\alpha^{-2}\bi{u}_k$, could also tend to zero.

The final coupling matrix will be obtained after a canonical transformation that diagonalises the submatrix $\mathsf{D}_\alpha^{-1}=\mathsf{C}_\alpha^{-1}+\lambda_{ij}\mathsf{C}_\alpha^{-1}\bi{u}_i\bi{u}_j^T\mathsf{C}_\alpha^{-1}$, and simultaneously the corresponding inductance submatrix. This can be achieved by rescaling the momenta $\bi{q}_\alpha$ with the square root of this matrix.  If indeed $\mathsf{D}_\alpha^{-1}$ is positive, the coupling matrix is finite.

We can follow here the steps of the analysis for the first Foster form, in which a free parameter is introduced by first dressing and then undressing the network variables with impedance variables, and in between rescaling and reordering the impedance variables. Assume that the initial network variables are collected in a vector $\bi{\Phi}_A$, while the impedance variables are $\bi{\Phi}_\alpha$.  We shift network variables with the change of variables
\begin{equation}
  \begin{pmatrix}
    \bi{x}_A\\  \bi{x}_\alpha
  \end{pmatrix}=
  \begin{pmatrix}
    \mathbbm{1}&-\bi{b}_i\bi{u}_i^T\\0&\mathbbm{1}
  \end{pmatrix}\begin{pmatrix}
    \bi{\Phi}_A\\ \bi{\Phi}_\alpha
  \end{pmatrix},\nonumber
\end{equation}
where $\left\{\bi{b}_i\right\}$ is a set of vectors to be determined later, that take the role of the $t$ parameter for the first Foster form. The capacitance matrix for the new variables reads
\begin{equation}
  \mathsf{C}_x=
  \begin{pmatrix}
    \mathsf{A}& \left(\mathsf{A}\bi{b}_i-\bi{a}_i\right)\bi{u}_j^T\\
    \bi{u}\left(\bi{b}_i^T\mathsf{A}-\bi{a}_i^T\right)& \mathsf{M}_\alpha
  \end{pmatrix},\nonumber
\end{equation}
where
\begin{equation}
  \mathsf{M}_\alpha= \mathsf{C}_\alpha+\left(\delta_{ij}-2\bi{b}_i^T\bi{a}_j+\bi{b}_i^T\mathsf{A}\bi{b}_j\right)\bi{u}_i\bi{u}_j^T.\nonumber
\end{equation}
We change variables again, in the form
\begin{equation}
  \begin{pmatrix}
    \bi{y}_A\\  \bi{y}_\alpha
  \end{pmatrix}=
  \begin{pmatrix}
    \mathbbm{1}&0\\0& \mathsf{M}_\alpha^{1/2}  \end{pmatrix}
  \begin{pmatrix}
    \boldsymbol{\varphi}_A\\ \boldsymbol{\varphi}_\alpha
  \end{pmatrix},\nonumber
\end{equation}
leading to the capacitance matrix
\begin{equation}
  \nonumber 
  \mathsf{C}_y=
  \begin{pmatrix}
    \mathsf{A}& \left(\mathsf{A}\bi{b}_i-\bi{a}_i\right)\bi{u}_j^T\mathsf{M}_\alpha^{-1/2}\\
    \mathsf{M}_\alpha^{-1/2}\bi{u}\left(\bi{b}_i^T\mathsf{A}-\bi{a}_i^T\right)& \mathbbm{1}
  \end{pmatrix}.
\end{equation}
We now undo the shift of the network variables, by
\begin{equation}
  \begin{pmatrix}
     \bi{z}_A\\  \bi{z}_\alpha
  \end{pmatrix}=
  \begin{pmatrix}
    \mathbbm{1}&\bi{b}_i\bi{u}_i^T\mathsf{M}_\alpha^{-1/2}\\ 0&\mathbbm{1}
  \end{pmatrix}
\begin{pmatrix}
    \boldsymbol{\sigma}_A\\ \boldsymbol{\sigma}_\alpha
  \end{pmatrix},\nonumber
\end{equation}
leaving a final capacitance matrix
\begin{equation}
  \mathsf{C}_z=
  \begin{pmatrix}
    \mathsf{A}& -\bi{a}_i\bi{u}_i^T\mathsf{M}_\alpha^{-1/2}\\ \mathsf{M}_\alpha^{-1/2}\bi{u}_i\bi{a}_i^T&\mathbbm{1}-\left(\bi{b}_j^T\mathsf{A}\bi{b}_k-\bi{a}_j^T\bi{b}_k-\bi{b}_j^T\bi{a}_k\right) \mathsf{M}_\alpha^{-1/2}\bi{u}_j\bi{u}_k^T\mathsf{M}_\alpha^{-1/2}
  \end{pmatrix}.\nonumber
\end{equation}
It now behoves us to invert this final capacitance matrix and demand that the inverse capacitance matrix presents no coupling between impedance modes. By construction this then entails that the corresponding submatrix is the identity matrix, and the possible coupling between impedance modes due to  inductance can be eliminated by diagonalising the corresponding inductance matrix. The condition of no coupling is seen to be achieved with the choice $\bi{b}_j=\mathsf{A}^{-1}\bi{a}_j$. This provides us with the  matrix $\mathsf{M}_\alpha=\mathsf{C}_\alpha+\left(\delta_{ij}-\bi{a}_i^T\mathsf{A}^{-1}\bi{a}_j\right)\bi{u}_i\bi{u}_j^T$. In order for the procedure to work, we require that this matrix be positive. Furthermore, the coupling matrix $\bi{u}_i^T\mathsf{M}_\alpha^{-1}\bi{u}_j$ should have finite components.

This presentation actually suggests a different approach. The complete succession of changes of variables is a point transformation that can be compacted to
\begin{equation}
  \begin{pmatrix}
    \bi{z}_A\\ \bi{z}_\alpha
  \end{pmatrix}=\begin{pmatrix}
    \mathbbm{1}&0\\0& \mathsf{M}_\alpha^{1/2}  \end{pmatrix}
  \begin{pmatrix}
    \bi{\Phi}_A\\ \bi{\Phi}_\alpha
  \end{pmatrix}.\nonumber
\end{equation}
Thus, now consider this change of variables, with a positive operator $\mathsf{M}_\alpha$ to be determined. The corresponding capacitance matrix, starting from (\ref{eq:multiportcapacitance}), reads now
\begin{equation}
  \mathsf{C}_z=\begin{pmatrix}
    \mathbbm{1}&0\\0& \mathsf{M}_\alpha^{-1/2}  \end{pmatrix} \mathsf{C}\begin{pmatrix}
    \mathbbm{1}&0\\0& \mathsf{M}_\alpha^{-1/2}  \end{pmatrix}=
  \begin{pmatrix}
    \mathsf{A}& -\bi{a}_i\bi{u}_i^T\mathsf{M}_\alpha^{-1/2}\\-\mathsf{M}_\alpha^{-1/2}\bi{u}_i\bi{a}_i&\mathsf{M}_\alpha^{-1/2}\left(\mathsf{C}_\alpha+\bi{u}_i\bi{u}_i^T\right)\mathsf{M}_\alpha^{-1/2}
  \end{pmatrix}.\nonumber
\end{equation}
In order for there to be no coupling amongst impedance modes in the Hamiltonian, we require that, on inverting this matrix, the corresponding submatrix be proportional to the identity operator.  Structurally, the matrix $\mathsf{C}_z$ is similar to the capacitance matrix presented in (\ref{eq:multiportcapacitance}), and the inverse can be computed in a similar fashion. In fact, the condition we require for $\mathsf{M}_\alpha$ is tantamount to
\begin{equation}
  \mathsf{M}_\alpha^{1/2}\left(\mathsf{C}_\alpha^{-1}+\lambda_{ij}\mathsf{C}_\alpha^{-1}\bi{u}_i\bi{u}_j^T\mathsf{C}_\alpha^{-1}\right)\mathsf{M}_\alpha^{1/2}=\mathbbm{1},\nonumber
\end{equation}
where the coefficients $\lambda_{ij}$ are precisely those found earlier in (\ref{eq:inverse-cap-coeffmatrices}).
That is, $\mathsf{M}_\alpha=\mathsf{D}_\alpha^{-1}=\left(\mathsf{C}_\alpha^{-1}+\lambda_{ij}\mathsf{C}_\alpha^{-1}\bi{u}_i\bi{u}_j^T\mathsf{C}_\alpha^{-1}\right)^{-1}$.
We can now see that the long process of Paladino et al. is actually nothing else than the standard canonical analysis.

\section{Applications}
All the formal manipulations above and in the appendices below should not obscure the final objective: to provide model building tools for real devices, in which new phenomena can be uncovered, and which pave the road to more powerful quantum simulators and computers. As stated in the introduction, one of the crucial aspects of the study of multimode system quantization is to achieve faster switching times in qubits, by increasing coupling. This is usually studied in the context of spin-boson Hamiltonians, to which many of the models above can be connected. We first study generic statements about spin-boson models and convergence in transmission lines connected to qubits. Then we compute explicitly three models that have connections to existing experimental devices or proposals thereof. 
\subsection{Spin-Boson Models}
\label{sec:generic-behaviour}

In all the preceding results for transmission lines (TL), section \ref{sec:netw-with-transm}, the Hamiltonians have TL-Network interaction terms of four types, namely
\begin{equation}
  \frac{C_{g_i}}{N_\alpha}\left(\bi{q}_i^T \mathsf{A}^{-1}_i\bi{a}_i\right)\sum_n u_n(x_i) Q_n\,,\qquad - \frac{\boldsymbol{\phi}_i^T\bi{b}_i}{L_{g_i}}\sum_n  u_n(x_i) \Phi_n,\nonumber
\end{equation}
or the same two, but substituting $u_n(x_i)$ with $\Delta u_n(x_i)$.

Let us now concentrate on inductive couplings. Were we to integrate out the transmission lines, their effect on the evolution of the network variables is best codified in the spectral functions
\begin{equation}
  \label{eq:spectrals}
  J_i^I(\omega)=\frac{\pi}{2L_g} \sum_n \frac{\left[u_n(x_i)\right]^2}{N_\alpha \omega_n} \delta(\omega-\omega_n),
\end{equation}
where $I$ stands for inductive, 
and correspondingly for the $\Delta u_n(x_i)$ case. Here the subindex $i$ corresponds to the relevant boundaries of the transmission lines (including possible insertion points). Compare with the last term of (\ref{eq:Ham_TL_LCcoup_2Networks3}).

The asymptotic behaviour of $u_n(x_i)$ is a consequence of the structure of the corresponding operators. Even though the underlying operators are not of the Sturm--Liouville type, see \ref{Walter_appendix}, and therefore the Sturmian theorems are not applicable, one can extract the asymptotic behaviour of $k_n$ from the spectral equations, and from here, by substitution, the asymptotic behaviour in $n$ of $u_n(x_i)$. For the cases we analyse, the eigenfunctions must have the form $\sin\left[k_n(x-x_0)\right]$, and the secular equations are generically of the form
\begin{equation}
  \label{eq:genericsecular}
  \tan(\xi)= \frac{ a \xi}{\xi^2-b},
\end{equation}
where $\xi$ is $k$ times some length $x_i-x_0$. The asymptotic solution of this equation (\ref{eq:genericsecular}) is
\begin{equation}
  \xi_n\sim n\pi + \frac{a}{n\pi}+O\left(n^{-3}\right).\nonumber
\end{equation}
We have $u_n(x_i)\sim\sin(\xi_n)$, and thus $u_n(x_i)\sim (-1)^n a/n\pi $. Since $\omega_n=v k_n$ for some propagation speed $v$, the large $n$ behaviour of the summands is $1/n^3\sim 1/\omega^3$. We see that indeed the spectral density falls with a negative power of the frequency, and that this is the generic behaviour for all systems of transmission lines linearly coupled to lumped element networks. This model has an intrinsic ultraviolet cutoff, and there is no need for further regularisation nor renormalization.

Passing now to capacitive couplings, the analysis of Appendix \ref{sec:capac-induct-coupl} suggests that the effect of the transmission line on the evolution of the network would be encoded in a quantity proportional to $\sum_n\left[u(x_i)\right]^2\omega_n^3\delta\left(\omega-\omega_n\right)$, which, according to our analysis, would present a linear  divergence in that the large $n$ behaviour of the summands is $\sim n$, thus implying that $J(\omega)\sim\omega$ for large $\omega$.

However, let us contextualise these models. In the cases of interest to us, the network will include combinations of Josephson junctions in order to have regimes in which to operate qubits. That entails reducing the operator $\bi{q}_i\cdot \mathsf{A}^{-1}_i\bi{a}_i$, essentially, and effectively, to a Pauli matrix multiplied by some constant. In so doing there is inevitably an energy scale, and thus a frequency, involved in the reduction process, that pertains to the network side. Comparing to the classical analysis of Appendix \ref{sec:capac-induct-coupl}, this introduces an asymmetry that curtails our formal manipulation of adding a time derivative to the Lagrangian: we do not move the derivative in $\bi{q}_i$ to the $Q_n$s, and thus we not add a $\omega_n$ factor. Summarising, the coupling after the reduction to a qubit will be of the form
\begin{equation}
  \sigma^{(i)}_x\sum_n u_n(x_i)Q_n,\nonumber
\end{equation}
and the relevant spectral function will be (up to a global constant)
\begin{equation}
  J_i^{SC}(\omega)=A\sum_n \left[u_n(x_i)\right]^2\omega_n \delta\left(\omega-\omega_n\right).\nonumber
\end{equation}
Here the superindex $SC$ stands for ``spin-capacitive'' coupling. Following the analysis above for large $n$, we see that $J_i^{SC}(\omega)$ tends to zero as $1/\omega$ for large $\omega$. That is to say, in this kind of model with capacitive coupling there is a natural Drude cutoff structure in the qubit regime.

Notice that for the inductive coupling case, when the flux field can be substituted by a spin variable, this  argument is not relevant. The relevant spectral function will indeed be (\ref{eq:spectrals}), with decay $\sim\omega{^{-3}}$.

We now construct the spin-boson model in two cases. First the charge qubit coupled to a semi-infinite  transmission line, in (\ref{sec:an-example:-half}). Then the flux qubit galvanically coupled to an infinite transmission line, in \ref{sec:an-example-galvanic}.
\subsection{Example 1: Charge qubit coupled to semi-infinite TL}
\label{sec:an-example:-half}

Let us consider the Lagrangian of an extension of the circuit shown in Fig. \ref{fig:TL_Ccoup_CQubit}, where the length of the line is extended $L\to\infty$, and an additional inductive coupling ($L_g$) has been introduced,
\begin{eqnarray}
L&=&\frac{C_J}{2}\dot{\Phi}_J^2+U(\Phi_J)+\frac{C_g}{2}\left(\dot{\Phi}_J-\partial_t\phi(0,t)\right)^2\nonumber\\ &&+\int_0^\infty\mathrm{d}x\,\left[\frac{c}{2}\left(\partial_t\phi\right)^2-\frac{1}{2l}\left(\partial_x\phi\right)^2\right]-\frac{1}{2L_g}\left(\Phi_J-\phi(0,t)\right)^2.\nonumber
\end{eqnarray}
Applying our techniques, and after the dust has settled, we have the Hamiltonian
\begin{eqnarray}
H&=&\frac{1}{2C_J}Q_J^2+\frac{1}{2L_g}\Phi_J^2+U(\Phi_J)+\int_0^\infty\mathrm{d}k\,\left[\frac{1}{2c}q_k^2+\frac{k^2}{2l}\phi_k^2\right]\nonumber\\ &&+\frac{1}{c}\frac{C_g}{C_g+C_J}Q_J\int_0^\infty\mathrm{d}k\,u_k(0)q_k-\frac{1}{L_g}\Phi_J\int_0^\infty\mathrm{d}k\,u_k(0)\phi_k,\nonumber
\end{eqnarray}
where we have used the normalisation
\begin{equation}
\int_0^\infty\mathrm{d}x\,u_k(x)u_q(x)+ \alpha u_k(0)u_q(0)=\delta(k-q),\nonumber
\end{equation}
such that
\begin{equation}
u_k(x)= \frac{1}{\sqrt{2\pi\left[k^2+\left(\alpha k^2- 1/\beta\right)^2\right]}}\left[\left(k+ i \alpha k^2 - i/\beta\right) e^{ikx} + \left(k- i \alpha k^2 + i/\beta\right) e^{-ikx}\right],\nonumber
\end{equation}
with the useful choice $\alpha= C_gC_J/c(C_g+C_J)$ and $\beta=L_g/l$, see \ref{subsec:infin-length-transm}. Notice that
\begin{equation}
u_k(0)=\sqrt{\frac{2}{\pi}}\frac{k}{\sqrt{k^2+\left(\alpha k^2-1/\beta\right)^2}}.\nonumber
\end{equation}
Observe that for small $k$ $u_k(0)\sim\sqrt{2/\pi}\,\beta k\to0$, if $\beta$ is finite. On the other hand, if there is no inductive coupling of the transmission line to the network, $L_g\to\infty$, then $1/\beta=0$ and $u_k(0)\sim\sqrt{\pi/2}$ for $k\to0$. Looking now at the large $k$ behaviour, notice that $u_k(0)\sim\sqrt{2/\pi}/\alpha k$ if there is capacitive coupling ($C_g\neq0$). If, on the other hand, $C_g=0$, then $u_k(0)$ tends to a constant for large $k$.

In some parameter regimes of the network hamiltonian
\begin{equation}
H_N=\frac{1}{2C_J}Q_J^2+\frac{1}{2L_g}\Phi_J^2+U(\Phi_J)\nonumber
\end{equation}
it is possible to limit the analysis to a finite dimensional subspace of energy eigenstates of the network. For definiteness assume that a two-dimensional energy eigenspace is enough to describe the most relevant phenomenology of the system, and denote an orthonormal basis of this subspace as $\left\{|+\rangle,|-\rangle\right\}$. Assume furthermore that the expectation values of $Q_J$ and $\Phi_J$ in those basis states are zero (to avoid dealing with an operator valued offset in the effective Hamiltonian). Then the effective quantum Hamiltonian is
\begin{eqnarray}
H_{\mathrm{eff}}&=&\hbar\Omega \sigma_z+\int_0^\infty\mathrm{d}k\,\left[\frac{1}{2c}q_k^2+\frac{k^2}{2l}\phi_k^2\right]\nonumber\\
 && +\frac{\alpha {\cal Q}}{C_J} \,\bi{n}_Q\cdot\boldsymbol{\sigma}\int_0^\infty\mathrm{d}k\,u_k(0)q_k-\frac{{\cal F}}{\beta l}\,\bi{n}_\Phi\cdot\boldsymbol{\sigma}\int_0^\infty\mathrm{d}k\,u_k(0)\phi_k,\nonumber
\end{eqnarray}
where ${\cal Q}=\left|\langle+|Q_J|-\rangle\right|$, ${\cal F}=\left|\langle+|\Phi_J|-\rangle\right|$, and $\bi{n}_Q$ and $\bi{n}_\Phi$ are unitary vectors in the $x-y$ plane. Write
\begin{equation}
  q_k=\sqrt{\frac{\hbar\omega_k c}{2}}\left(a_k+a_k^{\dag}\right)\quad\mathrm{and}\quad \phi_k=i\sqrt{\frac{\hbar}{2\omega_k c}}\left(a_k-a_k^{\dag}\right),\nonumber
\end{equation}
where $\omega_k=v_p k=k/\sqrt{lc}$.
The relevant spectral functions are therefore
\begin{equation}
  J^C(\omega)\propto \frac{\omega^3}{\left(\omega^2-\alpha\omega_\alpha^2/\beta\right)^2+\omega^2\omega_\alpha^2},\nonumber
\end{equation}
with $\omega_\alpha=v_p/\alpha$, and
\begin{equation}
  J^L(\omega)
\propto  \frac{\omega}{\left(\omega^2-\alpha\omega_\alpha^2/\beta\right)^2+\omega^2\omega_\alpha^2}.\nonumber
\end{equation}
Observe that both in the presence and in the absence of inductive coupling the leading behaviour at small frequencies will be ohmic ($J(\omega)=J^C(\omega)+J^L(\omega)\sim\omega$ for small $\omega$). On the other hand, at large frequencies the leading behaviour is determined by the capacitive coupling, with $J^Q(\omega)\sim\omega^{-1}$.
\subsection{Example 2: Flux qubit galvanically coupled to infinite TL}
\label{sec:an-example-galvanic}

\begin{figure*}[ht]
	\centering{\includegraphics[width=0.85\textwidth]{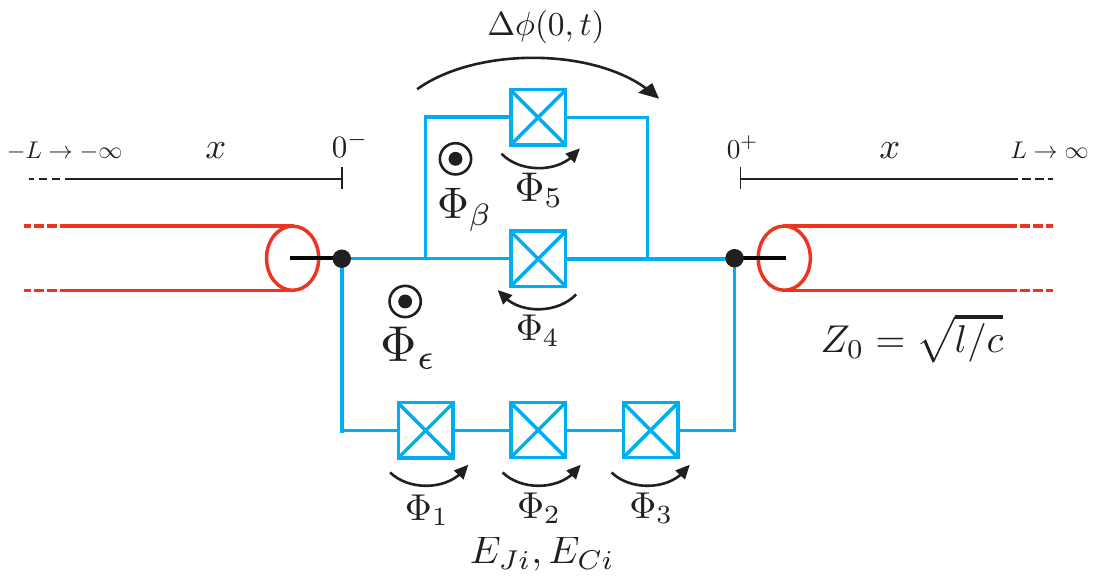}}
	\caption{\label{fig:Flux_qubit_GC} Circuit models for a capacitively-shunted flux qubit inserted in the transmission line.}
      \end{figure*}

In this example, we follow the circuit layout of a flux qubit galvanically, and  tunably, coupled to a transmission line with a SQUID-loop shared between the two \cite{Pol_2017}. It is represented in Fig. \ref{fig:Flux_qubit_GC}.
Clearly the variables depicted are redundant, since they fulfill  the fluxoid quantization condition on the separate loops:
\begin{equation}
\begin{split}
\varphi_{1} + \varphi_{2} + \varphi_{3} + \varphi_{4} + 2 \pi f_{\epsilon} &= 0 ,\\
\varphi_{4} + \varphi_{5} + 2 \pi f_{\beta} &= 0, \\
\varphi_{4} + \frac{\Delta \Phi \left( 0 , t \right)}{\varphi_{0}} &= 0,
\end{split}\nonumber
\end{equation}
where $f_{\epsilon}= \Phi_{\epsilon} / \Phi_{0}$, $f_{\beta}= \Phi_{\beta} / \Phi_{0}$ are the magnetic frustration in each loop, being $\Phi_0$ the magnetic flux quantum. We shall use $\varphi_{1}$, $\varphi_{2}$, $\varphi_{4}$ as the independent degrees of freedom, where the fluxes are related to the phase variables through $\varphi_i=2 \pi \Phi_i/\Phi_0$. Starting from the Lagrangian that includes all these elements, with Josephson junction potentials,  linearising the terms with inductive couplings, and redefining variables in a suitable manner, we are led to an effective Lagrangian with the form of Eq. (\ref{eq:Lag_TL_LCgalvcoup_Networks_2}), in which we set $\boldsymbol{\Phi}_{\mathrm{ext}}$ to zero.

      There is one crucial difference with respect to the analogous Lagrangian (\ref{eq:Lag_TL_LCgalvcoup_Networks_2}) from \ref{sec:line-galv-coupl}, namely that we take transmission lines of infinite length. We substitute the boundary conditions (\ref{eq:EVP_TL_LCgalvcoup_Networks_eq3}) by a normalizability condition. It proves convenient not to demand reality of the generalized eigenfunctions $u_k(x)$, identified by wavenumber $k$. Nonetheless, they can be selected to have $u_k^*(x)=u_{-k}(x)$. Since the flux on the lines is a real magnitude, the coefficients of its expansion
      \begin{equation}
        \Phi(x,t)=\int_{-\infty}^\infty dk\,\Phi_k(t)u_k(x)\nonumber
      \end{equation}
      fulfill the relation $\Phi_k^*=\Phi_{-k}$. It is also relevant to notice that the conjugate momentum $Q_k$ is given by $c\dot{\Phi}_{-k}$ plus additional terms.
      
Following the same steps as in the previous section we derive the Hamiltonian
\begin{eqnarray}
\fl \qquad\qquad H= \frac{1}{2}\mathbf{q}^T(\mathsf{A}^{-1}+ \frac{C_A^2}{\alpha c} \mathsf{A}^{-1}\mathbf{a}\mathbf{a}^T \mathsf{A}^{-1})\mathbf{q}+\frac{1}{2}\boldsymbol{\phi}^T\mathsf{B}^{-1}\boldsymbol{\phi}+V(\boldsymbol{\phi})\nonumber\\
 +\frac{C_A}{c} (\mathbf{q}^T\mathsf{A}^{-1} \mathbf{a})\int^{\infty}_{-\infty} dk \,~  Q_{-k}  \Delta u_k(0)  -\frac{1}{L_B} (\boldsymbol{\phi}^T\mathbf{b})\int^{\infty}_{-\infty} dk \,~    \Phi_k  \Delta u_k(0)  \nonumber\\
 +\int^{\infty}_{-\infty} dk \,~ \left[ \frac{|Q_k|^2}{2c}+\frac{k^2|\Phi_k|^2}{2l} \right].
\label{eq:Ham_TL_LCgalvcoup_Networks_fluxonium}
\end{eqnarray}
As expected, we have made the choices
\begin{equation}
\begin{split}
\alpha&=\frac{C_{gA}-C_A^2\mathbf{a}^T \mathsf{A}^{-1}\mathbf{a}}{c}= \frac{ C}{c} \left( r_{4} + r_{5} + \frac{r_{1} r_{2} r_{3}}{ r_{1} r_{2} + r_{2} r_{3} + r_{3} r_{1} } \right),\nonumber\\
\beta&=L_{gB}/l = \frac{\varphi_{0}^{2}}{l E_{J} \left[ r_{4} + r_{5} \cos{\left( 2 \pi f_{\beta} \right)} \right]},\nonumber
\end{split}
\end{equation}
in order to eliminate the mode-mode coupling terms. Here we have introduced parameters that pertain to the experiment of reference, for later comparison.
Notice that in Eq. (\ref{eq:Ham_TL_LCgalvcoup_Networks_fluxonium})  $Q_k$ is accompanied by $\Delta u_{-k}(0)$. Now, given the relation  $\Phi_k^*=\Phi_{-k}$ and the analogous  $Q_k^*=Q_{-k}$, we define creation and annhilation operators related to the charge and flux operators 
\begin{eqnarray}
	\qquad\qquad Q_k &\equiv& \sqrt{\frac{\hbar \omega_k c}{2}}(\tilde{a}_{-k} + \tilde{a}_k^{\dagger}),\nonumber\\
	\qquad\qquad \Phi_k &\equiv& i\sqrt{\frac{\hbar}{2 \omega_k c}}(\tilde{a}_k - \tilde{a}_{-k}^{\dagger}),\nonumber
\end{eqnarray}
Hence the interacting part of the Hamiltonian reads
\begin{equation*}
\begin{split}
H_{\text{int}}=& \frac{C_A}{c} (\mathbf{q}^T\mathsf{A}^{-1} \mathbf{a})\int^{\infty}_{-\infty} dk \,~  Q_{-k}  \Delta u_k(0)  -\frac{1}{L_B} (\boldsymbol{\phi}^T\mathbf{b})\int^{\infty}_{-\infty} dk \,~    \Phi_k  \Delta u_k(0)  \\
=& \frac{\gamma}{c} \left(q_{1}r_{2}+q_{2}r_{1}\right)  \int^{\infty}_{-\infty} dk \,~ \sqrt{\frac{\hbar \omega_k c}{2}} \Delta u_k(0) (\tilde{a}_{k} + \tilde{a}_{-k}^{\dagger}) \\
&+i \frac{E_{J}}{\varphi_{0}^{2}} ~ r_{3}  \cos{\left( 2 \pi f_{\epsilon} \right) } \left( \phi_{1} + \phi_{2} \right) \int^{\infty}_{-\infty} dk \,~    \sqrt{\frac{\hbar}{2 \omega_k c}} \Delta u_k(0)  (\tilde{a}_k - \tilde{a}_{-k}^{\dagger}),
\end{split}
\end{equation*}
again with parameters pertaining to the different elements of the system under consideration and $\gamma=r_{3}/(r_{1} r_{2} + r_{2} r_{3} + r_{3} r_{1})$.
\begin{figure*}[h]
	\centering{\includegraphics[width=0.75\textwidth]{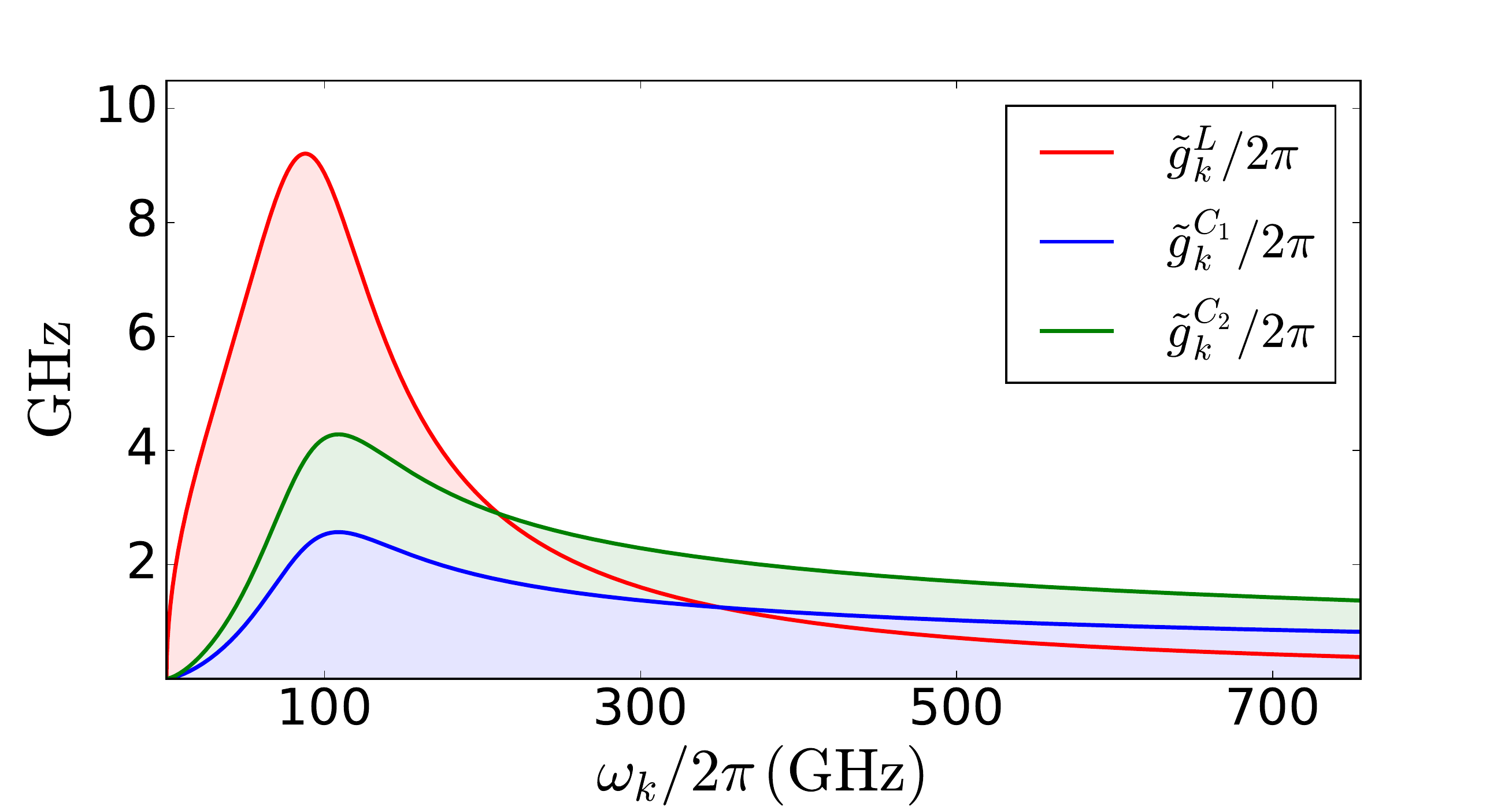}}
	\caption{\label{fig:Flux_qubit_GC_Inf_TL_gk} \textbf{Capacitive and inductive coupling constants against their associated mode frequency $\omega_{k}$ for the circuit in Fig. \ref{fig:Flux_qubit_GC}}. We have used the values of the experiment in \cite{Pol_2017} and plotted $\tilde{g}_k=g_k/\sqrt{\beta}$ in frequency units. The inductive coupling constant behaves  as $g_{k}^{L}\propto\omega_k^{1/2}$ when $\omega_k$ tends to 0, and  as $g_{k}^{L}\propto\omega_k^{-3/2}$ for large $\omega_k$. On the other hand, the typically neglected capacitive couplings behaves as $g_{k}^{C_i}\propto\omega_k^{3/2}$ for $\omega_k\to0$ and  as $g_{k}^{C_i}\propto\omega_k^{-1/2}$ for $\omega_k\to\infty$. The natural ultraviolet cutoffs for the inductive and capacitive coupling constants are located at $\Omega_k^{(c_L)}=87.9 \,\mathrm{GHz}$ and $\Omega_k^{(c_C)}=108.8 \,\mathrm{GHz}$ respectively.}
\end{figure*}

As stated above, $u_k^*=u_{-k}$, which implies  $\Delta u_k^*(0)=\Delta u_{-k}(0)$. Let thus define $\Delta u_k(0)=|\Delta u_k(0)|\exp(i \sigma_k)$, from which $\sigma_{-k}=-\sigma_k$, and $|\Delta u_k(0)|$ is independent of the sign of $k$. We introduce a new set of canonical operators $a_k= \exp(i\sigma_k)\tilde{a}_k$ and  $a_k^{\dag}= \exp(-i\sigma_k)\tilde{a}_k^{\dag}$. The interaction part of the Hamiltonian now reads, in terms of these operators,
\begin{equation*}
\begin{split}
H_{\mathrm{int}}&= \frac{\gamma}{c} \left(q_{1}r_{2}+q_{2}r_{1}\right) \int^{\infty}_{-\infty} dk \,~ \sqrt{\frac{\hbar \omega_k c}{2}} \left|\Delta u_k(0)\right| (a_{k} + a_{-k}^{\dagger}) \\
&+i \frac{E_{J}}{\varphi_{0}^{2}} ~ r_{3}  \cos{\left( 2 \pi f_{\epsilon} \right) } \left( \phi_{1} + \phi_{2} \right) \int^{\infty}_{-\infty} dk \,~    \sqrt{\frac{\hbar}{2 \omega_k c}} |\Delta u_k(0)|  (a_k - a_{-k}^{\dagger})\\
&= \int^{\infty}_{-\infty} dk \,~ g_k^{C_i}n_{i} (a_{k} + a_k^{\dagger}) +i \int^{\infty}_{-\infty} dk \,~ g_k^{L}\varphi_{i} (a_k - a_{k}^{\dagger}),
\end{split}
\end{equation*}
where $g_k^{C_1}=r_2\gamma v_p \sqrt{\pi Z_0/R_Q} \sqrt{k}|\Delta u_k(0)|$,  $g_k^{C_2}=r_1g_k^{C_1}/r_2$, and $g_k^{L}=(E_J/\hbar) r_3 \cos(2\pi f_\epsilon)\sqrt{Z_0/\pi R_Q} |\Delta u_k(0)|/\sqrt{k}$. As before, $R_Q$ is the quantum of resistance and $v_{p}$ the velocity of propagation in the line. 

The infrared and ultraviolet behaviour of these couplings are readily determined from the normalization of the $u_k$ functions. In this case, the coupling vector that fulfills the boundary conditions and normalization is
\begin{equation}
|\Delta u_k(0)|=\sqrt{\frac{2}{\pi }} \beta  \frac{|k|}{\sqrt{\beta  k^2 \left(\beta +4 \alpha  \left(\alpha  \beta  k^2-2\right)\right)+4}}.
\end{equation}
In particular, the relevant limiting behaviours of $|\Delta u_k(0)|$ are $\beta |k|/\sqrt{2\pi}$ as $k\to0$ and $1/(\alpha|k|\sqrt{2\pi})$ as $k\to \infty$. We see that the capacitive coupling  $g_k^{C_i}\sim k^{3/2}$ tends to zero faster than the inductive one $g_k^{L}\sim k^{1/2}$ as $k\to0$. Furthermore, we recover again a natural ultraviolet cutoff, in that $g_k^{C_i}\sim k^{-1/2}$ and $g_k^{L}\sim k^{-3/2}$ for large $k$, see Fig. \ref{fig:Flux_qubit_GC_Inf_TL_gk}. In other words, the inductive coupling dominates at low energies ($J(\omega)\sim J^L(\omega)$), whereas the capacitive coupling dominates at high energies ($J(\omega)\sim J^C(\omega)$).

\subsection{Example 3: Charge qubits coupled to a 2-port impedance}

\begin{figure*}[ht]
	\centering{\includegraphics[width=1.0\textwidth]{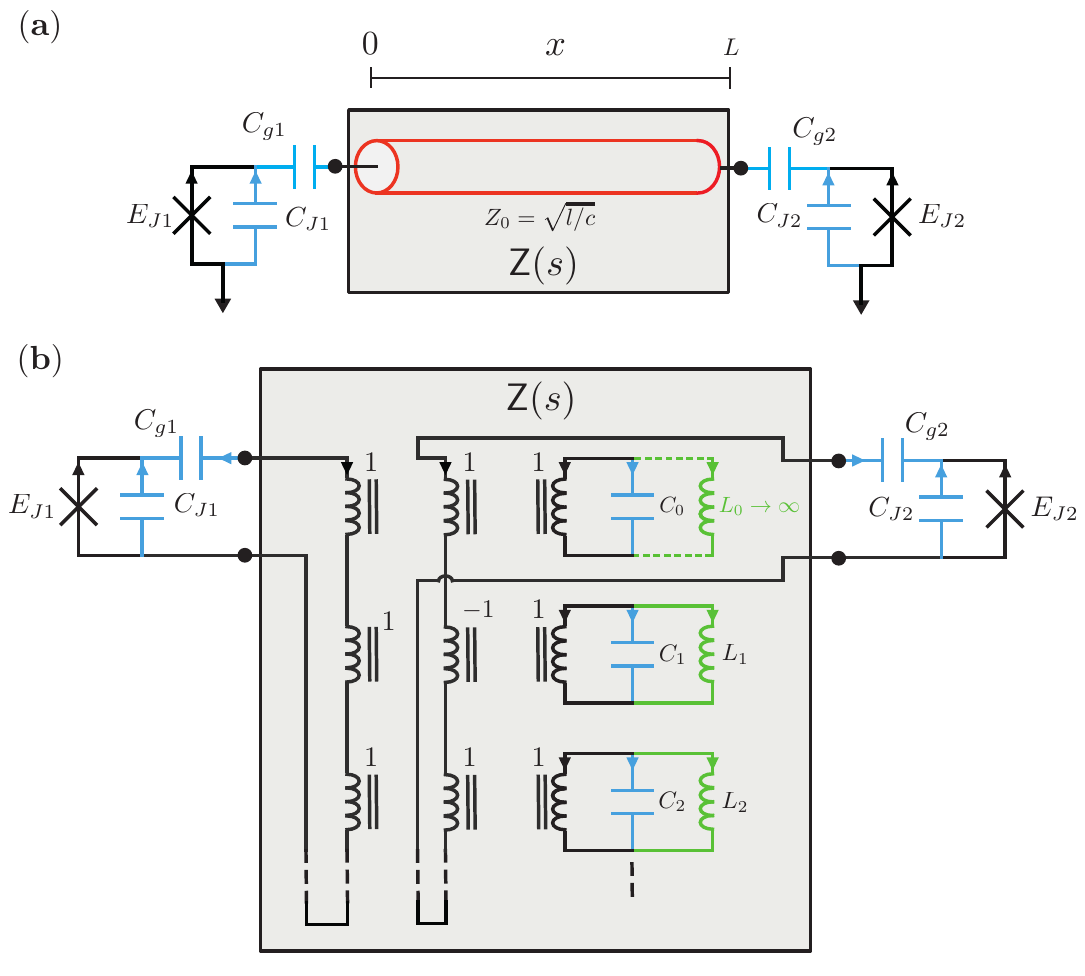}}
	\caption{\label{fig:2CQ2PortTL_v2} \textbf{(a) Transmission line resonator capacitively coupled to two Josephson junctions at its ports is  modelled as a (b) 2-port impedance synthesized with a lossless multiport Foster expansion}. The two port impedance $\mathsf{Z}(s)$ is realized with an infinite series of stages of LC circuits coupled through a Belevitch transformer to its ports. The first stage lumped element capacitor is  $C_0 = c L$, where $c$ is the capacitance per unit length of the line and $L$ its length. The rest of the capacitors are $C_\alpha=C_0/2$ with $\alpha>0$. The inductors are $L_\alpha=2Ll/\alpha^2\pi^2$ for $\alpha>0$. The formal expansion of $\mathsf{Z}$ for the open transmission line resonator does not contain an inductor in its first stage. We introduce a virtual one in order to simplify the circuit theory analysis to derive a Hamiltonian, and then take the limit of $L_0\rightarrow\infty$ (infinite impedance of the branch) at the end. In both circuits, the arrows represent the convention used for positive current directions at each element.}
\end{figure*}

Let us finish this section with an example of a multiport impedance described by infinite degrees of freedom coupled to an anharmonic network of two degrees of freedom. In Fig. \ref{fig:2CQ2PortTL_v2}(a) we see the theoretical model of a transmission line resonator of length $L$ capacitively coupled through $C_{g\sigma}$ to two Josephson junctions of $(E_{J\sigma},C_{J\sigma})$, with $\sigma=\{1,2\}$. This circuit could be analysed as an eigenvalue problem with the theory developed in \ref{sec:mult-netw-coupl} and Appendix \ref{Walter_appendix}. However, we are going to derive the quantized Hamiltonian following the alternative presented in \ref{sec:multi-port-impedance} in order to illustrate this second method. 

To start we need to use the lumped-element equivalent circuit by which  the  transmission line, open at both ends, can be expanded. See to this point Fig. \ref{fig:2CQ2PortTL_v2}(b). Indeed, this circuit is the multiport generalization of the $1^{\mathrm{st}}$-Foster expansion used in \cite{Gely_2017_DivFree} to describe the one port transmission line resonator. The classical response of the open transmission line can be encoded in a two-port impedance matrix

\begin{equation}
\mathsf{Z}(s)=Z_0\begin{pmatrix}
\mathrm{coth}\left(s\sqrt{lc}L\right) &  \mathrm{csch}\left(s\sqrt{lc}L\right)\\
\mathrm{csch}\left(s\sqrt{lc}L\right) & \mathrm{coth}\left(s\sqrt{lc}L\right)
\end{pmatrix},\nonumber
\end{equation}
where $l$ and $c$ are the inductance and capacitance per unit length, $L$ is the length of the line and $Z_0=\sqrt{l/c}$ its characteristic impedance. This matrix belongs to the family of lossless positive real matrices (LPR) that can be synthesized by a passive network of inductors, capacitors and ideal transformers \cite{Newcomb_1966_LinearMPortSynthesis}, see the grey box in Fig. \ref{fig:2CQ2PortTL_v2} (b). For more details on this expansion, we refer the reader to \cite{Newcomb_1966_LinearMPortSynthesis} and Appendix \ref{sec:Mport_synthesis}. Notice that we have introducted a virtual inductance $L_0$ (whose limit is taken later as $L_0\rightarrow\infty$) to ease the network theory analysis. We can now directly apply the method developed by  Solgun and  DiVincenzo  \cite{Solgun_2015} to compute Hamiltonians of anharmonic networks coupled to a more general lossy environment described by a Brune multiport impedance. Following this reference the  equations of motion are readily derived for our set of degrees of freedom, chosen to be  the flux differences at the junctions and inductors $\bi{\Phi}=(\bi{\Phi}_J^T,\bi{\Phi}_L^T)^T=(\Phi_{J1},\Phi_{J2},\Phi_{L0},\Phi_{L1},...)^T$. The central  idea  is the elimination of the flux variables associated to the ideal transformers, since they do not store energy. The capacitance matrix of the circuit is cast into the form of (\ref{eq:multiportcapacitance}), and reads
\begin{equation}
\mathsf{C}=
\begin{pmatrix}
C_{J1}+C_{g1}& 0&\sqrt{C_{g1}}\bi{u}_1^T \\
0&C_{Jb}+C_{g2}&\sqrt{C_{g2}}\bi{u}_2^T \\
\sqrt{C_{g1}}\bi{u}_1& \sqrt{C_{g2}}\bi{u}_2 &  \mathsf{C}_\alpha+\sum_i C_{gi}\bi{u}_i\bi{u}_i^T
\end{pmatrix},\nonumber
\end{equation}
where the submatrix is $\mathsf{C}_{\alpha}= \text{diag}(C_0, C_1, ...)$, with $C_\alpha=C_0/2$ for $\alpha>0$, the infinite norm coupling vectors are $\bi{u}_1=\sqrt{C_{g1}}(1,1,1,...)^T$ and $\bi{u}_2=\sqrt{C_{g2}}(1,-1,1,...)^T$, and, implicitly, $\bi{a}_1=\sqrt{C_{g1}}(1,0)^T$ and $\bi{a}_2=\sqrt{C_{g2}}(0,1)^T$. The inductance matrix is
\begin{equation}
	\mathsf{L}^{-1}=\begin{pmatrix}
	0 & 0 & 0 \\
	0 & 0 & 0\\
	0 & 0 & \mathsf{L}_\alpha^{-1} 
	\end{pmatrix},\nonumber
\end{equation}
with $\mathsf{L}_\alpha^{-1}=\text{diag}(L_0^{-1},L_1^{-1},...)$, and $L_\alpha=2Ll/\alpha^2\pi^2$ with $\alpha\in (1,2,...)$. The Hamiltonian of the circuit can be directly derived inverting its capacitance matrix with the formula (\ref{eq:inverse-multiport-c})
\begin{equation}
H=\frac{1}{2}\bi{Q}^{T}\mathsf{C}^{-1}\bi{Q}+\frac{1}{2}\bi{\Phi}^{T}\mathsf{L}^{-1}\bi{\Phi}-\sum_{\sigma=1,2}E_{J\sigma}\cos{\varphi_{J\sigma}}.\label{eq:2CQ2PortTL_Ham}
\end{equation}
Let us consider a truncated model with $N$ inductors, and later take the limit $N\to\infty$.
The direct coupling between the two anharmonic degrees of freedom can be calculated through the upper-left submatrix, for which we need to compute the auxiliary matrices 
\begin{eqnarray}
	\mu&=&\begin{pmatrix}
	\sum_\alpha C_\alpha^{-1}&\sum_\alpha (-1)^{\alpha+1} C_\alpha^{-1}\\
	\sum_\alpha (-1)^{\alpha+1} C_\alpha^{-1}&\sum_\alpha C_\alpha^{-1}
	\end{pmatrix}\nonumber\\
	&=&\begin{pmatrix}
	(2N+1)C_{g1}/C_0&\pm 2\sqrt{C_{g1}C_{g2}}/C_0\\
	\pm 2\sqrt{C_{g1}C_{g2}}/C_0&(2N+1) C_{g2}/C_0
	\end{pmatrix}\nonumber
\end{eqnarray}
and 
\begin{equation}
	\nu= -\begin{pmatrix}
	C_{g1}/C_{\Sigma 1}& 0\\
	0&C_{g2}/C_{\Sigma 2}
	\end{pmatrix},\nonumber
\end{equation}
with $C_{\Sigma \sigma}=C_{J\sigma} + C_{g\sigma}$ for $\sigma=\{1,2\}$, and where the $+$ ($-$) sign appears when we truncate to an even (odd) number of inductors  $N$ in the circuit. The upper-left part of the inverse capacitance matrix is diagonal in the limit of infinite stages ($N\rightarrow\infty$), since  
\begin{equation}
	\lim\limits_{N\rightarrow\infty}\beta=\begin{pmatrix}
	C_{\Sigma 1}/C_{J1}&0\\
	0&C_{\Sigma 2}/C_{J2}
	\end{pmatrix}.\nonumber
\end{equation}
This result would have also been retrieved using the continuous wave flux field expansion of \ref{sec:mult-netw-coupl}, where the infinite dimensional coupling vectors would have been purely orthogonal. Thus, we can write the inverse matrix as 
\begin{equation}
\lim\limits_{N\rightarrow\infty}\mathsf{C}^{-1}\rightarrow
\begin{pmatrix}
1/C_{J1}& 0&\rho_{11}\frac{C_{g1}}{C_{\Sigma 1}}\bi{u}_1^T\mathsf{C}_\alpha^{-1} \\
0&1/C_{J2}&\rho_{22}\frac{C_{g2}}{C_{\Sigma 2}}\bi{u}_2^T\mathsf{C}_\alpha^{-1}\\
\rho_{11}\frac{C_{g1}}{C_{\Sigma 1}}\mathsf{C}_\alpha^{-1}\bi{u}_1& \rho_{22}\frac{C_{g2}}{C_{\Sigma 2}}\mathsf{C}_\alpha^{-1}\bi{u}_2 &  \mathsf{D}_\alpha^{-1}
\end{pmatrix},\nonumber
\end{equation}
where $\rho$ is the matrix defined in (\ref{eq:inverse-cap-coeffmatrices}). We restate now the terms in (\ref{eq:2CQ2PortTL_Ham})
\begin{equation}
H=\frac{Q_{J\sigma}^2}{2C_{J\sigma}}-E_{J\sigma}\cos{\varphi_{J\sigma}}+\rho_{\sigma \sigma}\frac{C_{g\sigma}}{C_{\Sigma \sigma}}Q_{J\sigma}{\bfeta}_{\sigma}^T\bi{Q}_{L}+\frac{1}{2}\bi{Q}_L^{T}{\mathsf{D}}_\alpha^{-1}\bi{Q}_L+\frac{1}{2}\bi{\Phi}_L^{T}{\mathsf{L}}_\alpha^{-1}\bi{\Phi}_L,\nonumber
\end{equation}
with coupling vectors ${\bfeta}_\sigma={\mathsf{C}}_\alpha^{-1}{\bi{u}}_\sigma$, and where we have used Einstein's summation rule for repeating greek letters. We take now safely the limit $L_0\rightarrow\infty$ and eliminate the free degree of freedom $(\Phi_{L_0},Q_{L_0})$, truncating the vectors $\tilde{\bi{\Phi}}_L=(\Phi_{L_1},\Phi_{L_2},...)^T$, $\tilde{\bi{Q}}_L=(Q_{L_1},Q_{L_2},...)^T$ and $\tilde{\bfeta}_{\sigma}=(\eta_{\sigma 1}, \eta_{\sigma 2},...)^T$. Additionally, the matrices $\tilde{\mathsf{D}}_\alpha$ and  $\tilde{\mathsf{L}}_\alpha^{-1}$ take after ${\mathsf{D}}_\alpha$ and  ${\mathsf{L}}_\alpha^{-1}$. In fact they are the untilded versions,  pruned of their first row and column. As previously discussed, we can perform the rescaling and rotation of the harmonic variables $\bi{x}_L=D_0^{-1/2}\mathsf{U}\tilde{\mathsf{D}}_\alpha^{1/2}\tilde{\bi{\Phi}}_L$ and $\bi{p}_L=D_0^{1/2}\mathsf{U}^T\tilde{\mathsf{D}}_\alpha^{-1/2}\tilde{\bi{Q}}_L$, where $D_0$ is a capacitance constant that mantains the units of the conjugated variables. We finally reach the normal mode structure
\begin{equation}
H=4E_{C\sigma}n_{J\sigma}^2-E_{J\sigma}\cos{\varphi_{J\sigma}}+\rho_{\sigma \sigma}\frac{C_{g\sigma}}{C_{\Sigma \sigma}}Q_{J\sigma}\bxi_{\sigma}^T \bi{p}_{L}+\frac{\bi{p}_L^2}{2 D_0}+\frac{D_0}{2}\bi{x}_L^T \mathsf{\Omega}_\alpha^2\bi{x}_L,\nonumber
\end{equation}
where $\mathsf{\Omega}_\alpha^2= D_0\mathsf{U}\tilde{\mathsf{D}}_\alpha^{-1/2}\tilde{\mathsf{L}}_\alpha^{-1}\tilde{\mathsf{D}}_\alpha^{-1/2}\mathsf{U}^T$ is a diagonal matrix, and ${\bxi}_{\sigma}=D_0^{-1/2} \mathsf{U}^T \mathsf{D}_\alpha^{1/2} {\bfeta}_\sigma$ are the coupling vectors with finite norm even in the limit $N\rightarrow\infty$. 
\begin{figure*}[h]
	\centering{\includegraphics[width=0.8\textwidth]{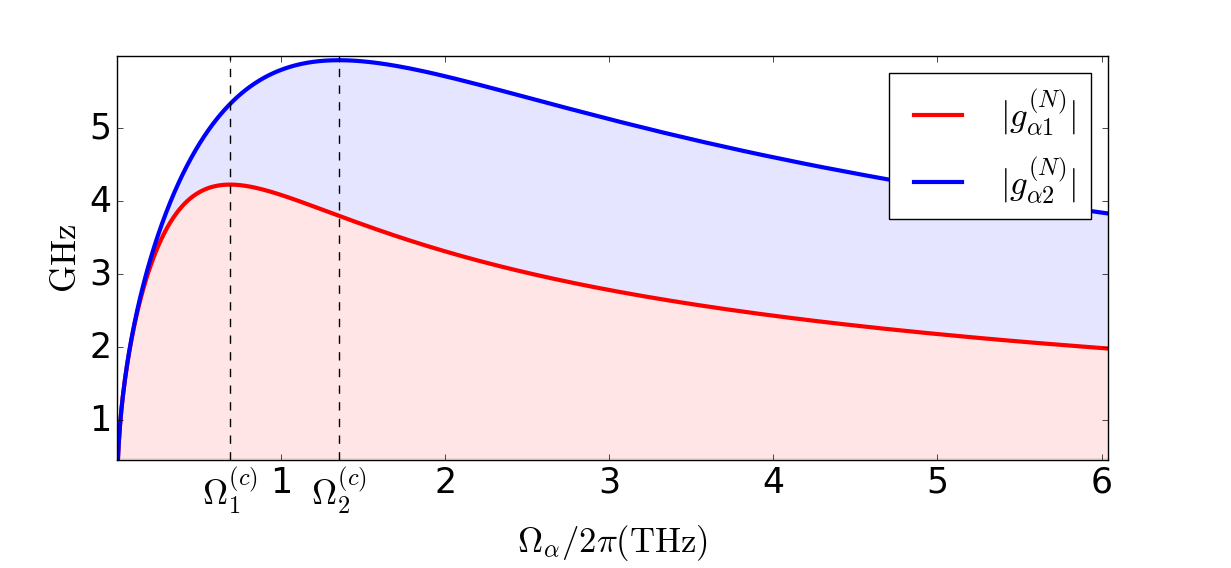}}
	\caption{\label{fig:2CQ2PortTL_gn} \textbf{Capacitive coupling $g_{\alpha\sigma}$ against its mode frequency $\Omega_{\alpha}$ for the circuit in Fig. \ref{fig:2CQ2PortTL_v2}}. We have used the same values to those of Device A in \cite{Bosman_2017_MMUSC} for the first junction, $C_{g1}=40.3\,\mathrm{fF}$, $C_{J1}=5.13 \,\mathrm{fF}$, while the second junction has values $C_{g2}=C_{g1}/2$ and $C_{J2}=C_{J1}/2$. The transmission line has again the parameters  $c=249\,\mathrm{pF/m}$, $l=623\,\mathrm{nH/m}$, while the length has been doubled $L=9.4\,\mathrm{mm}$, so  that the fundamental mode with frequency $\Omega_{1}= 4.26 \,\mathrm{GHz}$ enter into the more desirable experimental regime of $(4-6)\, \mathrm{GHz}$. In red (blue), $g_{\alpha1}$ ($g_{\alpha2}$), where the number of modes used in the numerical diagonalization has been truncated to $N=6000$: a natural cutoff of the coupling constants appear for the resonator modes at frequency $\Omega_{1}^{(c)}=685.5\,\mathrm{GHz}$ ($\Omega_{2}^{(c)}=1.35\,\mathrm{THz}$).}
\end{figure*}
We promote the conjugate variables to operators, i.e. $[\hat{x}_\alpha,\hat{p}_\alpha]=i\hbar$ and $[\hat{\varphi}_J,\hat{n}_J]=i$, and rewrite the harmonic sector in terms of $a_\alpha$ and $a_\alpha^\dagger$ in an anologous manner to previous sections to reach
\begin{equation}
H=4E_{C\sigma}\hat{n}_{J\sigma}^2-E_{J\sigma}\cos{\hat{\varphi}_{J\sigma}}+\hbar g_{\alpha\sigma}\hat{n}_{J\sigma}(a_\alpha+a_\alpha^\dagger)+\hbar \Omega_{\alpha}a_\alpha^\dagger a_\alpha,\nonumber
\end{equation}
where the coupling constants are redefined as $g_{\alpha \sigma}= \rho_{\sigma \sigma}\frac{C_{g\sigma}}{C_{\Sigma \sigma}}\sqrt{\frac{\Omega_{\alpha}D_0}{4\pi R_{Q}}}$ and $\Omega_{\alpha}$ are the square root of the diagonal entries in $\mathsf{\Omega}_\alpha^2$. In Fig. \ref{fig:2CQ2PortTL_gn} we see an example with realistic parameters where the different coupling and junction's capacitances $(C_{J\sigma}, C_{g\sigma})$ translate into different saturation frequencies for the coupling parameters. Again, both coupling parameters increase (decay) as $g_{\alpha \sigma}\propto \sqrt{\Omega_{\alpha}}$ ($ 1/\sqrt{\Omega_{\alpha}}$) in the low (high) frequency limit.

\section{Conclusions}
In this work, we have critically analysed a number of approaches to the quantization of superconducting circuits with an infinite dimensional environment, with particular interest on the issue of divergences in the capacitive coupling constants.

We have first examined the coupling of transmission lines to networks with a finite number of modes, and we have shown that there is an intrinsic (soft) ultraviolet cutoff, $k_\alpha=1/\alpha$. This length parameter $\alpha$ arises as due to an optimal dressing of the transmission line modes, where the optimality is precisely determined by the requirement that the final Hamiltonian description be that of an infinite number of independent harmonic modes coupled to the finite network. The underlying mathematical construction is that of singular value problems for second order differential operators, with boundary values that include the singular value itself. We have provided, in an appendix, a new proof of the associated expansion theorem. This approach has provided us with tools to predict the large frequency behaviour of the coupling. Furthermore, the study of transmission line modes by this means provides us with an expression for the Hamiltonian in the continuum. In it, one reads the $A^2$ term directly, dependent on the length parameter $\alpha$.

Passing then to other environments with infinite number of modes, we have made apparent the source of issues that appear in many approaches when, after truncation to $N$ modes, one takes the limit $N\to\infty$. We again use as the central criterion that the final Hamiltonian description be that of an infinite set of independent modes, coupled to the relevant system. We have shown that this is achieved with a canonical transformation that is indeed determined by this criterion. This technique is also applicable to multiport situations.

One of the main objectives of the quantization of superconducting circuits is their use for quantum information tasks. Thus, it is relevant to consider how spin-boson models arise from our presentation. We have focused on the characterization of the spectral densities and their ultraviolet behaviour. In particular, we have shown that the spectral density for qudits capacitively coupled to transmission lines falls off with a soft power law cutoff. An alternative way of stating the same result is that the qubit couplings $g_n$ decay as $n$ grows. The generic behaviour for transmission lines coupled capacitively to qubits is $g_n\sim \omega_n^{-1/2}$.  We have also provided explicit computations for models of experimental relevance, which portray this behaviour explicitly.

Furthermore, we have given a detailed argument for the validity of standard approximations in the literature, in which the behaviour $g_n\sim \omega_n^{1/2}$ is assumed for capacitive coupling to transmission lines. We have shown that this is indeed a correct assumption for the lowest frequency sector, and that it is valid for truncations to a finite number of modes if not too many are assumed present.

Looking to the future, we have presented detailed arguments at each point, with the objective that this work can be  a reference for initial development of new useful models. The solid mathematical foundations in which we rely for the expansion in modes will undoubtedly be useful in other physical contexts as well, and we will be exploring further in this point. Finally, we have presented a prediction for the maximum coupling achievable with a transmission line, that has not yet been measured.

\section*{Acknowledgments}
We thank Mikel Sanz, Pol Forn-D\'iaz, Mario F. Gely and Gary A. Steele for useful discussions. We would like to thank Giovanni Viola for very thoughtful feedback. A. P-R thanks Firat Solgun for helpful discussions and explanations on the Solgun-DiVincenzo's multiport quantization method used in the third circuit example. The authors acknowledge funding from Spanish MINECO/FEDER FIS2015-69983-P, UPV/EHU grant EHUA15/17, and the Basque Government IT986-16 and Ph.D. Grant No. PRE-2016-1-0284.

\appendix

\begin{appendices}
\section*{Appendix}
\section{Mathematical details: Networks with Transmission Lines}
\label{Walter_appendix}
The main tools used in the paper have been 1) mapping the Transmission Line Lagrangians from a field presentation to a mode description, that takes into account the coupling at \emph{points} with other circuit elements, and 2) manipulations of vectors $\bi{u}$ and matrices $\bi{u}\bi{u}^T$ as in the finite vector case even for full mode expansions.

Both these aspects can be justified in a Hilbert space context (even though more general presentations could be possible) by addressing there the interesting problem of boundary conditions that incorporate the singular value, as we presently see.

In this appendix we first consider finite transmission lines and then half-line transmission lines. With lesser detail we signal two other configurations, and then we give a general description of the recipe we have applied.

\subsection{Finite Length Transmission Lines}
\label{sec:finite-length-transm}

Consider the following singular value problem:
\begin{eqnarray}
  \qquad\quad \qquad - u''(x)  &=& k^2 u(x)\,\quad x\in(0,L),\label{eq:appwalt1}\\
  \qquad\qquad \qquad u(L)&=&0,\\
 \qquad \frac{1}{\beta}u(0)- u'(0)&=&  \alpha k^2 u(0).\label{eq:appwalt3}
\end{eqnarray}
$\alpha$ and $\beta$ are constants with dimension of length.

The presence of the singular value $k^2$ in the boundary condition at $0$ means that this is not a Sturm--Liouville problem, and the usual oscillation and expansion theorems are therefore not applicable. Nonetheless, it is easy to establish a secular equation, which the singular values must fulfill, and identify the corresponding functions. In the example at hand, the functions are
\begin{equation}
  \nonumber
  u_n(x)=N_n\sin\left[k_n(L-x)\right], 
\end{equation}
for singular value $k_n$, where the normalization factor $N_n$ will be fixed later, and the secular equation reads
\begin{equation}
  \label{eq:appsecular}
  \frac{\alpha}{L}\left(kL\right)^2\sin(kL) - (kL) \cos(k L) - \frac{L}{\beta} \sin(k L)=0.
\end{equation}

Using this secular equation it is easy to establish that the functions $u_n(x)$ are orthogonal with respect to the inner product
\begin{equation}
  \label{eq:appinnerprod}
  \langle u,v\rangle = \int_0^L \mathrm{d}x\,\bar{u}(x) v(x) +\alpha \bar{u}(0) v(0),
\end{equation}
with positive $\alpha$.

\subsubsection{Expansion theorem}
\label{sec:appexpansion}
We now need to establish an expansion theorem, that guarantees that any function $f$ defined in the interval $(0,L)$ can be written as a superposition $f=\sum_n f_n u_n$ with coefficients $f_n$, and in such a way that the value at the endpoint is also recovered. This will be achieved by identifying problem (\ref{eq:appwalt1}-\ref{eq:appwalt3}) with an eigenvalue problem for a self-adjoint operator, following the idea presented in \cite{Walter_1973_EVP}. We present here an alternative proof of that expansion theorem.

Thus, consider the Hilbert space ${\cal H}=L^2\left[(0,L)\right]\oplus \mathbb{C}_\alpha$, with elements $U=(u,a)$, where $u\in L^2\left[(0,L)\right]$ and $a$ is a complex number. The definition as a direct sum entails the inner product
\begin{equation}
  \label{eq:biginnerprod}
  \langle U,V\rangle = \int_0^L \mathrm{d}x\,\bar{u}(x) v(x) +\alpha \bar{a} b,
\end{equation}
for elements $U=(u,a)$ and $V=(v,b)$. Again, $\alpha$ is a positive length.

Let us now define an operator $A$, of inverse length squared dimension,  with domain
\begin{equation}
  \nonumber
  D(A)= \left\{U=(u,a)\in {\cal H}|u,u'~\mathrm{are}~ AC[0,L]\,,\,u(L)=0\,,\,a=\lim_{x\to 0}u(x)\right\}
\end{equation}
where $AC[0,L]$ denotes absolutely continuous in the interval, 
and acting on elements of its domain as
\begin{equation}
  \nonumber
  AU=(-u'',-u'(0)/\alpha+u(0)/\alpha\beta).
\end{equation}
$\beta$ is another positive length.

It is easy to check that this is a symmetric operator, i.e., $\langle U,AV\rangle=\langle AU,V\rangle$ for all $U,V\in D(A)\subset {\cal H}$. It is also easy to check that it is a monotone (accretive, dissipative) operator: for all $U\in D(A)$
\begin{equation}
  \label{eq:monotone}
  \langle U,AU\rangle= \int_0^L \mathrm{d}x\,|u'|^2+ \frac{1}{\beta}|u(0)|^2\geq 0.
\end{equation}

We now prove that $A$ is maximal monotone. That is, we prove that the range of $1+L^2A$ is the whole Hilbert space ${\cal H}$. In other words that for all $F\in {\cal H}$ there exists $U\in D(A)$ such that $U+L^2AU=F$. We insert the length squared $L^2$ for dimensional reasons, but we need not use precisely the length of the interval; multiplying this length by any number provides us with the same result. The problem now consists in showing the existence of solutions $u$, such that both $u$ and $u'$ are absolutely continuous, for the problem
\begin{eqnarray}
  \label{eq:appexistence}
 \qquad\qquad \qquad \qquad -L^2 u''+u&=& f,\nonumber\\
   \qquad\qquad \qquad \qquad \quad \quad\quad u(L)&=&0,\\
  \qquad \left(1+\frac{L^2}{\alpha\beta}\right)u(0)- \frac{L^2}{\alpha}u'(0)&=& a, \nonumber                                                                     
\end{eqnarray}
where $F=(f,a)$, with $f\in L^2[(0,l)]$ and $a$ a complex number. Make the change of variable
\begin{equation}
  \nonumber
  u(x)=v(x)+ \frac{a\sinh\left(1-x/L\right)}{\sinh(1)\left(1+L^2/\alpha\beta\right)+\cosh(1) L/\alpha}.
\end{equation}
We now have to prove the existence of a solution $v$, such that $v,v' \in AC[0,L]$, for the following problem with inhomogeneous term $f\in L^2[(0,l)]$:
\begin{eqnarray}
  \nonumber
  \qquad\qquad\qquad\qquad -L^2 v''+v&=& f,\nonumber\\
  \qquad\qquad\qquad\qquad\qquad\,\,\, v(L)&=& 0,\nonumber\\
  \qquad\left(1+\frac{L^2}{\alpha\beta}\right)v(0)- \frac{L^2}{\alpha}v'(0)&=& 0.\nonumber
\end{eqnarray}
This existence has been well established from the Sturm--Liouville case (see for instance \cite{reed1975ii}), whence the existence (and uniqueness) of solutions for problem (\ref{eq:appexistence}) is obtained, and thus maximal monotony of $A$. It follows that it is self-adjoint (see, for instance, proposition 7.6 in \cite{Brezis_2010_Functional}), and the spectral theorem of self-adjoint operators provides us with the expansion theorem we desired. Namely, the operator $A$ has a discrete spectrum $\{k^2_n\}_{n=0}^\infty$, solutions of (\ref{eq:appsecular}), $k_n$ being inverse lengths, with eigenvectors $U_n=\left(u_n(x),u_n(0)\right)$ where
\begin{equation}
  \nonumber
  u_n(x)= N_n \sin\left[k_n(L-x)\right],
\end{equation}
and the normalisation can be chosen as
\begin{equation}
  \nonumber
  N_n^2=\left[\frac{L}{2} +\left(\frac{\alpha}{2}+ \frac{1}{2\beta k_n^2}\right)\sin^2(k_nL)\right]^{-1}
\end{equation}
for the eigenvectors $\left\{U_n\right\}_{n=0}^\infty$ to form an orthonormal basis, $\left\langle U_n,U_m\right\rangle=\delta_{nm}$. Notice that we have chosen a real basis, and we will use this fact in the formulae that follow in this subsection \ref{sec:finite-length-transm}.
An element $F=(f,a)\in {\cal H}$ admits an expansion $F=\sum_{k=0}^\infty \langle U_k,F\rangle U_k$. Consider now an element $f\in L^2[(0,L)]$, with continuous representative $\tilde{f}$,  and extend it to ${\cal H}$ as $F=(f,\tilde{f}(0))$. We thus obtain the expansion we desired:
\begin{equation}
  \nonumber
  f(x)= \sum_{n=0}^\infty\langle U_n,F\rangle u_n(x) = \sum_{n=0}^\infty u_n(x)\left[\alpha u_n(0)\tilde{f}(0)+\int_0^L \mathrm{d}\xi\,u_n(\xi) f(\xi)\right].
\end{equation}

\subsubsection{Secondary inner product}
\label{sec:second-inner-prod}
We shall now prove (with the normalisation here used) equation (\ref{eq:TL_LCcoup_Network_ortho_2}). In what follows, the first step is integration by parts, the second makes use of the fact that $-u''_m(x)=k_m^2 u_m(x)$, the third relates the computation to the orthogonality $\langle U_n,U_m\rangle =\delta_{nm}$, and the last one introduces the boundary condition of (\ref{eq:appwalt1}):
\begin{eqnarray}
  \nonumber
  \int_0^L\mathrm{d}x\,u_n'(x)u_m'(x)&=& -\int_0^L\mathrm{d}x\,u_n(x)u_m''(x)-u_n(0)u_m'(0)\nonumber\\
                                     &=& k_m^2 \int_0^L\mathrm{d}x\,u_n(x)u_m(x) -u_n(0)u_m'(0)\nonumber\\
                                     &=& k_m^2\left[\langle U_n,U_m\rangle -\alpha u_n(0)u_m(0)\right]-u_n(0)u_m'(0)\nonumber\\
                                     &=& k_m^2 \delta_{nm} -u_n(0)\left[\alpha k_m^2 u_m(0)+u_m'(0)\right]\nonumber\\
  &=& k_m^2 \delta_{nm} -\frac{1}{\beta}u_n(0)u_m(0).
\end{eqnarray}
Alternatively, let us consider the quantity
\begin{equation}
  \label{eq:betainner}
  \left\langle U_n,F\right\rangle_{1/\beta}=\int_0^L\mathrm{d}x\,f'(x)u'_n(x) + \frac{1}{\beta}a u_n(0),
\end{equation}
for $F=(f,a)\in{\cal H}$. It is in fact well defined by integration by parts, and it thus follows that
\begin{equation}
  \label{eq:betaalphaequiv}
  \left\langle U_n,F\right\rangle_{1/\beta}= k_n^2\left\langle U_n,F\right\rangle,
\end{equation}
where the right hand side is the inner product we have used above, namely (\ref{eq:biginnerprod}).

We shall now extend this operation to an inner product in ${\cal H}$. As we have shown above, \ref{sec:appexpansion}, an element $F\in{\cal H}$ can be expanded as $F=\sum_nF_nU_n$, and the inner product reads $\langle F,G\rangle=\sum_n\bar{F}_nG_n$. Then we define $\langle F,G\rangle_{1/\beta}$ as
\begin{equation}
  \label{eq:betainnergen}
  \left\langle F,G\right\rangle_{1/\beta}=\sum_{n=0}^\infty k_n^2 \bar{F}_nG_n.
\end{equation}
Notice that this is an inner product: it is not degenerate because $k_n^2>0$ for all $n$ (v. Eq. (\ref{eq:monotone})). In fact, this is the inner product associated with the natural quadratic form induced by $A$, as presented in Eq. (\ref{eq:monotone}).

We immediately obtain an orthonormal basis $V_n=U_n/k_n$ with respect to this inner product, and we are led to the corresponding expansion theorem,
\begin{equation}
  \label{eq:betaexpansion}
  F=\sum_{n=0}^\infty \langle V_n,F\rangle_{1/\beta} V_n.
\end{equation}
\subsubsection{Sum rules}
\label{sec:appsumrules}

The expansion in ${\cal H}$ indicated above provides us with sum rules that prove very useful in our analysis of circuits. First, consider the special element of ${\cal H}$ given by $U=(0,1)$. Since it admits an expansion, it is the case that
\begin{equation}
  \label{eq:firstsumrule}
  1= \alpha\sum_{n=0}^\infty u_n^2(0).
\end{equation}
Notice that, asymptotically,
\begin{equation}
  \label{eq:asymptotics}
  k_n\sim \frac{n\pi}{L}+ \frac{L}{\alpha n \pi}+O(n^{-2}),
\end{equation}
which ensures convergence, since $u_n(0)\sim (-1)^n 2/\alpha n\pi$.

The sum rule (\ref{eq:firstsumrule})  is explicitly proven as follows:
\begin{eqnarray}
  \label{eq:ruleproof}
  (0,1)&=& \sum_{n=0}^\infty \langle U_n,(0,1)\rangle U_n\nonumber\\
       &=& \sum_{n=0}^\infty \alpha u_n(0) U_n\nonumber\\
  &=& \left(\sum_{n=0}^\infty \alpha u_n(0) u_n(x), \alpha\sum_{n=0}^\infty u_n^2(0)\right)
\end{eqnarray}

In the text we have used a different notation, namely $|\bi{u}|^2=N_\alpha/\alpha c$. The vector $\bi{u}$ is related to the sequence $\{u_n(0)\}$ by an overall normalization factor $\sqrt{N_\alpha /c}$.

Let us now obtain another sum rule by expanding $(0,1)\in{\cal H}$ in the basis $V_n$, orthonormal with respect to $\langle \bullet,\bullet\rangle_{1/\beta}$. Clearly $\left\langle V_n, (0,1)\right\rangle_{1/\beta}=v_n(0)/\beta=u_n(0)/\beta k_n$, and it follows that
\begin{equation}
  \label{eq:secondsumrule}
  1=\frac{1}{\beta}\sum_{n=0}^\infty \frac{u_n^2(0)}{k_n^2}.
\end{equation}

\subsection{Infinite Length Transmission Lines}
\label{subsec:infin-length-transm}
Let us now consider that the interval is only semibounded, i.e. $(0,\infty)$. As we shall see, the corresponding operator $A$ does not have a discrete spectrum, but is nonetheless self-adjoint, and an expansion theorem, in the form of an integral transform, does hold. Thus, we now examine the problem
\begin{eqnarray}
  \label{eq:appwaltinf}
 \qquad\qquad\quad \qquad - u''(x)&=& k^2 u(x)\,\quad x\in(0,\infty),\\
 \qquad\qquad \frac{1}{\beta}u(0)- u'(0)&=&  \alpha k^2 u(0),
\end{eqnarray}
where, again,
$\alpha$ and $\beta$ are positive constants with dimension of length, and we require (square) normalizability of $u$. Clearly there are no strong solutions to this problem. Setting aside functional details, it can nonetheless be checked that the following $u_k(x)$ functions are generalised orthonormal with respect to the inner product
\begin{equation}
  \label{eq:appinnerprodinfty}
  \langle u,v\rangle = \int_0^\infty \mathrm{d}x\,\bar{u}(x) v(x) +\alpha \bar{u}(0) v(0)\,:
\end{equation}
\begin{equation}
  \label{eq:genorthornor}
  u_k(x)= \frac{1}{\sqrt{2\pi\left[k^2+\left(\alpha k^2- 1/\beta\right)^2\right]}}\left[\left(k+ i \alpha k^2 - i/\beta\right) e^{ikx} + \left(k- i \alpha k^2 + i/\beta\right) e^{-ikx}\right]. 
  \end{equation}
  Observe that we have again chosen real $u_k(x)$, and furthermore that $u_{-k}(x)=-u_k(x)$, which allows us to restrict ourselves to positive $k$.
  They are generalised orthonormal in the sense that
  \begin{equation}
    \label{eq:orhtoinfty}
    \langle u_k,u_q\rangle= \delta(k-q).
  \end{equation}
  In order to prove this statement directly it is convenient to use the distributional identity
  \begin{equation}
    \label{eq:distribeq}
    \int_0^\infty\mathrm{d}x\, e^{i(k-q) x}= \pi \delta(k-q) + i \mathrm{P}\frac{1}{k-q},
  \end{equation}
  where $\mathrm{P}$ denotes principal part.

\subsubsection{Expansion theorem}
\label{sec:expansion-theorem-cont}

We shall follow the scheme presented in \ref{sec:appexpansion} to prove a corresponding expansion theorem.
We shall make use of the well known fact that, in physics terms, the free Hamiltonian in the half-line is well defined and self-adjoint once we fix at the origin the condition
\begin{equation}
  \label{eq:bchamhalf}
  \cos(\theta)u(0)+\sin(\theta) l u'(0)=0,
\end{equation}
with $\theta$ and angle and $l$ the unit length. Notice that for the non-relativistic free particle there is no natural length. We are free to choose any length, and in fact nothing crucial in our proof will depend on the choice adopted. Thus, we select $l=\sqrt{\alpha\beta}$.
Let us denote the free Hamiltonian in the half-line with this choice of length and condition (\ref{eq:bchamhalf}) $H_\theta$, and its domain by $D(H_\theta)\subset L^2\left[(0,\infty)\right]$. In fact, we shall also make the choice
\begin{equation}
  \label{eq:thetachoice}
  \tan(\theta)= -\frac{1}{2}\sqrt{\frac{\beta}{\alpha}},
\end{equation}
for later convenience.

Let us now introduce an operator $A$ acting on $D(A)\subset {\cal H}= L^2\left[(0,\infty)\right]\oplus \mathbbm{C}_\alpha$, given by
\begin{equation}
  \label{eq:domainf}
  D(A)=\left\{U=(u,a)\in{\cal H}|u\in H^2[(0,\infty)],\,a=\lim_{x\to0}u(x)\right\}.
\end{equation}
$H^2[(0,\infty)]$ is the Sobolev space $H^2[(0,\infty)]=W^{2,2}[(0,\infty)]$ (for details, see for instance \cite{Brezis_2010_Functional} or \cite{LionsMagenes_2011}).
  The inner product in ${\cal H}$ is of course
  \begin{equation}
    \label{eq:innerinf}
    \left\langle U,V\right\rangle=\left\langle(u,a),(v,b)\right\rangle=\int_0^\infty\mathrm{d}x\,\bar{u}(x)v(x)+ \alpha\bar{a}b.
  \end{equation}

  The operator $A$ acts on its domain as
  \begin{equation}
  \label{eq:actionAinf}
  AU=(-u'',-u'(0)/\alpha+u(0)/\alpha\beta).
\end{equation}
Both $u(0)$ and $u'(0)$ are understood as limits.
It is again easy to prove that it is a symmetric operator, and that it is monotone. In order to prove that is maximal monotone we examine the invertibility of $1+ \alpha\beta A$ in ${\cal H}$. This requires us studying the problem
\begin{eqnarray}
  \label{eq:appexistenceinf}
  \,\,\qquad\qquad\qquad\quad-\alpha\beta u''+u&=& f,\nonumber\\
  \qquad\qquad\qquad 2u(0)- \beta u'(0)&=& a,
\end{eqnarray}
for $f\in L^2[(0,\infty)]$, $a\in\mathbbm{C}$, and $u$ square summable. Make the change of variable
\begin{equation}
  \label{eq:chofvarinf}
  u(x)=\frac{a}{2+\sqrt{\beta/\alpha}} e^{-x/\sqrt{\alpha\beta}}+v(x).
\end{equation}
We now have to study the problem
\begin{eqnarray}
  \label{eq:vexistenceinf}
  \,\,\qquad\qquad\qquad\quad -\alpha\beta v''+v&=& f,\nonumber\\
  \qquad\qquad\qquad 2v(0)- \beta v'(0)&=& 0,
\end{eqnarray}
and the boundary condition at the origin is $v(0)+\tan(\theta)\sqrt{\alpha\beta}v'(0)=0$, with the choice (\ref{eq:thetachoice}). As $H_\theta$ is selfadjoint and positive, the existence (and uniqueness) of the solution of this problem is guaranteed, whence the existence (and uniqueness) of the solution of (\ref{eq:appexistenceinf}) for all $F\in{\cal H}$ follows. As a consequence, there is a unique self-adjoint extension of $A$ as defined above, and we obtain the expansion theorem we desire: for all $F\in{\cal H}$ we have $F=\int_0^\infty\mathrm{d}k\,\langle U_k,F\rangle U_k$, using physics notation, where $U_k=(u_k(x),u_k(0))$, with $u_k(x)$ defined in (\ref{eq:genorthornor}). Restricting ourselves to $f\in L^2[(0,\infty)]$, we obtain
\begin{eqnarray}
  \label{eq:infexpansion}
  \qquad\qquad\quad f_k&=& \int_0^\infty\mathrm{d}x\, u_k(x) f(x) + \alpha u_k(0) f(0)\nonumber\\
   \quad\,\,\mathrm{and}\quad f(x)&=& \int_0^\infty\mathrm{d}k\, f_k u_k(x),
\end{eqnarray}
with the caveats more usual for Fourier transforms.

\subsubsection{Secondary inner product}
\label{sec:second-inner-prod-inf}

We now proceed to construct a second expansion by considering a new inner product. As in the finite interval case, it is defined from the natural quadratic form induced by $A$,
\begin{equation}
  \label{eq:quadinf}
  \left\langle U,U\right\rangle_{1/\beta}=\left\langle U,AU\right\rangle=\int_0^\infty\mathrm{d}x\, |u'(x)|^2+ \frac{1}{\beta} |u(0)|^2.
\end{equation}
The extension of this inner product to ${\cal H}$ can be presented via polarization identities, via a Parseval identity, or alternatively by the spectral theorem, namely
\begin{equation}
  \label{eq:secondinnerinf}
  \left\langle F,G\right\rangle = \int_0^\infty\frac{\mathrm{d}k}{k^2}\bar{F}_kG_k,
\end{equation}
where
\begin{equation}
  \label{eq:fkexp}
  F_k=\left\langle U_k,F\right\rangle=\int_0^\infty\mathrm{d}x\,u_k(x)f(x)+ \alpha u_k(0) a
\end{equation}
for $F=(f,a)\in{\cal H}$, and analogously for $G$. The kernel of the integration is $1/k^2$, thus seemingly producing a non-integrable singularity at the origin. A more careful analysis implies analyzing $u_k(x)u_k(y)/k^2$, which is in fact regular at the origin $k=0$. It is therefore feasible to establish a (generalised) orthonormal basis with respect to this inner product, $\left\langle V_k,V_q\right\rangle_{1/\beta}=\delta(k-q)$, by $V_k=U_k/k$. A new expansion theorem follows, in the form
\begin{equation}
  \label{eq:secondexpinf}
  F=\int_0^\infty\mathrm{d}k\,\left\langle V_k,F\right\rangle_{1/\beta}V_k.
\end{equation}
  \subsubsection{Sum rules}
\label{sec:infsumrules}

Let us now consider sum rules, analogous to (\ref{eq:firstsumrule}). In particular, let us expand $F=(0,1)$ in the form  $F=\int_0^\infty\mathrm{d}k\,\langle U_k,F\rangle U_k$ . Clearly, $\langle U_k,F\rangle= \alpha u_k(0)$. Thus the expansion reads, for the second component of $F$,
\begin{eqnarray}
  \label{eq:contsumrule}
  1&=& \int_0^\infty\mathrm{d}k\, \alpha u_k^2(0)\nonumber\\
   &=& \frac{\alpha}{\pi} \int_o^\infty\mathrm{d}k\,\frac{2 k^2}{k^2+\left(\alpha k^2-1/\beta\right)^2}\nonumber\\
  &=& \frac{1}{\pi}\int_{-\infty}^{+\infty}\mathrm{d}q\,\frac{q^2}{\left(q^2-x^2\right)^2+q^2},
\end{eqnarray}
where $x=\sqrt{\alpha/\beta}$. Remember that we are using a real basis $u_k(x)$. This last integral can be computed explicitly by residues, and it is actually independent of the real variable $x$, thus providing us with an independent check of the expansion and the sum rule derived thereof, namely
\begin{equation}
  \label{eq:contsumrule2}
  \int_0^\infty\mathrm{d}k\,u_k^2(0)= \frac{1}{\alpha}.
\end{equation}
The sum rule analogous to (\ref{eq:secondsumrule}) follows from the expansion (\ref{eq:secondexpinf}), and  reads
\begin{equation}
  \label{eq:contsecondsumrule}
  \beta=\int_0^\infty\mathrm{d}k\,\frac{u_k^2(0)}{k^2}=\frac{\alpha}{\pi}\int_0^\infty\frac{\mathrm{d}q}{q^2+\left(q^2-\alpha/\beta\right)^2}.
\end{equation}
Again, this integral can be explicitly computed, and the sum rule is checked. Indeed, let
\begin{equation}
  g(q,x)=\frac{1}{q^2+(q^2-x^2)},\nonumber
\end{equation}
with which we compute
\begin{eqnarray}
  \label{eq:explicitint}
  I_g(x)&=&\frac{1}{\pi} \int_0^\infty\frac{\mathrm{d}q}{q^2+\left(q^2-x^2\right)^2}\nonumber\\
        &=& 2i\left[\lim_{q\to i(1+\sqrt{1-4x^2})/2}\left(q-\frac{i}{2}(1+\sqrt{1-4x^2})\right)g(q,x)\right.\nonumber\\
  &&\quad\left.+\lim_{q\to i(1-\sqrt{1-4x^2})/2}\left(q-\frac{i}{2}(1-\sqrt{1-4x^2})\right)g(q,x)\right]\nonumber\\
  &=&\frac{1}{x^2}.
\end{eqnarray}

\subsection{Other configurations}
\label{sec:other-configurations}

Let us now consider other configurations. We shall not give all the details, which can be filled out following the scheme in the previous subsections.

\subsubsection{Galvanic coupling}
\label{sec:galvanic-coupling}

Let the Hilbert space be $L^2[(-L,0)]\oplus L^2[(0,L)]\oplus\mathbbm{C}_\alpha$, with elements $F=(f_-,f_+,a)$. The inner product is
\begin{equation}
  \label{eq:galvanicinner}
  \left\langle (f_-,f_+,a),(g_-,g_+,b)\right\rangle=\int_{-L}^0\mathrm{d}x\,\bar{f}_- g_-+\int_0^L\mathrm{d}x\,\bar{f}_+ g_++ \alpha \bar{a}b.
\end{equation}

We define an operator $A$ with domain
\begin{eqnarray}
  \label{eq:galvanicformal}
  \fl
  D(A)&=&\left\{U=(u_-,u_+,\Delta)|u_-,u_-'\in AC[(-L,0)],\,u_+,u_+'\in AC[(0,L)],\right.\nonumber\\ &&\left. u_+(L)=u_-(-L)=0,\,u_-'(0)=u_+'(0),\,\Delta = u_+(0)-u_-(0)\right\},
\end{eqnarray}
acting on its domain as $AU=(-u_-'',-u_+'',- u'(0)/\alpha+\Delta/\alpha\beta)$, with $\Delta=u_+(0)-u_-(0)$.
It is clear that it is a  symmetric operator, and it is positive since
\begin{equation}
  \label{eq:uaugalv}
  \left\langle U,AU\right\rangle=\int_{-L}^0\mathrm{d}x\,|u_-'|^2+\int_{0}^L\mathrm{d}x\,|u_+'|^2+\frac{1}{\beta}\left|\Delta\right|^2.
\end{equation}

The eigenvalue and eigenvector equation, $AU=k^2U$, give us the following boundary value problem:
\begin{eqnarray}
  \label{eq:eqgalvanichilb}
 \qquad\qquad\,\,\,\quad -u''_-&=&k^2 u_-,\nonumber\\
 \qquad\qquad\,\,\,\quad -u''_+&=&k^2 u_+,\nonumber\\
 \qquad\qquad\quad u'_-(0)&=&u'_+(0),\nonumber\\
 \qquad\qquad\quad u_+(L)&=&u_-(-L)=0,\nonumber\\
 \qquad\qquad\,\,-u'_-(0)&=& \left(\alpha k^2 -\frac{1}{\beta}\right)\left[u_+(0)-u_-(0)\right].
\end{eqnarray}
The secular equation reads
\begin{equation}
  \label{eq:seculargalvanic}
  k\cos(kL)=2\left(\alpha k^2- \frac{1}{\beta}\right)\sin(kL).
\end{equation}
Notice that in fact this operator is definite positive, and $0$ is not an eigenvalue.

It will be no surprise at this point that we define a secondary inner product as the natural quadratic form induced by $A$, with reference to (\ref{eq:uaugalv}), thus justifying Eq. (\ref{eq:TL_TL_LCgalvcoup_Networks_ortho_2}).

We can also obtain sum rules, such as
\begin{equation}
  \label{eq:galvanic-sum-rule}
  \frac{1}{\alpha}=\sum_{n=0}^\infty \frac{\left|\Delta_n\right|^2}{\langle U_n,U_n\rangle}.
\end{equation}

\subsubsection{Point insertion}
\label{sec:point-insertion}
Again, consider the Hilbert space ${\cal H}=L^2[(0,L)]\oplus \mathbbm{C}\oplus L^2[(L,\infty)]$, and an operator $A$ acting on $U=(u_-,u_+,a=u_+(L)=u_-(L))\in D(A)\subset{\cal H}$ as $(-u_-'',-u_+'',u(L)/\alpha\beta -(u_+'(L)-u_-'(L))/\alpha) $, where we demand that $u_+(L)=u_-(L)$, $u(0)=0$ and the values at the endpoints are understood as limits . The eigenvector and eigenvalue equation reads
\begin{eqnarray}
  \qquad\qquad\qquad\quad-u''_-&=&k^2 u_-,\nonumber\\
  \qquad\qquad\qquad\quad-u''_+&=&k^2 u_+,\nonumber\\
  \qquad\qquad\qquad\,u_-(L)&=&u_+(L),\nonumber\\
  \qquad\qquad\qquad\,u_-(0)&=&0,\nonumber\\
  \qquad\qquad\qquad\,u_-'(L)&=& u_+'(L)+\left(\alpha k^2 -\frac{1}{\beta}\right) u(L).\nonumber
\end{eqnarray}

\subsection{The recipe}
\label{sec:recipe}

Let us summarize in a general way the point of view presented in this section. We are considering Hilbert spaces of the form ${\cal H}=\left(\oplus_{k=1}^{N}L^2[I_k]\right)\oplus\left(\oplus_{j=1}^M \mathbbm{C}^{n_j}_{\alpha_j}\right)$. There are $N$ intervals $I_K$ and $M$ relevant boundaries. In fact we are considering  second order differential operators acting on single functions, whence for all $M$ relevant boundaries we have $n_j=1$. The number of relevant boundaries will depend on the system we model. The largest possible number corresponds to sequences of $N$ finite length transmission lines such that all endpoints are relevant boundaries, thus $M=2N$. We are considering the standard inner product in each interval, and the second order differential operation ${\cal L}:u\to -u''$ in each. More general second order Sturm--Liouville differential operators, ${\cal L}:u\to -(p u')'+q u'$ ($p>0$) can also be considered, and, by proper modifications of the weight, other second order differential operators. 

In this manner, the elements $U\in{\cal H}$ are of the form $U=\left(u_1,\ldots,u_N,a_1,\ldots,a_m\right)$, $V=\left(v_1,\ldots,v_N,b_1,\ldots,b_m\right)$. If required, we will explicitly denote the element a number component belongs to, as in $a_j^U$, $a_j^V$.  The inner product is determined by the direct sum structure as
\begin{equation}
  \nonumber 
  \left\langle U,V\right\rangle=\sum_{k=1}^N\int_{I_k}\mathrm{d}x\,\bar{u}_k(x)v_k(x)+\sum_{j=1}^M \alpha_j\bar{a}_jb_j.
\end{equation}
We shall consider operators $A$ that will act on the function components as Sturm--Liouville operators, ${\cal L}_k$ acting on the $k$-th component. Their domains will be determined by the finiteness or otherwise of the intervals. For finite intervals, typically we shall require $u_k$ and $u'_k$ to be $AC[I_k]$, absolutely continuous in the corresponding interval.

An endpoint is a \emph{relevant} boundary, i.e. there is a number component associated to it, if the domain of $A$ is not restricted by a boundary condition on $u$ at that endpoint. The definition of the domain of $A$ includes a condition on the number components ($a_j$) at relevant boundaries. A boundary can be associated to just one interval or it can be associated to two intervals, if it is a common endpoint. If a relevant boundary, with index $j$, is associated to just one interval, $I_k$, the corresponding number component $a_j$ will be determined by the (free) value of the limit of $u_k$ when tending to that boundary. If, however, the relevant boundary with index $j$ is associated, as their common endpoint $x_k$, to two intervals $I_k$ and $I_{k+1}$, there will be a condition for $U$ to belong to the domain in terms of the continuity of either $u$ or its derivative, i.e., either $\lim_{x\to x_k^-} u_k(x)=\lim_{x\to x_k^+} u_{k+1}(x)$ or similarly for the derivatives, and the number component $a_j$ will be determined by the common value or by the  jump in the functions. Finally, $AU=\left({\cal L}_iu_1,\ldots,{\cal L}_Nu_N,d_1/\alpha_1 +a_1/\alpha_1\beta_1,\ldots, d_M/\alpha_M+a_M/\alpha_M\beta_M\right)$, where $d_j$ is a linear combination of the limits of derivatives of the functions at the $j$-th relevant boundary. Again, we will explicity denote $d_j^U$, $d_j^V$, etc., if required. The eigenvalue and eigenvector equation is thus
\begin{eqnarray}
  \nonumber 
  \qquad\qquad\qquad\qquad{\cal L}_iu_i&=&k^2 u_i,\nonumber\\
  \qquad\qquad\qquad d_j+\frac{1}{\beta_j}a_j&=&\alpha_j k^2 a_j,\nonumber
\end{eqnarray}
together with some boundary conditions. For our procedure to be well defined, the boundary terms arising from
\begin{equation}
  \nonumber 
  \int_{I_k}\mathrm{d}x\,\left[\bar{v}_k(x) {\cal L}_ku_k(x)-\overline{\left({\cal L}_kv_k(x)\right)}u_k(x)\right]
\end{equation}
must be cancelled by
\begin{equation}
  \nonumber 
  \sum_{j=1}^M \left(\bar{a}_j^vd_j^u-\bar{d}_j^v a_j^u\right).
\end{equation}
This ensures symmetry. Furthermore, we require that  $\left\langle U,AU\right\rangle$ be lower bounded. By shifting some ${\cal L}_k$ by a constant (alternatively, by demanding that they all be positive) the demand is positivity. This can generally be achieved if the boundary terms
\begin{equation}
  \nonumber 
  \sum_{k=1}^N\left[-p_k\bar{u}_ku_k'\right]_{\partial I_k}
\end{equation}
are cancelled by $\sum_{j=1}^M\bar{a}_jd_j$ on the domain of $A$.

Once symmetry and positivity are ensured, it remains to be examined whether indeed we are led to a self-adjoint operator, whence an expansion theorem would follow. We have addressed this issue in the examples by studying the existence (and uniqueness) of solutions to $\left(1+L^2A\right)U=F$ for $F\in{\cal H}$. This amounts to studying the system of equations
\begin{equation}
  \nonumber 
  u_k+L^2{\cal L}_k u_k = f_k
\end{equation}
for $k=1,\ldots,N$ together with conditions
\begin{equation}
  \nonumber 
  a_j^U+\frac{L^2}{\alpha_j}\left(d_j^U+ \frac{a_j^U}{\beta_j}\right)=a_j^F
\end{equation}
for $j=1,\ldots,M$, and the required boundary conditions for elements $U\in D(A)$. Let $j$ be the index of  the only boundary relevant to the $k$-th element (for definiteness; extensions are straightforward); this entails the idea that the other endpoint of $I_k$ has associated homogeneous boundary conditions. We construct $u_k^0$, a solution of the problem
\begin{eqnarray}
  \nonumber 
  \qquad\qquad\qquad\qquad \quad \,\, u_k+L^2{\cal L}_ku_k&=&0,\nonumber\\
  \qquad\qquad\qquad a_j^U+\frac{L^2}{\alpha_j}\left(d_j^U+ \frac{a_j^U}{\beta_j}\right)&=&a_j^F,\nonumber\\
  \qquad\qquad\qquad \mathrm{other~B.C.}&&\nonumber
\end{eqnarray}
Notice that the relevant bounday condition involves $u_k$ linearly. Once this has been achieved, one makes the change of variables $u_k=u_k^0+v_k$, and we are led to the study of $v_k+L^2 {\cal L}_kv_k=f_k$ with homogeneous boundary conditions. For the systems we have considered, this has a unique solution, and the existence and uniqueness of solutions to   $\left(1+L^2A\right)U=F$  has been thus established. This, in turn, establishes that $F\in{\cal H}$ can be expanded as $F=\sum_{n=0}^\infty\langle U_n,F\rangle U_n$, with $U_n$ orthonormal eigenvectors of $A$.

\subsubsection{Side results}
\label{sec:side-results}

As a side product of the process we obtain sum rules, generically by expanding special elements of the Hilbert space, of the form $u_k=0$ and one of the $a_j$ set to one while the others are zero. This gives us
\begin{equation}
  \label{eq:genericsumrule}
  \frac{1}{\alpha_j}=\sum_{n=0}^\infty\left|a_j^{U_n}\right|^2
\end{equation}
for normalised $U_n$ eigenvectors.

Another side result is what we have termed the secondary inner product. We have relied in our proofs on the positivity of $\left\langle U,AU\right\rangle$ for $U\in D(A)$ (in fact based on physical reasons: it has to be associated to the harmonic approximation for small oscillations), and this provides us, by extension to the whole Hilbert space ${\cal H}$, with a positive definite quadratic form. From the expansion theorem, $F=\sum_n\langle U_n,F\rangle U_n$, denoting the coefficients of $F$ (resp. $G$) in the orthonormal basis $U_n$ with eigenvalues $k_n^2$ (orthonormal with respect to the initial product $\langle\cdot,\cdot\rangle$) as $F_n$ (resp. $G_n$), the new inner product  $\langle\cdot,\cdot\rangle_{1/\beta}$ is given as
  \begin{equation}
    \nonumber 
    \left\langle F,G\right\rangle_{1/\beta}= \sum_{n=0}^\infty\frac{1}{k_n^2} \bar{F}_nG_n.
  \end{equation}

\subsection{Alternative mathematical approaches}
\label{sec:altern-math-appr}

\subsubsection{Trace operator}
\label{sec:trace-operator}

We have restricted ourselves to the Hilbert space setting, due to the later application to quantization. Nonetheless, a number of questions regarding these expansions can also be analysed in terms of Sobolev spaces, for which the concept of \emph{trace} (in the sense of trace of an element $u\in W^{1,p}(\Omega)$ which is understood as the ``boundary function'' $u|_{\partial\Omega}$, see \cite{LionsMagenes_2011} ) appears. That context is natural in order to study geometrically the transformation from Lagrangian to Hamiltonian in cases such as those we consider (see \cite{Barbero_2015,Barbero_2017}  for such a viewpoint )

\subsubsection{Delta distribution}
\label{sec:delta-function}

An alternative approach, still in the Hilbert space context, is to consider the Hilbert space $L^2\left[[0,L);\mu\right]$, where the measure presents a point mass in the initial point. Even more, the idea can be extended to the case of measures with additional point masses, both in the interior and at the endopoints. This is, for instance, the concrete presentation that appears in \cite{Walter_1973_EVP}.  It is also related to the computations of \cite{Moein_2016,Moein_2017_CutFree}.

Undoubtedly this is a feasible route; we have preferred to set it aside to avoid problems in moving to the Hamiltonian formalism. It should be noted that constructing precisely the functional analytic details need not be trivial at all, though.

\section{Capacitive to inductive coupling}
\label{sec:capac-induct-coupl}

A common presentation of spin-boson Hamiltonians is related to the seminal work of Caldeira and Leggett \cite{CaldeiraLeggett_1981,CaldeiraLeggett_1983,Leggett_1984}. There a system, with coordinate $q$, is coupled to a bath of linear oscillators, with coordinates $x_\alpha$, and the coupling of interest is of the form $q\sum_\alpha c_\alpha x_\alpha$. Caldeira and Leggett thoroughly analyse other possibilities, in particular those of the form $\dot{q}\sum_\alpha c_\alpha x_\alpha$ (or equivalently $-q\sum_\alpha c_\alpha\dot{x}_\alpha$), and show with an example how to relate both forms. In fact, as they also point out, this is achieved through a canonical transformation, and there is no point transformation that can reproduce it. For completeness we present this canonical presentation here, and then we study  the mapping from a capacitive $\dot{q}\sum_\alpha c_\alpha \dot{x}_\alpha$ to the inductive form $q\sum_\alpha c_\alpha x_\alpha$, which we will see is a point transformation.

Thus, let us first consider the Lagrangian 
\begin{equation}
  \nonumber 
   L=\frac{m}{2}\dot{q}^2-V(q)+\frac{1}{2}\dot{\bi{x}}^T\mathsf{M}\dot{\bi{x}}-\frac{1}{2}\bi{x}^T\mathsf{\Lambda}^2\bi{x}-{q}\bi{c}^T\dot{\bi{x}},
\end{equation}
where there is a single system variable $q$, and a bath of harmonic oscillators, with position variables $\bi{x}$, are coupled to the system via a coupling vector $\bi{c}$ in the interaction term $-{q}\bi{c}^T\dot{\bi{x}}$. Let the corresponding canonical momenta be $p$ and $\bi{p}$. The Hamiltonian is
\begin{equation}
  \nonumber 
  H=\frac{p^2}{2m}+\frac{1}{2}\bi{p}^T\mathsf{M}^{-1}\bi{p}+q\bi{c}^T\mathsf{M}^{-1}\bi{p}+\frac{q^2}{2}\bi{c}^T\mathsf{M}^{-1}\bi{c}+V(q)+\frac{1}{2}\bi{x}^T\mathsf{\Lambda}^2\bi{x}.
\end{equation}
With the canonical transformation $\boldsymbol{\pi}=-\bi{x}$, $\boldsymbol{\xi}=\bi{p}$, and going back to the Lagrangian, one obtains
\begin{equation}
  \nonumber 
  \tilde{L}=\frac{m}{2}\dot{q}^2-V(q)-\frac{q^2}{2}\bi{c}^T\mathsf{M}^{-1}\bi{c}+\frac{1}{2}\dot{\boldsymbol{\xi}}^T\mathsf{\Lambda}^{-2}\dot{\boldsymbol{\xi}}-\frac{1}{2}\boldsymbol{\xi}^T\mathsf{M}^{-1}\boldsymbol{\xi}-q\bi{c}^T\mathsf{M}^{-1}\boldsymbol{\xi}.
\end{equation}
The coupling indeed is now of inductive form, as expected. Notice however that the variable $\xi$ has dimensions of momenta, and that has to be taken into account to determine the spectral density, for instance. On computing explicitly, the spectral density reads, formally,
\begin{equation}
  \nonumber 
  J(\omega)=\bi{c}^T\mathsf{M}^{-1}\mathsf{\Lambda}\mathsf{O}^T\delta\left(\omega-\mathsf{\Omega}\right)\mathsf{\Omega}^{-1}\mathsf{O}\mathsf{\Lambda}\mathsf{M}^{-1}\bi{c}.
\end{equation}
Here $\mathsf{O}$ is the orthogonal matrix diagonalizing $\mathsf{\Lambda}\mathsf{M}^{-1}\mathsf{\Lambda}$, namely
\begin{equation}
  \nonumber 
  \mathsf{O}\mathsf{\Lambda}\mathsf{M}^{-1}\mathsf{\Lambda}\mathsf{O}^T=\mathsf{\Omega}^2,
\end{equation}
with diagonal $\mathsf{\Omega}$. The definition we use for the spectral density $J(\omega)$ is best expressed in terms of the classical equations of motion for the variable $q$. The source term for  $q$ due to the dynamics of $\boldsymbol{\xi}$ is reexpressed, after solving the classical equations of motion for $\boldsymbol{\xi}$, in the form $\int_0^t\mathrm{d}\tau\int\mathrm{d}\omega J(\omega)\sin\omega(t-\tau)\,q(\tau)$, which provides the definition of $J(\omega)$. 

Dimensionally, $\mathsf{\Lambda}^2$ is mass times frequency squared, $\mathsf{M}$ is mass, $\mathsf{O}$ adimensional, so the terms bracketed by the $\bi{c}$ vectors (setting aside the delta) are frequency over mass.

This example shows that the identification of the spectral density is coupling dependent (as would be obvious from dimensional analysis). We now study capacitive coupling in this regard.  In particular, 
we will compute this process explicitly for the following general example:
\begin{equation}
  \label{eq:CLcapacitiveLagr}
  L=\frac{m}{2}\dot{q}^2-V(q)+\sum_\alpha \left[\frac{m_\alpha}{2}\dot{x}_\alpha^2-\frac{m_\alpha\omega_\alpha^2}{2}{x}_\alpha^2\right]-\dot{q}\sum_\alpha c_\alpha \dot{x}_\alpha.
\end{equation}
We shall use a more compact notation that also provides a slight generalisation, namely
\begin{equation}
  \nonumber 
  L=\frac{m}{2}\dot{q}^2-V(q)+\frac{1}{2}\dot{\bi{x}}^T\mathsf{M}\dot{\bi{x}}-\frac{1}{2}\bi{x}^T\mathsf{\Lambda}^2\bi{x}-\dot{q}\bi{c}^T\dot{\bi{x}}.
\end{equation}

In this case the  canonical transformations can be reduced to a point transformation applied to the initial Lagrangian, namely
\begin{equation}
  \label{eq:contacttrans}
  \bi{x}=\bi{x}_2+q\,\mathsf{M}^{-1}\bi{c},
\end{equation}
and the new Lagrangian reads
\begin{eqnarray}
  \nonumber 
  \tilde{L}&=& \frac{1}{2}\left(m+\bi{c}^T\mathsf{M}^{-1}\bi{c}\right)\dot{q}^2-V(q)-\frac{1}{2}\bi{c}^T\mathsf{M}^{-1}\mathsf{\Lambda}^2\mathsf{M}^{-1}\bi{c}\,q^2\nonumber\\
  && +\frac{1}{2}\dot{\bi{x}}_2^T\mathsf{M}\dot{\bi{x}}_2-\frac{1}{2}\bi{x}_2^T\mathsf{\Lambda}^2\bi{x}_2-q\bi{c}^T\mathsf{M}^{-1}\mathsf{\Lambda}^2\bi{x}_2.\nonumber
\end{eqnarray}
For the specific case of (\ref{eq:CLcapacitiveLagr}) we have
\begin{eqnarray}
  \nonumber 
  \tilde{L}&=&\frac12\left(m+\sum\frac{c_\alpha^2}{m_\alpha}\right)\dot{q}^2- V(q)-q^2\sum\frac{c_\alpha^2\omega^2_\alpha}{2m_\alpha}\nonumber\\
  && +\sum\left(\frac{m_\alpha}{2}\dot{\xi}_\alpha^2-\frac{m_\alpha\omega^2_\alpha}{2}\xi_\alpha^2\right)- q\sum c_\alpha\omega_\alpha^2\xi_\alpha,\nonumber
\end{eqnarray}
with the new variables $\xi_\alpha=x_\alpha-c_\alpha q/m_\alpha$, 
and one obtains a spectral density
\begin{equation}
  \nonumber 
  J(\omega)=\sum\frac{c_\alpha^2\omega^3_\alpha}{m_\alpha} \delta(\omega-\omega_\alpha).
\end{equation}

The transformation (\ref{eq:contacttrans})  is implemented  in the quantum case by a unitary transformation. That is,
\begin{equation}
  \nonumber 
  \hat{U}=\exp\left[-i\hat{q}\bi{c}^T\mathsf{M}^{-1}\hat{\bi{p}}/\hbar\right]=\exp\left[-\frac{i}{\hbar}\sum_\alpha\frac{c_\alpha}{m_\alpha}qp_\alpha\right].
\end{equation}

\section{Hamiltonian Formalism with Transmission Lines}
\label{sec:hamilt-form}

\subsection{The continuum Hamiltonian}
\label{sec:cont-hamilt}
In order to compare with some of the literature, it might be convenient to write the continuum version of some of the Hamiltonians presented here. We shall carry out this task, fully explicitly, in the case of Hamiltonian (\ref{eq:Ham_LCcoup_Network2}), using the notation of subsection \ref{subsubsec:chap2_mixed_linear_coupling}. In order to obtain this expression, notice that we have started from a Lagrangian with a continuum part, namely (\ref{eq:Lag_TL_LCcoup_Network}). Next we have expressed the flux field $\Phi(x,t)$ as the infinite sum $\Phi(x,t)=\sum_n\Phi_n(t)u_n(x)$, where $u_n(x)$ are the function component of $U_n=(u_n(x),u_n(0))\in {\cal H}=L^2[(0,L)\oplus \mathbbm{C}_\alpha$. These basis vectors are orthogonal according to
\begin{equation}
  \nonumber 
  \left\langle U_n,U_m\right\rangle= c\left(\int_0^L\mathrm{d}x\,u_n(x)u_m(x)+ \alpha\, u_n(0) u_m(0)\right)=N_\alpha \delta_{nm}.
\end{equation}
The dimension of $N_\alpha$, namely capacity, has been chosen  so that the function elements $u(x)$ are adimensional. Furthermore $\alpha$ has dimension of length, as we stated previously.  Since the dimension of the flux field is voltage times time, we see that the dimension of $\Phi_n$ is $\left[\Phi\right]/[u_n]= V\cdot T$, with $V$ standing in for voltage, and $T$ for time. Later, we have $C$ for capacity.

The variables $Q_n=\partial L/\partial\dot\Phi_n$ have dimension $\left[Q_n\right]=[L]\cdot T/[\Phi_n]= C\cdot V^2\cdot T/ \left(V\cdot T\right)=C V$, i.e., charge.
In order to obtain a Hamiltonian with continuum component, we construct a function of time  that takes values in ${\cal H}$, $Q(t)$, as
\begin{equation}
  \nonumber 
  Q(t)=\sum_{n=0}^\infty Q_n(t) U_n=\left(Q(x,t),Q(0,t)\right).
\end{equation}
Observe that
\begin{eqnarray}
  \nonumber 
  \left\langle Q(t),Q(t)\right\rangle&=&\sum_{n=0}^\infty Q_n^2(t) N_\alpha\nonumber\\
  &=& c\left[ \int_0^L\mathrm{d}x\, Q^2(x,t)+\alpha\, Q^2(0,t)\right].\nonumber
\end{eqnarray}
We can thus substitute in the Hamiltonian (\ref{eq:Ham_LCcoup_Network2}) to obtain
\begin{eqnarray}
  \nonumber 
   H&=& \frac{1}{2}\bi{q}^T(\mathsf{A}^{-1}+ \frac{C_g^2}{\alpha c} \mathsf{A}^{-1}\bi{a}\bi{a}^T \mathsf{A}^{-1})\bi{q}+\frac{1}{2}\boldsymbol{\phi}^T\mathsf{B}^{-1}\boldsymbol{\phi}+V(\boldsymbol{\phi})\nonumber\\
  &&+\int_0^L\mathrm{d}x\,\left[\frac{c}{2N^2_\alpha}Q^2(x,t)+\frac{1}{2l}\left(\partial_x\Phi(x,t)\right)^2\right]\nonumber\\
  &&+\frac{c\alpha}{2N^2_\alpha} Q^2(0,t)+ \frac{C_g}{N_{\alpha}} (\bi{q}^T\mathsf{A}^{-1} \bi{a})Q(0,t)\nonumber\\ 
  &&+ \frac{1}{2L_g}\Phi^2(0,t)-\frac{1}{L_g} (\boldsymbol{\phi}^T\bi{b})\Phi(0,t).\nonumber
\end{eqnarray}
Here we have made use of Eq. (\ref{eq:TL_LCcoup_Network_beta_fix})  to substitute $\beta=L_g/l$. Bear in mind that a definite choice for $\alpha$ has been made, $\alpha=C_g(1-C_g\bi{a}^T\mathsf{A}^{-1}\bi{a})/c$. On first sight it might look as if this expression for the Hamiltonian had a major flaw, namely that it explicitly depends on the arbitrary constant $N_\alpha $ we have introduced. In fact, this constant fixes the unit of charge we use, and since only the combination $Q(x,t)/N_\alpha$ appears, there is no free parameter in the Hamiltonian.

\subsection{Canonical variables}
\label{sec:canonical-variables}

In all our analysis we have constructed Lagrangian functions which are quadratic in the derivatives, whence the Hamiltonian is derived in the standard way. As is well known, Hamiltonian dynamics is not fully determined by the Hamiltonian. Additionally the Poisson bracket is necessary to produce the relevant vector field. Starting from the Lagrangian (in general cases) this Poisson bracket is fully determined, and, in most cases, it is taken for granted, since the standard procedure introduces the canonical momenta. We have actually followed this route, in that the Poisson bracket has been systematically taken to be
\begin{equation}
  \nonumber 
  \left\{F,G\right\}=\sum_{i=1}^{N_{\mathrm{disc}}}\left(\frac{\partial F}{\partial \phi_i}\frac{\partial G}{\partial q_i}-\frac{\partial F}{\partial q_i}\frac{\partial G}{\partial \phi_i}\right)
  +\sum_{n=0}^\infty \left(\frac{\partial F}{\partial \Phi_n}\frac{\partial G}{\partial Q_n}-\frac{\partial F}{\partial Q_n}\frac{\partial G}{\partial \Phi_n}\right),
\end{equation}
for $F$ and $G$ functions of the canonical variables, $F\left(\bi{q},\bi{Q};\boldsymbol{\phi},\boldsymbol{\Phi}\right)$, and similarly for $G$. In keeping with the main text, we denote with $\boldsymbol{\phi}$ the variables associated with lumped element networks, and with $\boldsymbol{\Phi}$ those associated to transmission lines. $\bi{q}$ and $\bi{Q}$ are the corresponding canonical moments.

Consider now one transmission line (i.e. one interval of the real line, $I$), with flux field $\Phi(x,t)$ and charge field $Q(x,t)$. They are not canonically conjugate in general:
\begin{equation}
  \nonumber 
  \left\{\Phi(x,t),Q(x',t)\right\}=\sum_{n,m}\left\{\Phi_n(t),Q_m(t)\right\}u_n(x)u_m(x')=\sum_{n} u_n(x)u_n(x').
\end{equation}
When the expansion functions form a real orthonormal basis with respect to the inner product $\langle f,g\rangle=\int_I\mathrm{d}x\bar{f}(x)g(x)$, the expansion theorem can be reexpressed as the equality $\sum_n u_n(x)u_n(x')=\delta(x-x')$, thus proving $\Phi(x,t)$ and $Q(x,t)$ are canonically conjugate in such a case.

Let us assume that $u_n(x)$ are the function components of an orthonormal basis $U_n$ with respect to an inner product of the form we consider, $\left\langle (u,a),(v,b)\right\rangle=\int_I\bar{u}v+ \alpha\bar{a}b$. Then, \emph{formally}, we obtain
\begin{equation}
  \label{eq:deltas}
  \delta(x-x')=\left[1+\alpha\delta(x')\right]\sum_n u_n(x)u_n(x').
\end{equation}
Let us prove this statement. The expansion theorem, in the form presented in \ref{Walter_appendix}, informs us that $f(x)=\sum f_n u_n(x)$, where $f_n=\langle U_n,F\rangle$, and $F=(f(x),f(0))k$ (setting $0$ as the relevant boundary), i.e., $f_n=\int_I f u_n +\alpha f(0)u_n(0)$. Hence
\begin{eqnarray}
  \nonumber 
  f(x)&=& \alpha f(0)\sum_n u_n(0)u_n(x)+\int_I\mathrm{d}x'\,\left[\sum_n u_n(x)u_n(x')\right] f(x')\nonumber,\\
          &=&\int_I\mathrm{d}x'\,\left[1+\alpha\delta(x')\right]\left[\sum_n u_n(x)u_n(x')\right] f(x'),\nonumber
\end{eqnarray}
thus concluding (\ref{eq:deltas}).

Now define, formally,
\begin{equation}
  \nonumber 
  \tilde{Q}(x,t)=\left[1+\alpha\delta(x)\right] Q(x,t).
\end{equation}
By explicit computation one obtains (formally!)
\begin{equation}
  \nonumber 
  \left\{\Phi(x,t),\tilde{Q}(x',t)\right\}= \left[1+\alpha\delta(x')\right]\sum_n u_n(x)u_n(x')=\delta(x-x').
\end{equation}
Prima facie, the variables $\Phi(x,t)$ and $\tilde{Q}(x,t)$ are canonically conjugate, and they could be used for (field) quantization. This is in fact (although the way to this point is very different) what was proposed in \cite{Moein_2016}. 
Note however that one cannot make sense of quantities such as $\tilde{Q}^2$ without some regularisation and prescription, since  $\tilde{Q}$ is only defined in a distributional way. In reference \cite{Moein_2016} this problem is avoided since in fact they can refer back to what we have denoted as untilded charge field.

For reference, let us write the Poisson bracket for $F$ and $G$ functionals of the flux and the charge field:
\begin{eqnarray}
 \nonumber 
\fl \qquad\quad \left\{F,G\right\}&=&\int_I\mathrm{d}x\,\left[\frac{\delta F}{\delta\Phi(x)}\frac{\delta G}{\delta Q(x)}-\frac{\delta F}{\delta Q(x)}\frac{\delta G}{\delta \Phi(x)}\right]\\
\fl                    &&-\alpha\int_I\mathrm{d}x\,\left(\sum_n u_n(x)u_n(0)\right)\left[\frac{\delta F}{\delta\Phi(x)}\frac{\delta G}{\delta Q(0)}-\frac{\delta F}{\delta Q(0)}\frac{\delta G}{\delta \Phi(x)}\right].\nonumber
\end{eqnarray}

\section{Inversion of infinite matrices}
\label{sec:infinite-matrix-inversion}

\subsection{Single port Impedance}
\label{sec:single-port-imped}

A crucial aspect of our analysis, in its different forms, is the inversion of infinite dimensional capacitance matrices, presented in block-matrix format. In the single port case we mostly analyze, in which we couple a transmission line or a more general single port non-dissipative, passive, linear impedance with infinite modes to a network with a finite number of degrees of freedom, the coupling submatrix is of rank one. This is made explicit in our presentation in that we write the capacitance coupling block-matrix as $-C_g\bi{a}\bi{u}^T$ (Eq. (\ref{eq:C_LCcoup_Network})), $-C_A\bi{a}\Delta\bi{u}^T$ (Eq. (\ref{eq:C_TL_LCgalvcoup_Networks})), $-C_g\bi{u}\bi{v}^T$ (Eq. (\ref{eq:C_TL_LCcoup_TL})).  In a manner reminiscent of the Sherman--Morrison formula, we find that for invertible $\mathsf{A}_1$ and $\mathsf{A}_2$ and rank one $\mathsf{D}$ the following generally holds:
\begin{eqnarray}
  \label{eq:sherman-morrison}
\fl \quad\quad\quad \begin{pmatrix}
    \mathsf{A}_1&\mathsf{D}\\ \mathsf{D}^\dag&\mathsf{A}_2
  \end{pmatrix}^{-1}=&
  \begin{pmatrix}
    \mathsf{A}_1^{-1}&0\\0&\mathsf{A}_2^{-1} \end{pmatrix}\\
\fl \quad\quad\quad&+\frac{1}{1-\mathrm{Tr}\left(\mathsf{D}\mathsf{A}_2^{-1}\mathsf{D}^\dag\mathsf{A}_1^{-1}\right)}
  \begin{pmatrix} \mathsf{A}_1^{-1}\mathsf{D}\mathsf{A}_2^{-1}\mathsf{D}^\dag\mathsf{A}_1^{-1}&
    -\mathsf{A}_1^{-1}\mathsf{D}\mathsf{A}_2^{-1}\\  -\mathsf{A}_2^{-1}\mathsf{D}^\dag\mathsf{A}_1^{-1}&  \mathsf{A}_2^{-1}\mathsf{D}^\dag\mathsf{A}_1^{-1}\mathsf{D}\mathsf{A}_2^{-1}
  \end{pmatrix}.\nonumber
\end{eqnarray}
 This formula can be checked directly, making use that for a rank one operator $\mathsf{D}:{\cal V}_1\to{\cal V}_2$, and if the adjoint and the traces exist,
 \begin{equation}
   \nonumber 
   \mathsf{D}\mathsf{A}\mathsf{D}^\dag\mathsf{B}\mathsf{D}= \mathrm{Tr}\left[\mathsf{D}\mathsf{A}\mathsf{D}^\dag\mathsf{B}\right]\mathsf{D}\qquad \mathrm{and}\qquad
   \mathsf{D}^\dag\mathsf{B}\mathsf{D}\mathsf{A}\mathsf{D}^\dag= \mathrm{Tr}\left[\mathsf{D}\mathsf{A}\mathsf{D}^\dag\mathsf{B}\right]\mathsf{D}^\dag,\nonumber
 \end{equation}
 where $\mathsf{A}:{\cal V}_1\to{\cal V}_1$ and $\mathsf{B}:{\cal V}_2\to{\cal V}_2$.

 The condition for (\ref{eq:sherman-morrison}) to hold is that the trace $\mathrm{Tr}\left(\mathsf{D}\mathsf{A}_2^{-1}\mathsf{D}^\dag\mathsf{A}_1^{-1}\right)$ exist and be different from one.

\subsection{Multiport Impedance}
\label{sec:multiport-impedanceapp}

The analogous inversion for the multiport case is necessarily more involved. We have presented an explicit case in section \ref{sec:mult-netw-coupl}, and we look at the more general situation in \ref{sec:multi-port-impedance}. The essential result, as in the Woodbury--Sherman--Morrison case, is that perturbing an invertible operator with an operator of finite rank should produce, for the new inverse, again a perturbation of the same rank.  The (necessarily formal) inversion formula we require goes as follows. Let
\begin{equation}
  \nonumber 
  \mathsf{C}
    \begin{pmatrix}
\mathsf{A}&\mathsf{D}\\ \mathsf{D}^T&\mathsf{C}_1
\end{pmatrix}
\end{equation}
be a real matrix, with $\mathsf{A}$ and $\mathsf{C}_1$ invertible and $\mathsf{D}$ of finite rank. Its inverse if it, together with the correct subelements, exists, is given by
\begin{eqnarray}
  \nonumber 
 \fl 
  \mathsf{C}^{-1}=
                     \begin{pmatrix}
                       \mathsf{A}^{-1}&0\\0&\mathsf{C}_1^{-1}
                     \end{pmatrix}+\\
 \fl \,\,\,\, \begin{pmatrix} \mathsf{A}^{-1}\mathsf{D}\mathsf{C}_1^{-1}\left(\mathbbm{1}-\mathsf{D}^T\mathsf{A}^{-1}\mathsf{D}\mathsf{C}_1^{-1}\right)^{-1}\mathsf{D}^T\mathsf{A}^{-1}& -\mathsf{A}^{-1}\mathsf{D}\mathsf{C}_1^{-1}\left(\mathbbm{1}-\mathsf{D}^T\mathsf{A}^{-1}\mathsf{D}\mathsf{C}_1^{-1}\right)^{-1}\\ -\mathsf{C}_1^{-1}\mathsf{D}^T\mathsf{A}^{-1}\left(\mathbbm{1}-\mathsf{D}\mathsf{C}_1^{-1}\mathsf{D}^T\mathsf{A}^{-1}\right)^{-1}&\mathsf{C}_1^{-1}\mathsf{D}^T\mathsf{A}^{-1}\left(\mathbbm{1}-\mathsf{D}\mathsf{C}_1^{-1}\mathsf{D}^T\mathsf{A}^{-1}\right)^{-1}\mathsf{D}\mathsf{C}_1^{-1}
  \end{pmatrix}.\nonumber
\end{eqnarray}
The proof of this formula is by direct substitution. Both in the infinite and in the finite dimension case, and
as we analyze for multiport impedances in \ref{sec:multi-port-impedance}, the invertibility of the capacitance matrix is conditioned on the existence of the inverse $\left(\mathbbm{1}-\mathsf{D}\mathsf{C}_1^{-1}\mathsf{D}^T\mathsf{A}^{-1}\right)^{-1}$. In the infinite dimension case there can be further subtleties.

This formula reduces to (\ref{eq:sherman-morrison}) when the rank of $D$ is one. To see this one has to realize that $\left(\mathbbm{1}-\mathsf{D}\mathsf{C}_1^{-1}\mathsf{D}^T\mathsf{A}^{-1}\right)^{-1}$ is led by $\mathsf{C}_1^{-1}$, and analogously $\left(\mathbbm{1}-\mathsf{D}\mathsf{C}_1^{-1}\mathsf{D}^T\mathsf{A}^{-1}\right)^{-1}$ only appears preceded by $\mathsf{A}^{-1}$. This allows the reduction to the quantity $\left[1-\mathrm{Tr}\left(\mathsf{D}\mathsf{C}_1^{-1}\mathsf{D}^\dag\mathsf{A}^{-1}\right)\right]^{-1}$.

\section{Zeros and infinities for  $1^{\mathrm{st}}$-Foster Form Impedance}
\label{App:1st_Foster}

We observe that formally the coupling vector $\bi{e}_\alpha$ in the capacitance matrix (\ref{eq:Paladino_Cmat}) has infinite norm when the impedance has to be represented with an infinite set of stages. This by itself is not an issue.  However, in the computation of the inverse capacitance matrix the crucial quantity $\bi{e}_\alpha^T \mathsf{C}_\alpha^{-1} \bi{e}_\alpha$ does appear, and whenever this tends to infinity the inverse is poorly defined, if not altogether nonsensical.

Let us be explicit, and apply to the capacitance matrix
\begin{equation}
\mathsf{C}_{x}=\begin{pmatrix}
a & b\bi{e}_\alpha^T\\
b\bi{e}_\alpha & \mathsf{M}_{\alpha}
\end{pmatrix}.\nonumber 
\end{equation}
with  $\mathsf{M}_{\alpha}=\mathsf{C}_\alpha+d \bi{e}_\alpha\bi{e}_\alpha^T$, the formal inversion formula of Eq. (\ref{eq:sherman-morrison}). It is indeed applicable, since the coupling matrix is indeed rank one. Notice now that the formal expression $\mathrm{Tr}\left(\mathsf{D}\mathsf{A}_2^{-1}\mathsf{D}^{\dag}\mathsf{A}_1^{-1}\right)$ becomes $(b^2/a)\bi{e}_\alpha^T\mathsf{M}_\alpha^{-1}\bi{e}_\alpha$. It is thus incumbent on us to compute the inverse $\mathsf{M}_\alpha^{-1}$. It is clearly of the form $\mathsf{C}_\alpha^{-1}+\gamma\mathsf{C}_\alpha^{-1}\bi{e}_\alpha\bi{e}_\alpha^T\mathsf{C}_\alpha$, where the coefficient $\gamma$ must obey the equation
\begin{equation}
  \nonumber 
  \gamma+d+d \gamma \bi{e}_\alpha^T \mathsf{C}_\alpha^{-1} \bi{e}_\alpha,
\end{equation}
i.e., formally,
\begin{equation}
\label{eq:gamma-value}
  \gamma=\frac{-d}{1+d \bi{e}_\alpha^T \mathsf{C}_\alpha^{-1} \bi{e}_\alpha}.
\end{equation}
Still formally, we see that
\begin{equation}
  \nonumber 
  \frac{b^2}{a}\bi{e}_\alpha^T\mathsf{M}_\alpha^{-1}\bi{e}_\alpha= \frac{b^2}{a}\bi{e}_\alpha^T \mathsf{C}_\alpha^{-1} \bi{e}_\alpha \left(1- \frac{d \bi{e}_\alpha^T \mathsf{C}_\alpha^{-1} \bi{e}_\alpha}{1+d \bi{e}_\alpha^T \mathsf{C}_\alpha^{-1} \bi{e}_\alpha}\right)=\frac{b^2}{a}\frac{\bi{e}_\alpha^T \mathsf{C}_\alpha^{-1} \bi{e}_\alpha}{1+d \bi{e}_\alpha^T \mathsf{C}_\alpha^{-1} \bi{e}_\alpha}.
\end{equation}
This quantity can therefore tend to a finite limit in the pathological situation we consider, namely $b^2/ad$. Notice however that $\gamma$ from Eq. (\ref{eq:gamma-value}) tends to zero. Carrying out with the analysis, the final result is that in the limit in which $\bi{e}_\alpha^T \mathsf{C}_\alpha^{-1} \bi{e}_\alpha$ tends to infinity the inverse capacitance matrix presents no coupling whatsoever between impedance and network models.

Reconsider now the quantity $b^2/ad$. As stated in the previous Appendix \ref{sec:infinite-matrix-inversion}, it has to be different from one for the inversion to be possible, in the pathological situation we consider. Looking back to the original parameters of section \ref{sec:1st-foster-expansion}, in particular the matrix (\ref{eq:Paladino_Cmat}), we have that it reads $C_B/C_\Sigma=C_B/(C_A+C_B)$. If the capacity $C_A$, external to the one port impedance, were 0, the capacitance matrix would not be invertible in the case $\bi{e}_\alpha^T \mathsf{C}_\alpha^{-1} \bi{e}_\alpha\to\infty$. Notice that, in the same limit and for those parameters, we have
\begin{equation}
  \label{eq:1st_Foster_sum_rule_eMe}
  \bi{e}_\alpha^T\mathsf{M}_\alpha^{-1}\bi{e}_\alpha\to \frac{1}{C_B}.
\end{equation}
The result that there is an overcounting of degrees of freedom if $C_A=0$ and $\bi{e}_\alpha^T \mathsf{C}_\alpha^{-1} \bi{e}_\alpha\to\infty$  was implicit in the divergence of the charge energy found in the Hamiltonian in \cite{Gely_2017_DivFree} for the specific case of a transmission line resonator coupled to a charge qubit, when  the capacitance of the Josephson junction $C_J$ was taken to zero. Furthermore, we have shown the relation (\ref{eq:1st_Foster_sum_rule_eMe}) which will be useful in the  derivation of the corresponding Hamiltonian.

\section{$2^{\mathrm{nd}}$-Foster Form Admittance Quantization}
\label{App:2nd_Foster}

In this section, we derive a Hamiltonian of an anharmonic flux variable coupled to an admittance $Y(s)$ decomposed in the $2^{\mathrm{nd}}$-Foster form. This expansion has been widely used to describe the effect of a general environment seen by a harmonic oscillator \cite{Devoret_1995_QFluct}. We study the differences and similarities to the previous section describing the circuit analysed by Paladino et al. \cite{Paladino_2003}. 

\begin{figure*}[ht]
	\centering{\includegraphics[width=0.8\textwidth]{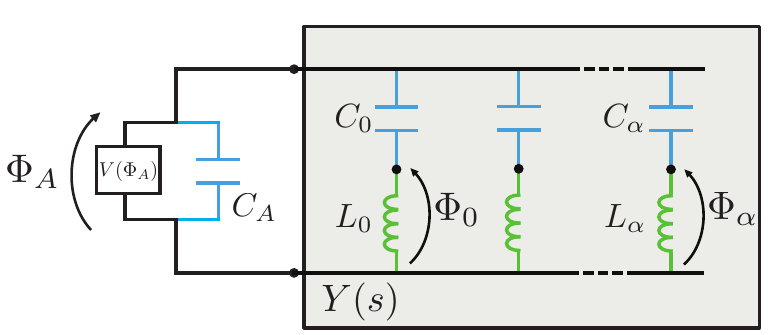}}
	\caption{\label{fig:2nd_Foster_form} \textbf{$2^{\mathrm{nd}}$-Foster form admittance coupled to anharmonic variable}. One port admittance $Y(s)$ modelled by  parallel $L_\alpha,\,C_\alpha$ oscillators is coupled to a flux variable non-anharmonic potential shunted by a capacitance. With the choice of flux variables in the inductors, we derive a Hamiltonian with capacitive coupling.}
\end{figure*}

First of all and contrary to the first Foster form, it must be noticed that the expansion of the admitance in this circuit allows only the description of electromagnetic environments with poles at frequency $s=0$, i.e. $\lim\limits_{s \rightarrow 0}Z(s)=\infty$. Secondly, this configuration has internal variables already separated from the network variables it is connected to, see Fig. \ref{fig:2nd_Foster_form}. Choosing as the internal degrees the flux diferences in the inductors (capacitors) we can derive a Hamiltonian with capacitive (inductive) coupling to the external variables. Here, we perform the analysis with the more cumbersome capacitive coupling in contrast with the inductive coupling done in \cite{Devoret_1995_QFluct}, in order to compare it with the previous calculation of the above section in  Appendix \ref{App:1st_Foster}. 
The Lagrangian of the system at hand can be directly written as
\begin{eqnarray}
L&=& \frac{C_A}{2} \dot{\Phi}_A^2+\sum_{\alpha}\left[\frac{C_\alpha}{2} \left(\dot{\Phi}_A-\dot{\Phi}_{\alpha}\right)^2-\frac{1}{2L_{\alpha}}\Phi_{\alpha}^2\right]-V(\Phi_A)\nonumber\\
&=&\frac{1}{2}\dot{\bi{\Phi}}^T \mathsf{C}\dot{\bi{\Phi}}-\frac{1}{2}\bi{\Phi}^T \mathsf{L}^{-1}\bi{\Phi} - V(\Phi_A),\nonumber
\end{eqnarray}
where $\bi{\Phi}^T=(\Phi_A,\bi{\Phi}_{\alpha}^T)=(\Phi_A, \Phi_0, \Phi_1, ...)$, the capacitance matrix reads
\begin{equation}
\mathsf{C}=\begin{pmatrix}
C_{\Sigma} & -\bi{c}_\alpha^T\\
-\bi{c}_\alpha & \mathsf{C}_{\alpha}
\end{pmatrix},\nonumber
\end{equation}
where we defined $C_{\Sigma}=C_A+\sum_{\alpha}C_\alpha$, $\mathsf{C}_\alpha=\mathrm{diag}(C_1, C_2...)$, the variable $\Phi_A$ couples to $\bi{\Phi}_{\alpha}$ through the vector $\bi{c}_\alpha=(C_1, C_2,...)^T$, and the inductance matrix is 
\begin{equation}
\mathsf{L}^{-1}=\begin{pmatrix}
0 & 0\\
0 & \mathsf{L}_{\alpha}^{-1}
\end{pmatrix}.\nonumber
\end{equation}
with $\mathsf{L}_\alpha^{-1}=\mathrm{diag}(L_1^{-1}, L_2^{-1}...)$. In contrast with the analysis of the $1^{\mathrm{st}}$-Foster expansion, we can directly invert the capacitance matrix
\begin{equation}
\mathsf{C}^{-1}=\begin{pmatrix}
C_A^{-1} & C_A^{-1}\mathbf{1}_\alpha^T\\
C_A^{-1}\mathbf{1}_\alpha & \mathsf{C}_{\alpha}^{-1}+C_A^{-1} \mathbf{1}_\alpha \mathbf{1}_\alpha^T
\end{pmatrix},\nonumber
\end{equation}
where again $\mathbf{1}_\alpha=(1,1,...)^T$ is the vector with the dimension of the Hilbert space describing the admitance to derive a Hamiltonian,
\begin{equation}
	H=\frac{1}{2}\bi{q}^T\mathsf{C}^{-1}\bi{q}+\frac{1}{2}\boldsymbol{\phi}^T\mathsf{L}^{-1}\boldsymbol{\phi}+V(\Phi_A).\nonumber
\end{equation}
Here the conjugate variables to the fluxes are the charges $\bi{q}=\partial L/\partial \boldsymbol{\phi}$. To simplify the analysis, we make a first rescaling of the variables $\bi{p}=(q_A,\bi{p}_\alpha^T)^T=\left(q_A,\left[C_0^{1/2}\mathsf{C}_\alpha^{-1/2}\bi{q}_\alpha\right]^T\right)^T$, with its corresponding change in fluxes $\boldsymbol{\psi}=(\Phi_A,\boldsymbol{\psi}_\alpha^T)^T=\left(\Phi_A,\left[C_0^{-1/2}\mathsf{C}_\alpha^{1/2}\boldsymbol{\phi}_\alpha\right]^T\right)^T$ such that the Hamiltonian transforms into
\begin{equation}
H_I=\frac{1}{2}\bi{p}^T\mathsf{C}_I^{-1}\bi{p}+\frac{1}{2}\boldsymbol{\psi}^T\mathsf{L}_I^{-1}\boldsymbol{\psi}+V(\psi_A),\nonumber
\end{equation}
with 
\begin{equation}
\mathsf{C}_I^{-1}=\begin{pmatrix}
C_A^{-1} & C_A^{-1}\bi{e}_\alpha^T\\
C_A^{-1}\bi{e}_\alpha & \mathsf{M}_\alpha^{-1}
\end{pmatrix},\nonumber
\end{equation}
where the coupling vector $\bi{e}_\alpha=C_0^{-1/2}\mathsf{C}_\alpha^{1/2}\mathbf{1}_\alpha$ and $\mathsf{M}_\alpha^{-1}=C_0^{-1}\mathbbm{1}+C_A^{-1} \bi{e}_\alpha \bi{e}_\alpha^T$. On the other hand, we have a nondiagonal inductance matrix 
\begin{equation}
\mathsf{L}_I^{-1}=\begin{pmatrix}
0 & 0\\
0 & (\mathsf{L}_\alpha^{I})^{-1}
\end{pmatrix}.\nonumber
\end{equation}
with the submatrix $(\mathsf{L}_\alpha^{I})^{-1}=C_0 \mathsf{C}_\alpha^{-1/2}\mathsf{L}_{\alpha}^{-1}\mathsf{C}_\alpha^{-1/2}$. We can diagonalize together the capacitance $\mathsf{M}_\alpha^{-1}$ and inductance $(\mathsf{L}_\alpha^{I})^{-1}$ submatrices with a rescaling and unitary transformation of the charges $\boldsymbol{\rho}=(q_A,\boldsymbol{\rho}_\alpha^T)^T=\left(q_A,\left[M_0^{1/2} \mathsf{U}\mathsf{M}_\alpha^{-1/2}\bi{p}_\alpha\right]^T\right)^T$ and their conjugate fluxes $\boldsymbol{\chi}=(\Phi_A,\boldsymbol{\chi}_\alpha^T)^T$, where $M_0$ is a finite constant with units of capacitance,
\begin{equation}
H_{II}=\frac{1}{2}\boldsymbol{\rho}^T\mathsf{C}_{II}^{-1}\boldsymbol{\rho}+\frac{1}{2}\boldsymbol{\chi}^T\mathsf{L}_{II}^{-1}\boldsymbol{\chi}+V(\Phi_A),\nonumber
\end{equation}
with the final capacitance and 
\begin{eqnarray}
\qquad\qquad\qquad\mathsf{C}_{II}^{-1}&=&\begin{pmatrix}
 C_A^{-1} & C_A^{-1}\bi{f}_\alpha^T\\
C_A^{-1}\bi{f}_\alpha & M_0^{-1} \mathbbm{1}
\end{pmatrix},\nonumber\\ 
\qquad\qquad\qquad \mathsf{L}_{II}^{-1}&=&\begin{pmatrix}
0 & 0\\
0 & (\mathsf{L}_\alpha^{II})^{-1}
\end{pmatrix},\nonumber
\end{eqnarray}
with the coupling vectors are $\bi{f}_\alpha=M_0^{-1/2}\mathsf{M}_\alpha^{1/2}\mathsf{U}^T\bi{e}_\alpha$, and the diagonal inductance submatrix is  $(\mathsf{L}_\alpha^{II})^{-1}=C_0^{-1}\mathsf{U}^{T}\mathsf{M}_\alpha^{1/2}(\mathsf{L}_{\alpha}^{I})^{-1}\mathsf{M}_\alpha^{1/2}\mathsf{U}$. We can rewrite the Hamiltonian as 
\begin{equation}
H=\frac{q_A^{2}}{2C_A}+V(\Phi_A)+\frac{1}{C_A}q_A\sum_{\alpha}f_{\alpha}\rho_{\alpha}+\sum_{\alpha}\frac{1}{2}\left[\frac{\rho_{\alpha}^2}{M_0}+M_0 \Omega_{\alpha}^{2}\chi_{\alpha}^{2}\right],\nonumber
\end{equation}
where we have defined the frequencies $\Omega_\alpha=1/\sqrt{M_0 L_\alpha^{II}}$. Analogously to the analysis of the $1^{\mathrm{st}}$-Foster form, the coupling vector to the variables describing the admitance has finite norm even when the number of harmonic variables tends to infinity, i.e. 
\begin{equation}
	\lim\limits_{|\bi{e}_\alpha|^2\rightarrow\infty}|\bi{f}_\alpha|^2=\lim\limits_{|\bi{e}_\alpha|^2\rightarrow\infty}M_0^{-1}\bi{e}_\alpha^T \mathsf{M}_\alpha \bi{e}_\alpha=C_A/M_0.\nonumber
\end{equation}

\section{Lossless Transmission Line Impedance}
\label{sec:Mport_synthesis}
A 2-port lossless transmission resonator can be characterized by its inductance $l$ and capacitance $c$ per unit length, and its total length $L$. Its impedance matrix 
\begin{equation}
\mathsf{Z}(s)=Z_0\begin{pmatrix}
\mathrm{coth}\left(s\sqrt{lc}L\right) &  \mathrm{csch}\left(s\sqrt{lc}L\right)\\
\mathrm{csch}\left(s\sqrt{lc}L\right) & \mathrm{coth}\left(s\sqrt{lc}L\right)
\end{pmatrix}.\nonumber
\end{equation}
is lossless and positive real (LPR) \cite{Newcomb_1966_LinearMPortSynthesis} in Laplace space. This is a generalization of the Foster reactance-function synthesis for the one-port circuit, and a simplified version of the Brune multiport impedance expansion used by Solgun and DiVincenzo \cite{Solgun_2015}. We can fraction-expand the formulae of the hyperbolic functions
\begin{eqnarray}
\qquad\qquad \coth\left(s\right)=\frac{1}{s}+\sum_{k=1}^{\infty}\frac{2s}{s^{2}+k^{2}\pi^{2}},\nonumber\\
\qquad\qquad \mathrm{csch}\left(s\right)=\frac{1}{s}+\sum_{k=1}^{\infty}\frac{2s(-1)^{k}}{s^{2}+k^{2}\pi^{2}},\nonumber
\end{eqnarray}
and find the decomposition of the impedance 
\begin{equation}
\mathsf{Z}(s)=s^{-1}\mathsf{A}_{0}+\sum_{k}^{\infty}\frac{s\mathsf{A}_{k}}{s^{2}+\Omega_{k}^{2}},\nonumber
\end{equation}
where  we have defined the matrices
\begin{eqnarray}
\qquad\qquad\,\,\, s^{-1}\mathsf{A}_{0}&=&\frac{1}{cL}\begin{pmatrix}
1&1\\
1&1
\end{pmatrix},\label{eq:Mport_synthesis_A0}\\
\qquad\qquad\qquad \mathsf{A}_{k}&=&\frac{2}{cL}\begin{pmatrix}
1&(-1)^{k}\\
(-1)^{k}&1
\end{pmatrix},\label{eq:Mport_synthesis_Ak}
\end{eqnarray}
and the frequencies $\Omega_{k}^2=\frac{k^{2}\pi^{2}}{lcL^{2}}$. Such an expansion is a consequence of $\mathsf{Z}=-\mathsf{Z}^{\dagger}$ and the PR property. Following Sec. (7) in \cite{Newcomb_1966_LinearMPortSynthesis}, it is easy to synthesize a lumped-element circuit that has this impedance to the desired level of accuracy. The matrix (\ref{eq:Mport_synthesis_A0}), which corresponds to the pole at $s=0$, can be decomposed into
\begin{eqnarray}
\qquad\qquad s^{-1}\mathsf{A}_{0}&=& \begin{pmatrix}
1\\
1
\end{pmatrix}\left[\frac{1}{scL}
\right] \begin{pmatrix}
1&1
\end{pmatrix},\label{eq:Mport_synthesis_A0decomp}
\end{eqnarray}
while the matrices (\ref{eq:Mport_synthesis_Ak}) with poles at the frequencies $\Omega_k$ can be expanded in 
\begin{equation}
\frac{s\mathsf{A}_{k}}{s^{2}+\Omega_{k}^{2}}= \begin{pmatrix}
	1\\
	(-1)^{k}
	\end{pmatrix}\left[\frac{2s/cL}{s^{2}+\Omega_{k}^{2}}
	\right] \begin{pmatrix}
	1&(-1)^{k}
	\end{pmatrix}.\label{eq:Mport_synthesis_Akdecomp}
\end{equation}

The first term in (\ref{eq:Mport_synthesis_A0decomp}) is implemented with a Belevitch transformer \cite{Belevitch_1950} of turn-ratios  $\mathsf{T}_{0}=\left[1\,\,1\right]$ and a capacitor of capacitance $C_{0}=cL$. Each term in (\ref{eq:Mport_synthesis_Akdecomp}) is synthesized via transformers $\mathsf{T}_{k}=\left[1\,\,(-1)^{k}\right]$ and a capacitor of capacitance $C_{k}=cL/2$ shunted by an inductor of inductance $L_{k}=2lL/k^2 \pi^2$. Connecting all the stages we finally arrive to the circuit equivalent sketched in Fig. \ref{fig:2CQ2PortTL_v2}.

\end{appendices}

\end{document}